# Risks & Benefits of LLMs & GenAI for Platform Integrity, Healthcare Diagnostics, Financial Trust and Compliance, Cybersecurity, Privacy & AI Safety: A Comprehensive Survey, Roadmap & Implementation Blueprint

*for Automated Review, Compliance Assurance, Moderation, Abuse & Fraud Detection, App Security, and Trust in Digital Ecosystems*

## Kiarash Ahi

*Founder, Virelya Intelligence Research Labs*
*Founder, Creative AI Apps*
*Ph.D., University of Connecticut*
*AI Product Leader, Siemens AG*
ahi@virelya.org | kiarash.ahi@uconn.edu

**Abstract**— Large Language Models (LLMs) and generative AI (GenAI) systems— such as ChatGPT, Claude, Gemini, LLaMA, Copilot, and Stable Diffusion, developed by OpenAI, Anthropic, Google, Meta, Microsoft, and Stability AI, respectively—are profoundly transforming digital platforms, marketplaces, and app ecosystems, while introducing significant challenges for cybersecurity and user privacy and opening new frontiers in high-stakes domains like healthcare diagnostics. This rapid acceleration has driven mobile app submissions from 1.8 million in 2020 to 3.0 million in 2024, with a projected 3.6 million by 2025. However, while empowering innovation, this technological shift presents a critical double-edged sword: concurrently introducing novel and rapidly escalating risks to platform integrity, financial trust and compliance, cybersecurity, user privacy, and opening new frontiers in high-stakes domains like healthcare diagnostics.

Our comprehensive analysis reveals alarming trends across diverse abuse vectors, including a projected surge in LLM-assisted malware from 2% in 2021 to 50% by 2025. We document a nearly tenfold rise in AI-generated Google reviews to 12.21% in 2023, projected to reach 30% by 2025. Additionally, we observe a 456% increase in AI-enabled scam reports and over a 1500% rise in AI-generated misinformation sites over the past year, alongside a projected 900% surge in deepfake fraud by 2025 compared to 2023 levels. In the financial sector, LLM-powered threats like synthetic identity fraud and sophisticated AI-generated scams are rapidly evolving, necessitating advanced defenses. Despite platforms' proactive use of AI to block millions of policy-violating apps and content, the scale and velocity of these threats underscore an urgent and unmet need for scalable integrity infrastructure to safeguard digital security and data privacy. Leading platforms such as Google Play, Apple App Store, Hugging Face Spaces, GitHub Copilot, OpenAI Plugin Stores, TikTok, Facebook, Amazon, Etsy, and Shopify now face unprecedented challenges in maintaining integrity at scale. Similarly, the integration of LLMs into clinical diagnostics presents unique challenges related to diagnostic accuracy, bias, and patient safety, necessitating robust governance.

Drawing on a review of over 400 academic papers, industry reports, and technical documents, this paper presents a comprehensive survey and data-driven analysis of the risks LLMs and GenAI pose to platform integrity and financial trust and compliance, and medical AI safety. Critically, we propose a strategic roadmap framework for using these same technologies to automate review and moderation through semantic code analysis, multimodal storefront validation, and intelligent policy auditing; detect abuse and fraud; enforce compliance across global jurisdictions (e.g., GDPR, CCPA , FinCEN, SEC, MiFID II); and enhance trust, user experience, and safety across digital ecosystems, financial systems, and clinical applications. Unlike prior work focused on isolated technical components or policy domains, our approach outlines a cross-functional architecture that integrates product, engineering, trust & safety, legal, and policy teams to operationalize AI-driven defenses. We ground our analysis in case studies of major platforms—including Google, Apple, Amazon, Meta, and Hugging Face—highlighting deployed LLM-powered systems, practical implementation insights, and lessons learned. Specifically, we examine how leading financial services platforms (e.g., JPMorgan Chase, Capital One, Stripe, Plaid, Revolut) are leveraging LLMs for synthetic identity detection, KYC/AML automation, regulatory parsing, and real-time financial scam detection, including the reported impact of reducing fraud loss rates by up to 21% and accelerating onboarding by 40–60%. Finally, we extend our proposed integrity framework to the domain of clinical diagnostics, introducing a novel multimodal AI system that interprets natural-language patient symptom descriptions using LLMs, aligns them with image-derived biomarkers, and delivers explainable treatment recommendations with physician oversight. We identify actionable best practices and emerging opportunities in explainable AI, federated review pipelines, and multi-agent compliance parsing.

We conclude that LLMs—when deployed with transparent governance and robust evaluation—can serve as a force multiplier for scalable integrity enforcement. To operationalize this vision, we propose Virelya: an envisioned framework and implementation blueprint for high-stakes domains like platform integrity, financial trust, and healthcare diagnostics. Drawing from successful paradigms in Electronic Design Automation (EDA), cybersecurity, and software quality assurance, Virelya is built upon an LLM Design & Assurance (LLM-DA) stack—an independent, cross-domain infrastructure layer for safety verification, compliance-as-code, and responsible deployment. It provides the integrated orchestration, trust, and governance capabilities needed to address the full spectrum of post-deployment challenges, offering features like advanced multi-LLM routing, agentic memory and planning, RAG evaluation, and audit/compliance tracking. This framework provides the operational blueprint for building trustworthy, compliant platforms and clinical systems in the generative AI era.

**Keywords**— Large Language Models (LLMs), Generative AI, Cybersecurity, Platform Integrity, Review Automation, Content Moderation, Abuse Detection, Fraud Prevention, Regulatory Compliance, Trust and Safety, Digital Marketplaces, Privacy, App Ecosystems, Federated Review Systems, Explainable AI, AI Governance, Synthetic Content, Developer Experience.

## Table of Contents

# I. INTRODUCTION

Platform ecosystems are foundational to modern communication, commerce, content creation, and productivity, serving as critical intermediaries in the digital economy [1]. These dynamic environments, encompassing vast marketplaces such as Google Play, Apple App Store, Microsoft Store, and Amazon Appstore, along with rapidly emerging generative AI (GenAI) platforms like Hugging Face Spaces, GitHub Copilot, and OpenAI Plugin Stores, host billions of users and trillions of digital interactions annually [2], [3], [4]. Ensuring trust, platform integrity, and comprehensive regulatory compliance has thus become a paramount and increasingly complex concern for operators of these digital domains [5]. These environments face escalating pressures from multiple fronts: an exponential surge in content and app submissions, the proliferation of increasingly sophisticated abuse strategies, and heightened regulatory scrutiny driven by a growing patchwork of global frameworks such as the General Data Protection Regulation (GDPR) [6], the California Consumer Privacy Act (CCPA) [7], and the Digital Services Act (DSA) [8]. Meeting these multifaceted challenges demands the development and deployment of scalable, intelligent, and adaptive review and governance mechanisms that can keep pace with both rapid technological advancements and the continuously evolving threat landscapes [9], [10], [11].

The advent of Large Language Models (LLMs) and GenAI systems—such as OpenAI's ChatGPT [12], Google's Gemini [13], Meta's LLaMA [14], text-to-image models like DALL·E and Stable Diffusion [15], and code generation tools like GitHub Copilot [16]—is fundamentally transforming this digital landscape. While these powerful models are enabling unprecedented faster development cycles, creative tooling, and highly personalized automation across a myriad of industries and user experiences [17], [18], they simultaneously introduce a significant new class of risks. These novel risks include the generation of insecure and vulnerable code [19], [20], the creation of highly deceptive storefronts and product listings [21], the widespread proliferation of synthetic content (e.g., deepfakes, AI-generated text, fake reviews) designed to mislead or defraud [22], [23], [24], the escalation of sophisticated AI-generated fraud at scale [25], and the development of cunning methods for scalable policy evasion that can bypass traditional detection systems [26]. Consequently, digital platforms, from established app stores to nascent generative AI marketplaces and sophisticated e-commerce sites, now face unprecedented challenges in maintaining platform integrity at scale while fostering innovation [27], [28]. Beyond these digital ecosystems, the profound capabilities of LLMs are also extending into high-stakes domains like healthcare, where the application of AI in clinical diagnostics promises transformative benefits but introduces equally significant safety and ethical considerations, including potential for diagnostic errors, bias, and privacy breaches [29], [30], [31].

This paper critically examines the dual-use nature of LLMs and generative AI within the intricate context of both digital platform integrity and clinical AI safety. We explore how these powerful technologies, while enabling remarkable productivity and creativity, can also be weaponized by malicious actors to undermine trust and safety [32], [33], or, in healthcare, lead to critical misdiagnoses if not governed responsibly. To counteract these emerging threats, we present defensive architectures that strategically leverage LLMs for critical safety operations. These include advanced static code analysis for identifying hidden vulnerabilities and malicious logic [34], sophisticated storefront validation to prevent misrepresentation and misleading claims [21], intelligent content moderation systems capable of discerning subtle policy violations in user-generated content [35], comprehensive compliance auditing against a complex web of global regulations [36], and robust, adaptive abuse and fraud detection mechanisms [25].

In 'Intelligent Policy Auditing,' LLMs can be trained on vast corpuses of legal and regulatory texts (e.g., GDPR, CCPA, DSA provisions) to not only parse and summarize policy documents but also to identify logical inconsistencies, omissions, or misalignments with declared app behaviors. This includes flagging ambiguous clauses, ensuring consistency across various sections, and generating compliance scores by mapping policy statements to established legal obligations.

For 'Multimodal Storefront Validation,' LLMs, in conjunction with other AI models, can cross-reference textual claims in descriptions and privacy policies with visual content (screenshots, promotional videos) and, crucially, with the actual runtime behavior of the app or service. This involves techniques like visual question answering (VQA) on screenshots to verify advertised UI elements, and semantic embedding comparisons between text and extracted features from dynamic analysis logs to detect inconsistencies, moving beyond simple keyword matching to discern semantic deception.

Specifically, LLMs can be fine-tuned to recognize patterns indicative of vulnerabilities, such as insecure data flows or improper API usage, by leveraging their understanding of both syntax and semantic intent. This moves beyond traditional rule-based or signature-driven static analysis by identifying novel attack vectors and polymorphic malware through advanced techniques like graph neural networks on Abstract Syntax Trees (ASTs) or deep learning on bytecode representations to pinpoint semantic vulnerabilities.

Crucially, this paper extends these integrity principles to clinical AI, proposing a novel multimodal system for diagnostics that interprets natural-language patient symptom descriptions using LLMs, aligns them with imaging-derived biomarkers, and delivers explainable diagnostic and treatment recommendations with essential physician-in-the-loop oversight. This application highlights the framework's adaptability to the highest-stakes environments, demanding unparalleled trustworthiness and transparency [37], [38].

We further propose a cross-functional operational model that emphasizes tightly orchestrated collaboration spanning product management, engineering, trust & safety operations, legal, and policy teams. This integrated approach is essential to effectively govern AI-driven workflows and ensure a holistic, proactive defense posture for platform security and broader AI safety [39], [40], [41].

The effectiveness of LLM-powered defenses, particularly for safety and ethical considerations, hinges on robust 'human-in-the-loop' oversight. This operational model demands clearly defined workflows for human review of flagged content, escalation protocols for complex cases, and continuous feedback mechanisms to refine model performance. Challenges include managing reviewer fatigue, ensuring consistency in human judgment, and effectively training human experts to interpret complex AI outputs, requiring a symbiotic relationship between AI and human intelligence.

Drawing from current industry initiatives—including Google's comprehensive SAFE Framework [42], their proactive App Defense Alliance (ADA) [43], and advanced Play Protect real-time scanning systems [44], as well as Apple's innovative LLM-powered review summarization and privacy enhancements [45]—we extract actionable best practices and identify emerging opportunities. These opportunities include advancements in explainable AI (XAI) to provide transparent rationales for moderation decisions [46], the development of federated review pipelines for enhanced privacy and distributed threat intelligence [47], and the implementation of multi-agent compliance parsing systems for adaptive and dynamic regulatory enforcement across diverse jurisdictions [48].

The concept of 'Adaptive Multi-Agent Compliance Parsing' involves deploying a network of specialized LLM-powered agents, each focused on a specific regulatory domain or legal jurisdiction. These agents can collaboratively interpret evolving regulations, detect emerging compliance gaps in real-time, and dynamically update enforcement rules. This decentralized yet coordinated approach allows for more agile responses to global regulatory shifts and facilitates cross-jurisdictional consistency checks.

We argue that with responsible deployment, continuous monitoring, and robust governance frameworks, LLMs can serve as a force multiplier, empowering platforms to scale their enforcement capabilities, counter evolving threats more effectively, and preserve user trust across rapidly advancing app ecosystems, generative AI marketplaces, and broader digital commerce platforms [49], [50]. While individual LLM-based integrity mechanisms have been studied in isolation [51], this paper provides the first unified, cross-platform roadmap—spanning app stores, generative marketplaces, digital commerce, and clinical diagnostics—for operationalizing LLMs in platform governance and high-stakes AI applications. This integrated approach highlights how shared challenges and solutions can be applied across diverse digital environments to foster a safer, more trustworthy, and compliant online experience for billions of global users [52]. Prior research often focuses on isolated domains—such as mobile app moderation, generative AI safety, or content review within e-commerce. This paper differentiates itself by synthesizing these traditionally siloed perspectives into an integrated blueprint. By aligning governance challenges and mitigation strategies across app stores, generative AI plugin marketplaces, digital commerce platforms, and the emerging field of AI-assisted medicine, it provides a unified model for AI-driven platform integrity. This level of cross-platform synthesis, paired with a cross-functional team model, has not been articulated in prior literature.

The development of this strategic roadmap framework was informed by a multi-faceted methodology, combining an extensive systematic literature review of academic and industry publications, in-depth analysis of major platform transparency reports and security initiatives, and synthesis of best practices derived from expert interviews with product leaders, engineering heads, and trust & safety professionals across leading digital ecosystems.

The remainder of this paper is structured as follows: **Section II** elaborates on the dual-use nature of LLMs, detailing both their transformative benefits and the novel risks they introduce across digital platforms and in sensitive applications like healthcare. **Section III** provides a comprehensive analysis of key threat vectors and security risks directly resulting from LLM-assisted development and content creation. In **Section IV**, we delve into the defensive applications of LLMs for reviewer automation and platform integrity, covering techniques like semantic code analysis and multimodal cross-validation. **Section V** outlines the critical role of cross-functional collaboration for platform integrity: building safe apps and digital ecosystems, emphasizing the integration of product, engineering, trust & safety, legal, and policy teams. **Section VI** presents in-depth case studies of major industry initiatives by platforms such as Google, Apple, Amazon, Meta, and Hugging Face, showcasing real-world LLM integration for integrity. **Section VII** discusses future directions and research opportunities, while **Section VIII** addresses the limitations and ongoing challenges in this evolving field. **Section IX** outlines the strategic landscape of the LLM ecosystem. **Section X** proposes the LLM Design & Assurance (LLM-DA) Stack as a cross-domain blueprint for responsible AI infrastructure. **Section XI** then extends this integrity framework to the critical domain of clinical diagnostics, detailing its application in multimodal mapping, diagnostic suggestions, and specialized governance. Finally, **Section XII** provides the conclusion and summarizes our key findings.

---

## II. THE DOUBLE-EDGED SWORD OF LLMs IN DIGITAL PLATFORMS AND APP ECOSYSTEMS

The proliferation of Large Language Models (LLMs) and generative AI systems has fundamentally reshaped the digital landscape, significantly accelerating app development by democratizing access to sophisticated software creation. Tools such as OpenAI's ChatGPT [12], Google's Gemini [13], and Microsoft's Copilot [16] enable even non-experts to generate complete app codebases, design user interfaces, craft compelling storefront content, draft privacy policies, and produce extensive marketing copy [17], [53]. This accessibility has undeniably driven a remarkable surge in the volume and diversity of app submissions across major platforms like Google Play and Apple App Store [54], [55]. Similar transformative trends are rapidly emerging across broader digital ecosystems, including specialized LLM plugin stores, burgeoning Gen-AI marketplaces (e.g., Hugging Face Spaces, OpenAI Plugin Stores), and traditional e-commerce platforms (e.g., Amazon, Etsy, Shopify) [56], [57].

In these environments, AI-generated content and automation workflows accelerate user-generated storefronts, service listings, and digital products, often with limited human oversight [58]. As indicated in Table 1 and illustrated in Fig. 1 (a), the number of mobile app submissions has shown a sharp upward trajectory. Starting from 1.8 million in 2020 and reaching 2.0 million in 2022, a notable acceleration was observed following the widespread introduction of LLM-based developer tools [54]. Submissions surged to 2.4 million in 2023, and further to 3.0 million in 2024, with a projected 3.6 million in 2025. This significant increase, particularly evident in the jump from 2.0 million in 2022 to a projected 3.6 million in 2025, highlights a paradigm shift from traditional, human-intensive development to highly automated, AI-assisted creation. This accessibility has not only driven a remarkable surge in the volume and diversity of app submissions, but also significantly reduced the time and expertise required for prototyping, iterating, and even multi-platform deployment. The rise of low-code/no-code paradigms, greatly enhanced by LLMs, further amplifies this acceleration and shifts development power into the hands of a broader user base [406]. In parallel, as indicated in Table 2 and illustrated in Fig. 1 (b), the share of malware generated by LLMs has grown from 2% in 2021 to a projected 35% by 2025, signaling an alarming trend in scalable, AI-powered cyber threats [74], [407].

*Table 1. Growth of mobile app submissions before and after the emergence of LLM-enhanced development tools.*

| Year | App Submissions (millions) |
|------|----------------------------|
| 2020 | 1.8 |
| 2021 | 1.9 |
| 2022 | 2.0 |
| 2023 | 2.4 |
| 2024 | 3.0 |
| 2025 (Projected) | 3.6 |

*Table 2. Annual malware detections (2021–2024) based on AV-TEST data and projected estimate for 2025. The table includes estimated counts and percentages of LLM-assisted malware, illustrating their accelerating share of global threats due to increased adoption of generative AI in cyberattacks.*

| Year | Annual Malware Detections (M) | LLM-Assisted Malware (%) | LLM-Assisted Malware (M) |
|------|-------------------------------|--------------------------|--------------------------|
| 2021 | 83.3 | 2 | 1.666 |
| 2022 | 104.5 | 5 | 5.23 |
| 2023 | 150.0 | 15 | 22.5 |
| 2024 | 184.3 | 30 | 55.29 |
| 2025 | 221.2 | 50 | 110.6 |

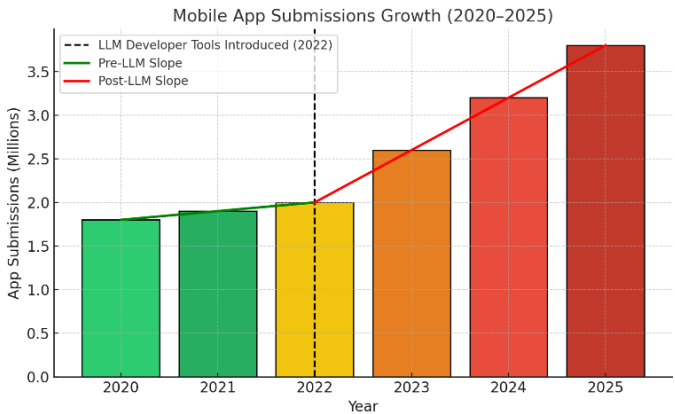

*(a)*

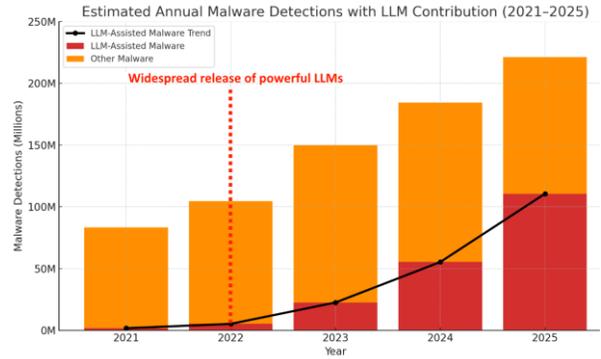

*(b)*

*Fig. 1. (a) Growth of Mobile App Submissions from 2020 to 2025, highlighting acceleration after LLM-based developer tools introduction. (b) Estimated annual global malware detections with LLM-assisted contribution (2021–2025). Stacked bars show total malware cases, with the red portion representing LLM-assisted threats. The black line highlights the rapid growth of AI-driven malware, rising from 2% to 50% of all detections over the five-year period.*

Accordingly, this profound empowerment represents a double-edged sword. While fostering unprecedented innovation, LLMs simultaneously lower the barriers for malicious actors to operate and scale abuse across digital platforms and app stores [32], [33], [66]. The very same generative capabilities that streamline legitimate application development can be repurposed by bad actors to create sophisticated polymorphic malware that evades traditional signature-based detection, generate highly deceptive storefronts designed to trick users, produce non-compliant privacy policies that mask illicit data practices, and forge convincing social engineering interfaces at an unprecedented scale [19], [22], [26].

While LLM-generated content existed in smaller capacities prior, a dramatic shift in the landscape of digital abuse became evident starting in late 2022. For instance, *as indicated in Table 3 and illustrated in Fig. 2,* the percentage of AI-generated Google reviews, which stood at a mere 1.42% in 2022, saw an explosive surge coinciding with the widespread release of powerful Large Language Models like ChatGPT. This figure jumped nearly tenfold to 12.21% in 2023 and is projected to reach 19% by the end of 2024 [59]. Using polynomial or exponential regression based on this trend, a reasonable projection for 2025 is in the range of 27%–30%. This rapid acceleration underscores how the accessibility and sophistication of generative AI tools quickly provided malicious actors with unprecedented capabilities to automate the creation of deceptive content at scale across digital platforms.

*Table 3. Share of Google reviews likely AI-generated from 2021 to 2025 (2025 projected). A sharp rise begins in 2023 with the release of LLMs like ChatGPT, with 2025 estimates reaching up to 30%.*

| Year | % of Google Reviews Likely AI-Generated |
|------|------------------------------------------|
| 2021 | 0.9 |
| 2022 | 1.42 |
| 2023 | 12.21 |
| 2024 | 19 |
| 2025 (projected) | 27%–30% |

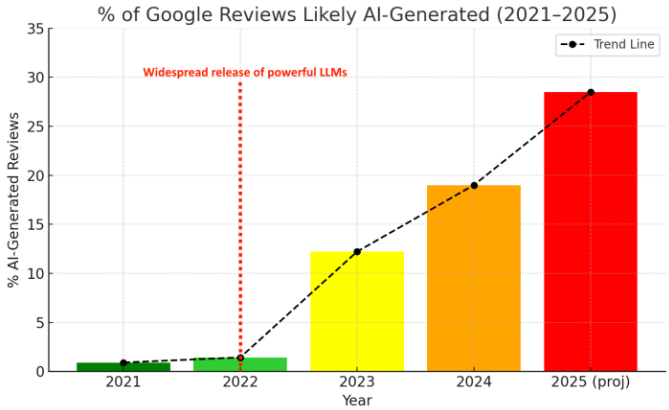

*Fig. 2. Percentage of Google reviews likely generated by AI from 2021 to 2025 (2025 projected). The data shows a marked surge beginning in 2023, coinciding with the public release of powerful LLMs like ChatGPT. The projected trend for 2025 suggests continued growth in synthetic review content, reaching as high as 30%.*

Beyond reviews, as indicated in Table 4 and illustrated in Fig. 3, the scope of LLM/GenAI-assisted digital abuse is rapidly expanding. For instance, AI-enabled scam reports saw a 456% increase between May 2023 and April 2024 [60], while AI-generated email threats surged 31-fold (over 3,000%) in 2023 alone [61]. Deepfake attacks are also projected to increase by more than 900% in 2025 compared with 2023 levels [62], [63]. Furthermore, the proliferation of AI misinformation sites has been dramatic, with over a 1,500% increase observed from May 2023 to April 2024 [64]. These trends, as summarized in Table 4, highlight a systemic expansion of AI-facilitated malicious activities across various digital domains.

*Table 4. Key Trends in LLM/GenAI-Assisted Digital Abuse*

| Abuse Type | Key Trend / Statistic | Source |
|------------|----------------------|--------|
| **Fake Reviews** | 3,333% increase in AI-generated Google reviews (2019-2025) | Originality.ai [59] |
| **AI-Enabled Scams** | 456% increase in scam reports (from May 2023 to April 2024) | TRM Labs [60] |
| **AI-Generated Email Threats** | 31-fold surge (or 3,000%+ increase) in 2023 | Trend Micro [61] |
| **Deepfake Attacks** | Projected 900%+ increase in global incidents (2024) | Deep Instinct [62], Sumsub [63] |
| **AI Misinformation Sites** | Over 1500% increase in AI news sites (from May 2023 to April 2024) | NewsGuard [64] |

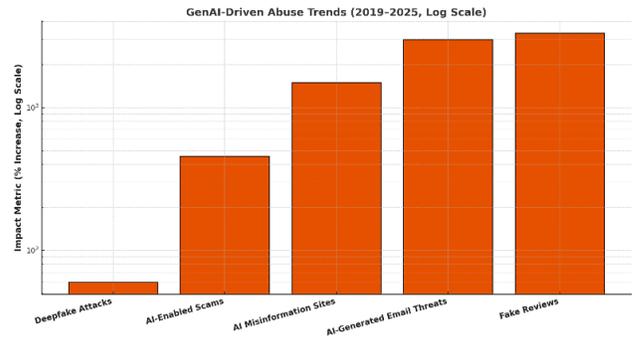

*Fig. 3. Log-scaled comparison of GenAI-driven abuse trends (2019–2025), covering AI-generated reviews, scams, deepfake incidents, misinformation sites, and email threats. Logarithmic scale highlights large disparities across abuse types.*

This inherent dual-use nature of LLMs—enabling both legitimate innovation and scalable abuse—is conceptually illustrated in Fig. 4 [65].

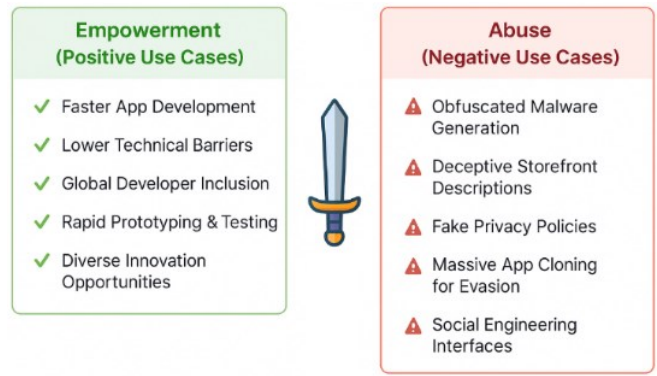

*Fig. 4. Dual-use nature of LLMs in app ecosystems: enabling both innovation and abuse at scale*

This section delves into the intricate dual impact of LLMs on app and digital platform ecosystems, emphasizing how they simultaneously enable widespread innovation while profoundly multiplying existing and introducing new vectors of risk.

## A. Lowered Barriers to App Development (Opportunity and Risk)

By profoundly simplifying complex tasks such as code generation, user interface (UI) design, content creation, and legal policy drafting, LLMs significantly lower the technical and operational threshold for individuals and organizations to conceptualize, build, and deploy digital applications [17], [18]. This democratization of sophisticated software creation fuels unprecedented innovation, fosters a greater diversity of digital offerings, and promotes global inclusion by enabling entrepreneurs from diverse backgrounds to participate in the digital economy [67].

Beyond solely security and compliance, LLM-driven review automation can also significantly enhance the legitimate developer experience by providing faster app approval times, clearer and more contextualized feedback on policy violations, and even proactive suggestions for improving security posture. However, this also necessitates careful design to minimize false positives and avoid creating unnecessary friction or frustration for developers due to opaque AI decisions. New startups and independent developers can now prototype, launch, and iterate products at record speed, contributing to increasingly vibrant and competitive app marketplaces [53], [68].

Yet, the inherent downside of this accessibility is critically important: developers, particularly those lacking deep expertise in cybersecurity, data privacy, or complex regulatory compliance, may inadvertently introduce significant vulnerabilities into their AI-generated or AI-assisted applications [19], [69], [70]. Common security pitfalls include missing or insufficient input validation, insecure data storage practices (e.g., unencrypted user data on device or servers), improper permission management (e.g., requesting excessive permissions beyond necessary app functionality), and the blind exposure to or integration of unsafe third-party SDKs without proper vetting [16], [71]. These issues often stem from an over-reliance on LLM outputs without critical human review, validation, or an understanding of underlying security principles [72]. Table 5 summarizes common vulnerabilities typically introduced by inexperienced LLM-driven development efforts.

*Table 5. Common vulnerabilities introduced by inexperienced LLM-driven app development.*

| Vulnerability | Description |
|---|---|
| **Missing Input Validation** | LLMs may generate input forms without enforcing sanitization, enabling injection attacks. |
| **Insecure Data Storage** | User data may be stored unencrypted on the device or servers. |
| **Improper Permission Management** | LLMs may request broad permissions beyond necessary app functionality. |
| **Unsafe SDK Integrations** | LLMs suggest popular SDKs without checking their privacy or security history. |
| **Exposure of Hardcoded Credentials** | Credentials accidentally embedded in code, making them retrievable. |
| **Lack of Rate Limiting** | Generated apps might omit throttling protections for APIs or login attempts. |
| **Weak Authentication Logic** | Simplistic authentication that fails to prevent unauthorized access. |
| **Unsecured API Communication** | HTTP communication without TLS, risking man-in-the-middle attacks. |

Beyond these technical vulnerabilities, the sheer volume and velocity of app submissions—a direct consequence of LLM-assisted development—significantly strain traditional, often manual, app review pipelines [54]. This immense burden makes it increasingly challenging for platforms to maintain consistent quality and safety standards, potentially leading to a higher incidence of harmful applications reaching end-users [73].

### B. New Vectors for Abuse and Vulnerabilities

Sophisticated threat actors are quickly adapting to the capabilities of LLMs, leveraging them to industrialize the creation and deployment of harmful applications and deceptive content [32],

[74]. This marks a significant shift from manual, labor-intensive abuse campaigns to highly automated, scalable operations. Common abuse patterns and novel vulnerabilities enabled by LLMs include:

**Obfuscated Malware Generation:** LLMs can rapidly create numerous code variants, including **polymorphic malware** that constantly changes its signature, making it exceedingly difficult for traditional static signature detection systems to identify [10], [75]. They can also assist in generating highly obfuscated malicious logic that blends seamlessly with benign code [76].

**Deceptive Storefront Content and Synthetic Media:** LLMs excel at crafting persuasive and legitimate-looking app descriptions, screenshots, promotional videos, and privacy policies that effectively mask underlying malicious behavior or misrepresent app functionality [21], [22]. Furthermore, generative AI can produce synthetic content such as deepfake videos of fake testimonials, AI-generated positive reviews, or fabricated user interfaces to enhance deception [23], [77].

**Automated Policy Circumvention:** Malicious actors can leverage LLMs to generate multiple subtle variations of an app or content, designed to bypass review heuristics or automated moderation systems. This "evasion by mutation" strategy makes it difficult for platforms to enforce policies consistently across a large volume of submissions [26], [78].

**Scalable Social Engineering:** LLMs can generate highly personalized and convincing social engineering interfaces, prompts, and phishing attempts within apps or content. They can simulate user interaction patterns, allowing malicious apps to feign normal behavior during dynamic analysis, thereby evading behavioral detection [25], [79].

Fig. 5 summarizes the principal types of LLM-enabled abuse tactics observed across app ecosystems, highlighting their strategic deployment against platform integrity.

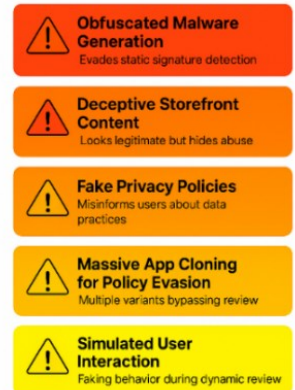

*Fig. 5. Common LLM-enabled abuse tactics observed in app ecosystems.*

These evolving dynamics necessitate a fundamental shift in how platform operators approach **trust and safety**. Traditional manual, rule-based, or simplistic signature-based review strategies are increasingly inadequate [73]. Instead, platforms must transition toward **AI-augmented, adaptive enforcement models** that can understand semantic intent, detect subtle behavioral anomalies, and continuously learn from new attack vectors [9], [28]. Beyond mobile applications, the risks extend widely: LLMs have been observed creating fake e-commerce storefronts on platforms like Amazon and Shopify [80], generating misleading LLM plugins on marketplaces such as OpenAI Plugin Stores and Hugging Face Spaces [56], and fabricating synthetic service listings that mimic legitimate patterns to evade review [57]. The capabilities of LLMs can also be turned defensively to audit complex artifacts like model cards, datasets, and documentation within platforms like Hugging Face Model Hub, helping to surface inconsistencies between claimed capabilities and actual behavior, as well as detecting policy-violating content embedded in training data or generated output samples [81]. This underscores the urgent need for a proactive and adaptable AI-driven defense [7], [8].

### C. How Platforms Can Harness LLMs Defensively

Rather than viewing LLMs solely as a source of amplified risk, platforms can strategically utilize the same advanced generative and analytical capabilities for defense, transforming them into powerful tools for **scaling trust and safety operations** [34], [49]. This proactive approach involves leveraging LLMs to augment human reviewers and automate detection processes, thereby serving as a critical force multiplier in the ongoing fight against platform abuse [28], [35]. Table 6 outlines major defensive capabilities that platforms can adopt by strategically leveraging LLM technologies.

Early initiatives by leading platforms such as Google and Apple clearly demonstrate the practical viability and significant impact of LLM and AI integration [42], [45], with Google, for example, further expanding its AI-powered, on-device scam detection in Messages and real-time app scanning capabilities through Google Play Protect in 2025 to counter evolving threats [82]. When properly trained, fine-tuned, and deployed with robust human oversight and feedback loops, LLMs can serve as a potent force multiplier for scaling trust and safety operations far beyond the capabilities of traditional methods [35], [52]. This strategic shift from reactive to proactive defense is crucial for maintaining the integrity and trustworthiness of digital ecosystems.

*Table 6. Defensive use cases for LLMs in app store safety operations.*

| LLM-Powered Capability | Strategic Benefit |
|---|---|
| **Static Code Analysis for Hidden Threats** | Detect polymorphic and obfuscated malware faster than traditional static analysis. |
| **Dynamic Behavior Monitoring** | Identify runtime evasion tactics and suspicious dynamic behaviors. |
| **Cross-Validation of Storefront and App Behavior** | Ensure consistency between what the app claims and what it actually does. |
| **Privacy Policy Auditing** | Spot missing disclosures and inconsistent data usage declarations. |
| **User Review Summarization and Abuse Detection** | Prioritize real-world abuse cases from user feedback at scale. |
| **Compliance Check Against Regional Regulations** | Automate GDPR, CCPA, and DSA compliance enforcement. |
| **Risk-Based Developer Trust Scoring** | Fast-track trusted developers, slow down high-risk new submitters. |

In the following sections, we will delve deeper into how LLMs can be strategically embedded into various stages of the app review lifecycle, from initial submission and static/dynamic analysis to continuous post-deployment monitoring. We will also explore their role in enhancing regulatory compliance enforcement, improving the developer experience through clearer communication, and refining sensitive content detection workflows, ultimately contributing to a safer digital environment for all users.

### III. Threat Vectors and Security Risks Introduced by LLM-Based Development

The introduction of LLMs into mobile app development pipelines has undeniably unlocked significant productivity gains and fostered unprecedented innovation [17], [53]. However, this transformative shift has simultaneously ushered in a new class of security, privacy, and compliance threats across app ecosystems and digital platforms [22], [26], [74]. These risks emerge not only from malicious actors deliberately abusing LLMs to generate harmful content or applications, but also from inexperienced or unwitting developers who inadvertently introduce vulnerabilities through automated code and content generation [72]. This section systematically analyzes the primary threat vectors and security risks directly resulting from LLM-assisted app development and content creation across various digital platforms.

Beyond intentional maliciousness, a significant, unintentional risk stems from LLMs' propensity for 'hallucination'—generating plausible-sounding but factually incorrect or nonsensical outputs. In the context of security, this can lead to the production of seemingly correct but critically flawed code, erroneous policy documentation, or deceptive content even by well-meaning developers, necessitating robust validation mechanisms and meticulous human-in-the-loop oversight to prevent the propagation of deep and subtle vulnerabilities or misrepresentations.

This section systematically analyzes the primary threat vectors and security risks directly resulting from LLM-assisted app development and content creation across various digital platforms.

### A. Insecure AI-Generated Code

One of the most significant and insidious threats introduced by LLMs is their propensity to generate **syntactically correct but often semantically insecure code** [19], [83], [84]. Developers relying heavily on model outputs, especially without deep security expertise or rigorous manual review, may inadvertently ship applications with critical vulnerabilities. These vulnerabilities include:

**Missing or Inadequate Input Validation**: LLMs may generate code that fails to properly sanitize or validate user inputs, creating fertile ground for various **injection attacks** (e.g., SQL Injection, Cross-Site Scripting (XSS), command injection) that can lead to data breaches or remote code execution [85], [86], [87].

**Improper Authentication and Authorization Flows**: AI-generated code might implement weak or flawed authentication mechanisms (e.g., simplistic password checks, insecure token handling) or incorrect authorization logic, enabling unauthorized access, privilege escalation, or account hijacking [88], [89].

**Hardcoded API Keys or Sensitive Credentials**: A common vulnerability is the accidental embedding of sensitive information, such as API keys, cryptographic secrets, or database credentials, directly within the application's source code, making them easily retrievable by attackers [90], [91].

**Excessive Permission Requests**: LLMs might suggest or generate code that requests broad, unnecessary device permissions (e.g., background location, microphone access, contact list) beyond what the app's core functionality requires, significantly expanding the attack surface and raising privacy concerns [69].

**Embedding Vulnerable Dependencies**: While LLMs can suggest popular third-party libraries and SDKs, they may not adequately vet these dependencies for known vulnerabilities (CVEs) or their privacy implications, thereby introducing supply chain risks into the application [71], [92], [93].

**Insufficient Data Encryption**: Generated code might neglect to implement robust encryption for user data at rest (on device or servers) or in transit, leaving sensitive information exposed to theft or unauthorized access [94].

**Lack of Rate Limiting and Brute-Force Protections**: Automated generation might omit crucial security measures like rate limiting for API endpoints or login attempts, making applications susceptible to brute-force attacks or denial-of-service (DoS) attempts [95].

**Unsecured API Communication**: LLMs might generate code that uses insecure communication protocols (e.g., HTTP instead of HTTPS/TLS) for transmitting sensitive data to backend servers, making the app vulnerable to man-in-the-middle (MITM) attacks and data interception [96].

A recent empirical study by Pearce et al. [16] found that a significant portion (40%) of code generated by GitHub Copilot contained security vulnerabilities, underscoring the urgent need for app stores and platforms to augment traditional static and dynamic analysis with more intelligent, LLM-aware detection capabilities to identify risks stemming from AI-produced software artifacts [19], [72]. Table 7 summarizes frequent coding risks and their potential impacts resulting from insecure LLM-generated code.

*Table 7. Common security vulnerabilities introduced by LLM-assisted code generation.*

| Vulnerability | Impact |
|---|---|
| **Missing Input Validation** | Enables injection attacks (SQLi, XSS) and other input manipulation exploits. |
| **Improper Authentication Flows** | Allows unauthorized access or privilege escalation by attackers. |
| **Hardcoded API Keys or Credentials** | Leaks sensitive credentials, enabling service hijacking or impersonation. |
| **Excessive Permission Requests** | Expands the attack surface by exposing unnecessary device capabilities. |
| **Embedding Vulnerable Dependencies** | Introduces known vulnerabilities into the app supply chain. |
| **Insufficient Data Encryption** | Exposes user data at rest to theft or unauthorized access. |
| **Insecure API Communication** | Risks man-in-the-middle attacks due to unprotected data transmission. |

## B. Storefront Misrepresentation and Synthetic Content

App metadata, including titles, descriptions, screenshots, promotional videos, and privacy labels, is increasingly being generated or enhanced by LLM and GenAI systems. While this can significantly improve marketing quality and reach for legitimate developers, it simultaneously enables deceptive actors to craft highly convincing and misleading storefronts at an unprecedented scale and speed [21], [57], [98]. Typical abuse patterns enabled by such generative capabilities include:

**False Advertising of Features or Functionality**: Apps might claim "no ads" while embedding aggressive advertising SDKs, or market themselves as "offline-only" while requiring persistent internet access for core features [97].

**Misleading Privacy Claims**: Developers might use LLMs to generate privacy labels that claim "no data collection" or "HIPAA-compliant" (for health apps) without any actual evidence or adherence to such standards, thereby deceiving users about data handling practices [14], [99].

**Fake Testimonials and Reviews**: LLMs can produce realistic-sounding positive reviews and user testimonials, inflating an app's perceived quality or trustworthiness and manipulating user acquisition [23], [100].

**Synthetic Visuals and Videos**: Generative AI models can create highly polished but entirely fabricated screenshots or promotional videos that misrepresent the app's actual user interface or functionality, leading to "bait-and-switch" scenarios [77], [101].

Fig. 6 visualizes examples of storefront misrepresentation patterns and their associated real risks, highlighting the various deceptive tactics employed by malicious actors. Detecting such inconsistencies requires sophisticated multimodal cross-validation of app behavior against storefront claims, moving beyond simple keyword matching to semantic understanding and behavioral analysis [21], [97]. This requires platforms to correlate information from various sources (text, images, code) to build a coherent understanding of an app's true nature.

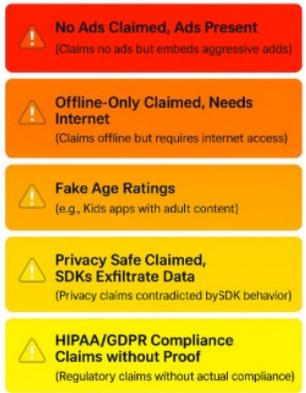

*Fig. 6. Common patterns of storefront misrepresentation enabled by LLMs.*

## C. Fake or Inconsistent Policy Documentation

The ability of LLMs to rapidly generate legal and compliance documentation, such as privacy policies, terms of service, and end-user license agreements (EULAs), is a double-edged sword [26], [102]. While this can assist developers in drafting complex documents quickly, it often results in policies that are either misaligned with actual app behaviors or are deliberately deceptive [14], [103]. Common issues arising from LLM-generated policy documentation include:

**Omission of Sensitive Data Collection**: Policies may explicitly state "no data collection" or fail to mention the collection of sensitive data categories (e.g., precise location, contacts, microphone audio, health data, children's data), even when the app's code actively requests and transmits such information [14], [104].

**Non-Disclosure of Third-Party SDKs and Data Sharing**: LLM-generated policies frequently omit any mention of third-party SDKs (e.g., ad networks, analytics providers, payment processors) embedded within the app that collect or share user data, creating a critical transparency gap [16], [105].

**Vague or Contradictory Language**: Policies might use overly generic or ambiguous language regarding data usage, sharing, or retention, making it difficult for users to understand their rights or for regulators to assess compliance. In some cases, different sections of the same document or linked documents may contain conflicting statements [103].

**Incorrect User Consent Flow Explanations**: Policies might describe robust user consent mechanisms or opt-out options (e.g., "users can easily opt out of data sharing") that are not actually implemented or are intentionally obfuscated within the app's user interface [106].

Table 8 provides real-world examples of common mismatches between AI-generated privacy policies and actual app behaviors, illustrating the challenges this poses for user privacy and regulatory oversight. Traditional manual policy review methods are fundamentally inadequate for catching such sophisticated and scalable issues. LLM-driven semantic analysis of policies, cross-referenced with dynamic analysis of app code and network behaviors, is becoming increasingly necessary to identify these inconsistencies at scale [36], [103].

*Table 8. Common inconsistencies between AI-generated privacy policies and actual app behaviors.*

| Mismatch Type | Example/Impact |
|---|---|
| Privacy Policy Omits Data Collection | Policy claims 'no data collection' but app requests contacts, location, microphone access. |
| Privacy Policy Omits Third-Party SDK Usage | Policy fails to mention ad networks or analytics SDKs embedded within the app. |
| Vague Language on Data Sharing | Policy says 'may share information' without specifying purposes or recipients. |
| Incorrect User Consent Flow Explanation | Policy claims users can opt out easily, but no in-app mechanism is provided. |
| Non-Disclosure of Sensitive Data Categories | Policy ignores sensitive data collection like health data, children's data, or financial info. |
| Conflicting Statements Across Documents | Terms of Service says one thing, while Privacy Policy contradicts it, creating compliance risks. |

## D. Unsafe SDK Integrations and Polymorphic Abuse

LLMs, when prompted for coding solutions or feature implementations, may suggest or integrate third-party SDKs that are either inherently risky, non-compliant with privacy regulations, aggressively monetizing user data, or vulnerable to supply chain attacks [16], [92]. Without adequate developer scrutiny, these SDKs can be blindly embedded, exposing applications to cascading privacy and security risks.

Many modern apps embed numerous third-party libraries for a variety of functionalities, including analytics, advertising, payments, social media integration, or user engagement. When LLMs are queried for "best SDKs" or "how to implement X feature" without proper security or privacy context, they may suggest libraries that are:

**Non-compliant with Privacy Regulations**: Exporting user data to restricted regions, failing to provide proper consent mechanisms, or collecting data beyond the scope of a privacy policy [69], [105].

**Aggressively Monetizing User Data**: SDKs designed primarily for advertising or data brokerage may excessively collect and sell user data without clear disclosure or control, leading to user backlash and regulatory fines [99], [107].

**Vulnerable to Supply Chain Attacks**: Obsolete, poorly maintained, or maliciously tainted SDKs can act as a backdoor, allowing attackers to inject malware, exfiltrate data, or compromise the app's integrity [92], [108].

Fig. 7 illustrates how unsafe SDK integrations can expose apps to cascading privacy and security risks, creating a complex web of dependencies that are difficult for platforms to fully audit manually. Platforms must develop robust capabilities to analyze apps' SDK dependencies carefully and warn developers about risky integrations. This task, given the sheer volume of apps and SDKs, can only be scaled effectively using LLM-assisted manifest and binary analysis tools that can identify embedded libraries, their declared permissions, and their observed network behaviors [71], [109].

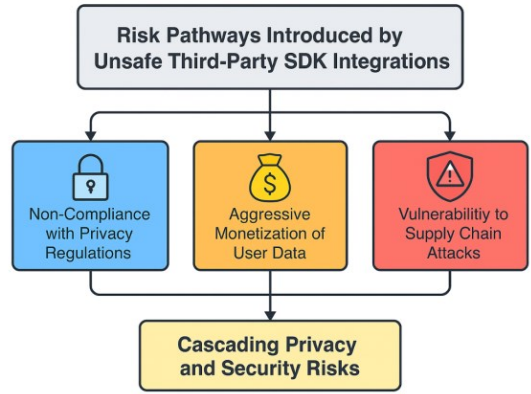

*Fig. 7. Risk pathways introduced by unsafe third-party SDK integrations.*

### E. Abuse Scaling with Polymorphic Variants

A particularly challenging threat vector enabled by LLMs is the ability for threat actors to generate **polymorphic app variants** [17], [74]. This involves subtly modifying code, permissions, metadata, or content across multiple app submissions to evade static detection signatures and traditional rule-based heuristics [10]. These variants may appear slightly different in their user interface (UI) or user experience (UX) but preserve their malicious core functionality, making them highly effective at bypassing initial automated checks and exhausting human reviewer capacity.

For instance, an attacker could use an LLM to:

Automate code obfuscation techniques, producing millions of unique malware samples from a single malicious payload [76].

Generate slightly different app descriptions and screenshots for functionally identical fraudulent apps, preventing their detection based on visual or textual similarity alone [98].

Randomize package names, class names, or API call sequences to avoid signature-based detection, while maintaining the malicious logic [75].

A simplified diagram of the polymorphic abuse pipeline is shown in Fig. 8, illustrating how LLMs serve as powerful tools for generating diverse, yet functionally similar, malicious artifacts. Platforms relying solely on shallow heuristics, hash matching, or fixed rules are ill-equipped to counter such scalable evasion strategies [73]. The solution lies in leveraging LLMs themselves to assist defenders by reasoning about deeper code semantics, behavioral patterns across app submissions, and the underlying intent of applications, even when surface features change [34], [110].

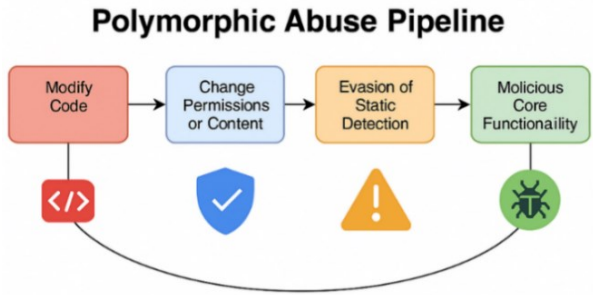

**Fig. 8. How LLMs enable polymorphic app variant generation for scalable abuse.**

### F. Regulatory Non-Compliance at Scale

The global digital regulatory landscape is becoming increasingly fragmented and complex, with new privacy, safety, and content governance laws emerging frequently across different jurisdictions [8], [10]. LLMs, by default, typically lack inherent jurisdictional awareness unless explicitly fine-tuned with massive, context-specific legal and regulatory datasets [26], [156]. As a result, apps generated with LLM assistance, or those operating in Gen-AI marketplaces, may inadvertently or deliberately violate major regulations such as:

**GDPR (EU)**: Apps may lack lawful bases for data processing, fail to implement proper consent mechanisms, or not honor user rights like the right to be forgotten or data portability [6], [36].

**CCPA (California)**: Failure to provide clear opt-out mechanisms for the sale of personal information or incomplete "Do Not Sell My Info" links [7].

**COPPA (U.S.)**: Non-compliance with regulations concerning the collection of personal information from children under 13, including failure to obtain verifiable parental consent [111].

**HIPAA (U.S.)**: Improper handling or disclosure of protected health information (PHI) in health-related applications, leading to severe penalties [112].

**Digital Services Act (EU)**: Non-transparent app ranking, unfair self-preferencing, inadequate reporting obligations for harmful content, or insufficient due diligence for products offered on the platform [8], [27].

**EU AI Act**: Emerging regulations focusing on the responsible development and deployment of AI systems, potentially imposing strict requirements on transparency, risk assessment, and human oversight for AI-powered apps and services [48].

These compliance challenges extend far beyond mobile apps. Any platform that allows LLM-assisted user contributions or storefronts—including generative AI marketplaces, social media platforms, or digital commerce sites—will face similar issues where disclosure, consent, and content moderation expectations are often jurisdiction-dependent [56], [57]. Manually verifying compliance across a diverse and rapidly changing regulatory environment is unsustainable [10].

As platforms increasingly adopt LLMs defensively to enhance integrity, attackers are also iterating their methods, creating an evolving **"LLM misuse detection arms race"** [32], [74], [113]. Malicious actors are fine-tuning their own generative models to produce evasive variants that can mimic compliant behavior or actively probe LLM-based detection thresholds. For instance, attackers may deploy adversarial prompt engineering to generate code or policies that appear benign to AI reviewers while embedding obfuscated logic or subtle policy violations [110], [114]. As shown in recent studies on prompt injection and evasion patterns [32], [66], [115], the cat-and-mouse dynamic now extends directly to LLM architectures themselves. This arms race necessitates not only technical vigilance but also continuous retraining of detection models with rapid feedback loops sourced from real-world submissions and flagged misuse incidents [78], [116]. Platforms that treat abuse detection as a static deployment rather than an adaptive, intelligence-driven system risk falling behind in this dynamic landscape. Novel approaches, such as **watermarking AI-generated content** to trace its origin or applying **traffic pattern analysis** (originally developed in Software-Defined Networking (SDN) contexts [42]) to detect coordinated abusive campaigns, may also inform new strategies for tracing and mitigating abusive behavior across app ecosystems [117], [118].

Table 9 provides a comprehensive mapping of common regulatory violations by LLM-assisted apps, highlighting the broad spectrum of compliance challenges. Without systematic, AI-augmented compliance validation, platforms risk hosting non-compliant applications and content, leading to severe regulatory penalties, significant financial losses, and a critical erosion of user trust [9], [49].

| Regulation Affected | Typical Violations in LLM-Generated Apps |
|---|---|
| GDPR (EU) | Missing lawful basis for data processing, inadequate consent, failure to honor deletion requests. |
| CCPA (California) | No clear opt-out option for data sale, incomplete 'Do Not Sell My Info' links. |
| COPPA (Children's Privacy, US) | Failure to obtain verifiable parental consent before collecting data from users under 13. |
| HIPAA (Health Data, US) | Improper handling or disclosure of protected health information (PHI) in health apps. |
| Digital Services Act (EU) | Non-transparent app ranking, unfair self-preferencing, inadequate reporting obligations. |

## G. Social Engineering and Trust Exploits

LLMs possess a remarkable ability to generate highly persuasive, contextually relevant, and emotionally resonant text and user flows, which malicious actors can weaponize for sophisticated social engineering attacks [25], [79]. These AI-driven exploits aim to manipulate users into divulging sensitive information, granting excessive permissions, making unauthorized purchases, or performing other harmful actions. Common abuse patterns leveraging LLMs for social engineering include:

**Deceptive Onboarding Flows**: Apps presenting seemingly legitimate onboarding processes that artfully coax users into granting excessive, unnecessary, or intrusive permissions (e.g., continuous background location tracking, access to call logs) through manipulative language or veiled benefits [104], [119].

**Phishing Overlays and Impersonation**: Malicious apps generating convincing in-app overlays or prompts that mimic legitimate login portals (e.g., banking apps, social media accounts) to steal credentials, or impersonating official platform communications to trick users into security-compromising actions [120], [121].

**Manipulative In-App Purchase Prompts**: Apps using persuasive language or emotional appeals to drive impulsive or unauthorized in-app purchases, often targeting vulnerable user groups or exploiting cognitive biases [122].

**Fake Customer Support and Chatbots**: LLMs can power highly realistic fake customer support chatbots or messaging interfaces within malicious apps designed to extract personal information or guide users to malicious websites [123].

**Credential Stuffing and Account Takeover (ATO) Attacks**: LLMs can be used to generate variations of stolen credentials or to automate attempts to bypass login protections, facilitating large-scale account takeovers [124].

A conceptual visualization of LLM-enabled social engineering attack patterns is shown in Fig. 9, highlighting the various touchpoints where user trust can be exploited. Detecting these nuanced psychological manipulations requires moving beyond simple keyword matching or syntactic analysis. LLM-powered abuse detection pipelines show significant promise in modeling and detecting these complex behavioral and semantic exploit vectors by analyzing user interaction patterns, linguistic cues, and emotional sentiment within the generated content [25], [79], [125]. Furthermore, advancements in behavioral biometrics and anomaly detection are becoming critical countermeasures [126].

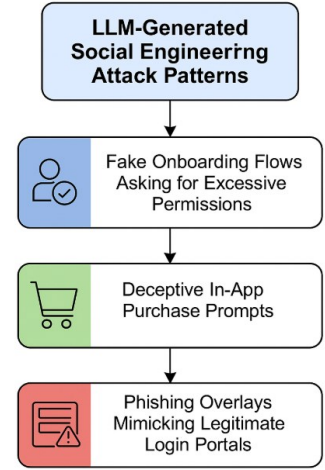

**Fig. 9. LLM-generated social engineering patterns in app user interfaces.**

## IV. LLMs for Reviewer Automation and Platform Integrity

To ensure this strategic roadmap offers not only breadth but also practical depth, each proposed capability is accompanied by implementation strategies, evaluation metrics, and real-world deployment references. Traditional mobile app review processes were originally designed for an era characterized by moderate app submission volumes and predominantly manually detectable abuse patterns [73], [127]. However, as app submissions have exploded in quantity and complexity—a phenomenon significantly driven by LLM-assisted development [54], [53]—the need for scalable, intelligent, and adaptive review workflows has become critically urgent [73]. Manual reviews alone, even with extensive human resources, are no longer sufficient for effectively ensuring platform safety, comprehensive regulatory compliance, and sustained user trust in the face of rapidly evolving threats [27], [128].

In this section, we comprehensively explore how LLMs can power the next generation of app review automation and overall platform integrity systems. This involves strategically augmenting human reviewers, drastically improving detection precision, and enabling more proactive and scalable enforcement across diverse digital ecosystems [28], [34]. It is crucial to note that these review challenges are not limited to mobile applications. Generative platforms hosting user-facing AI tools, such as Hugging Face Spaces or OpenAI Plugins, and digital commerce sites like Amazon and Etsy, increasingly rely on scalable review frameworks to ensure that uploaded models, user-generated content, prompts, and outputs consistently adhere to platform policies, safety standards, and legal requirements [56], [57], [129].

## A. Static Code Analysis Using LLMs

Conventional static code analyzers primarily operate based on hand-crafted rules, predefined patterns, and signature matching to identify known vulnerabilities or malicious code [130], [131]. While effective for well-understood threats, their limitations become apparent when confronted with novel, polymorphic, or highly obfuscated malware [75], [76]. However, LLMs offer a transformative capability: they can **reason about code at a higher semantic level**, understanding not just the syntax but also the underlying intent and potential behavioral implications of the code [12], [19], [21]. This enables several advanced applications for enhanced static code analysis:

**Detection of Polymorphic and Obfuscated Malware**: LLMs can analyze code patterns for structural and semantic anomalies that indicate malicious intent, even when surface features or traditional signatures change [75], [76]. This involves identifying unusual API calls, obfuscation techniques, or control flow manipulations indicative of malicious payloads [133].

**Identification of Latent Threats and Zero-Day Vulnerabilities**: By understanding the logical flow and purpose of code, LLMs can identify subtle programming errors or design flaws that could be exploited, potentially uncovering previously unknown vulnerabilities (zero-days) that evade signature-based tools [19], [83], [132].

**Detecting Privacy-Violating Behaviors**: LLMs can scan code for unauthorized access to sensitive user data, insecure communication channels, or covert data exfiltration routines, ensuring adherence to privacy policies and regulations [104], [94]. They can infer data flows and identify whether sensitive information is being handled securely [134].

**Spotting Misleading or Manipulative UI/UX Flows**: Beyond just code, LLMs can analyze code associated with user interface elements to detect patterns that suggest deceptive or manipulative flows, such as fake consent screens, hidden buttons, or misleading prompts designed for social engineering [119], [125].

Recent research, such as work by Chen et al. [19], has empirically demonstrated that LLMs fine-tuned on security-relevant codebases (e.g., repositories containing known vulnerabilities, malware samples) can significantly outperform traditional static analysis tools in terms of vulnerability detection accuracy, achieving higher precision and recall rates [83], [136]. The evolution of static code analysis pipelines through LLM integration is conceptually illustrated in Fig. 10, highlighting key tasks such as identifying obfuscated malicious logic, detecting privacy violations, and spotting misleading UI patterns. This semantic understanding capability is a game-changer for proactive security.

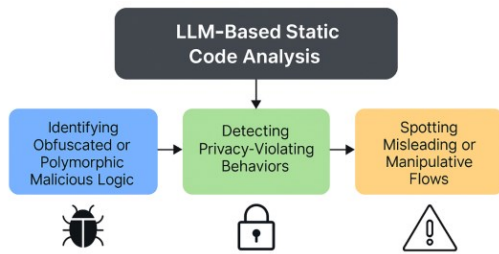

**Fig. 10. Evolution of static code analysis through LLM integration for detecting vulnerabilities.**

## B. Multimodal Cross-Validation of Storefront Claims

The integrity of a platform is not solely determined by the security of the underlying code but also by the accuracy and honesty of the information presented to users. LLMs are uniquely positioned to enable platforms to **cross-validate an application's declared functionalities and characteristics** (derived from its description, screenshots, promotional videos, and privacy labels) against its underlying code, observed runtime behaviors, and network activities [9], [15], [97], [135]. This multimodal reasoning capability is crucial for detecting and preventing **storefront misrepresentation** and deceptive practices [21], [98]. This powerful capability can detect cases where:

**Inconsistent Data Practices**: An app's storefront metadata explicitly claims "no data collection" or "does not access personal information," but its code is found to access contacts, precise location data, or microphone, or transmit such data over the network [104], [105].

**Misleading Functionality Claims**: An app prominently markets itself as "offline-only functionality" or "no internet required," yet dynamic analysis reveals it demands constant network access for core features, or contains aggressive advertising SDKs not disclosed [97], [107].

**False Compliance Assertions**: Storefronts or privacy labels claim "HIPAA-compliant" (for health apps) or "GDPR-ready" without the underlying code or behavior supporting such assertions, potentially exposing users and platforms to legal risks [99], [36].

**Synthetic Visuals and Text**: The app uses AI-generated screenshots or promotional videos that depict functionalities or user interfaces not present in the actual application, or employs AI-written fake reviews to artificially boost ratings [77], [23].

Common discrepancies between app storefront claims and detected behaviors are summarized in Table 10. This rigorous cross-validation process significantly reduces misrepresentation risks, enhances user protection, and fundamentally maintains ecosystem trust and transparency [21], [57]. The ability of LLMs to process and correlate information across different modalities (text, code, visual data, network logs) is central to this advanced detection capability [97], [137].

**Table 10. Common Storefront Claims vs Detected App Behavior**

| Storefront Claim | Detected App Behavior |
|---|---|
| No Ads Displayed | Integrates multiple aggressive ad SDKs |
| Offline-Only Functionality | Requires persistent network access |
| No User Data Collected | Accesses contacts, location, camera |
| HIPAA/GDPR Compliance | Privacy policy missing or contradictory |
| Minimal Permissions Required | Requests background location, microphone access |

## C. Policy and Document Review Automation

The sheer volume and complexity of legal and compliance documentation presents a significant bottleneck for global regulatory adherence. This includes privacy policies, terms of service, and user consent flows for mobile applications, alongside crucial documents like seller agreements on e-commerce platforms (e.g., Amazon) and community guidelines for social media networks (e.g., those governing user moderation and banning procedures). These documents are vital for conforming to frameworks such as GDPR [6], CCPA [7], COPPA [111], and emerging AI-specific regulations [138]. Large Language Models (LLMs) can profoundly enhance this process by semantically analyzing these documents to:

**Identify Missing or Ambiguous Disclosures**: LLMs can pinpoint omissions of legally required disclosures (e.g., data retention periods, cross-border data transfer details, specific third-party data recipients) or highlight language that is vague, contradictory, or intentionally misleading [14], [103], [106].

**Highlight Inconsistencies between Documentation and Observed App Behaviors**: By comparing the text of a privacy policy against the actual code and dynamic network behaviors of an app, LLMs can flag discrepancies. For instance, if a policy states no location data is collected, but the app requests and transmits GPS coordinates, the LLM can identify this critical mismatch [36], [104].

**Suggest Localized Policy Edits for Regional Regulatory Alignment**: LLMs can be fine-tuned with knowledge of various international privacy laws and suggest region-specific amendments or additions to policies to ensure compliance with diverse jurisdictional requirements (e.g., distinct consent requirements in Germany versus France under GDPR) [18], [156].

**Automate Compliance Checklists**: LLMs can be trained to automatically fill out compliance checklists or generate compliance reports based on the content of policies and observed app behavior, significantly reducing manual effort and improving auditability [103], [139].

The LLM-based automated policy document review pipeline is illustrated in Fig. 11, showing how LLMs analyze privacy policies, terms of service, and user consent flows to identify missing disclosures, highlight inconsistencies, and suggest localized edits for compliance. Research has shown that LLMs fine-tuned with extensive privacy law datasets can achieve high accuracy in flagging non-compliant clauses or data practices, enabling platforms to scale their legal and compliance efforts significantly [15], [156].

### D. Website and Metadata Correlation

Apps and items listed on digital marketplaces often link to external websites, promotional landing pages, social media profiles, or support forums. These linked resources represent an additional attack surface and a potential source of misrepresentation. LLMs can be leveraged to parse and semantically analyze these external web resources, comparing their content against the app's declared metadata, in-app behavior, and policy documentation [22]. This capability is vital for:

**Detecting Bait-and-Switch Tactics**: Identifying instances where the external website promotes features, pricing models, or functionalities that are not actually present in the submitted app, or vice-versa [22].

**Flagging Fraudulent Marketing**: Catching promotional content on external sites that misrepresents the app's capabilities, user base, or security certifications, often used in phishing or scam campaigns [98], [80].

**Identifying Undisclosed Data Collection**: Uncovering tracking scripts, aggressive advertising, or data collection practices on linked websites that contradict the app's stated privacy policy [105].

**Recognizing Brand Impersonation**: Detecting external sites that fraudulently imitate legitimate brands or services to trick users into downloading malicious apps or providing sensitive information [120].

**Verifying Compliance Claims**: Cross-referencing privacy seals, security certifications, or regulatory compliance claims made on external websites with the actual app and its documented policies [99].

The automated pipeline for website and metadata correlation detection is illustrated in Fig. 12, showing how LLMs parse external websites, compare them to app metadata, and trigger review escalations when misalignments occur. Such advanced detection protects users from deceptive marketing, fraudulent schemes, and potential security risks stemming from external digital assets [22], [57].

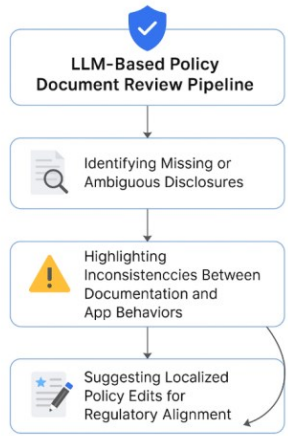

**Fig. 11. LLM-based automated policy document review pipeline for regulatory compliance**

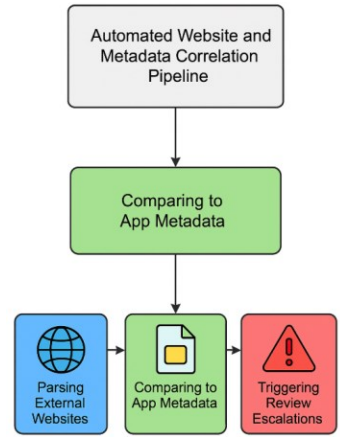

**Fig. 12. Pipeline for detecting mismatches between app metadata and linked website content using LLM-based cross-validation.**

### E. Automated Rejection Reasoning and Developer Feedback

A major point of frustration for developers submitting to app stores, sellers listing products on e-commerce sites, or users facing content moderation decisions on social media is receiving vague, generic, or non-actionable rejection reasons [23], [149], [141]. This lack of clarity prolongs iteration cycles, increases dissatisfaction, and escalates the support burden for platform operators [140]. LLMs can significantly improve this critical feedback loop by auto-generating clear, specific, and highly helpful feedback through semantic understanding of policy violations and analysis of the submitted content (e.g., code, product listings, or user-generated content):

**Summarizing Detected Policy Violations**: LLMs can succinctly explain why a submission (whether an app, product listing, or piece of user content) was rejected, referencing specific policy clauses and outlining the nature of the violation in plain language [23].

**Highlighting Relevant App Components**: Instead of a generic rejection, the feedback can pinpoint the exact code snippets, storefront elements (e.g., specific screenshots, lines in a description), or policy clauses responsible for the violation. This contextualization helps submitters quickly identify and address the issue [142].

**Suggesting Concrete Corrective Actions**: The LLM can propose specific, actionable steps developers can take to bring their app into compliance. For example: "Your app requests background location access without explicit user consent and lacks corresponding disclosures in your privacy policy. Please update your app to request runtime consent and revise your privacy policy accordingly, specifically adding a section on persistent background location data usage." Similarly, for an e-commerce seller, it might suggest: "Your product image violates guideline 3.4.c due to excessive text overlay. Please replace it with an image showing only the product." Or for a social media user: "Your post containing personal contact information violates our privacy policy (Section 2.1). Please edit the post to remove sensitive details." This moves beyond just flagging problems to providing solutions [23], [143].

Building on these capabilities, Fig. 13 illustrates an end-to-end pipeline for seamlessly integrating LLMs into platform review workflows. From ingesting app metadata, user reviews, and submitted code, LLMs can act as intelligent intermediaries. They assist in static/dynamic analysis, storefront cross-validation, and triaging abuse reports. Crucially, feedback loops update the models using compliance outcomes, developer appeals, and human reviewer corrections, enabling adaptive and explainable enforcement across submission, review, and post-deployment monitoring. This integrated architecture ensures that the entire process becomes more efficient, accurate, and developer-friendly [51], [144]. To illustrate the process of automated rejection reasoning and feedback in e-commerce, Fig. 14 depicts a flowchart outlining the integration of LLM-based analysis into the product review lifecycle.

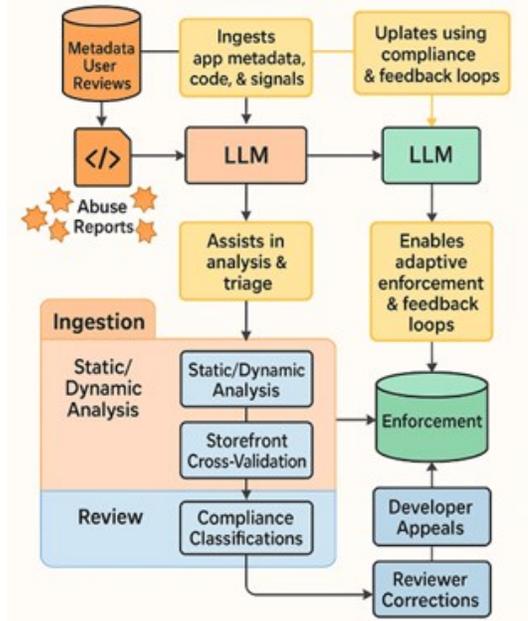

**Fig. 13. End-to-end integration of LLMs in app review pipelines.** *The system begins by ingesting metadata, user reviews, and submitted code. LLMs assist in static/dynamic analysis, storefront cross-validation, and triaging abuse reports. Feedback loops update models using compliance outcomes, developer appeals, and reviewer corrections, enabling adaptive and explainable enforcement across submission, review, and post-deployment monitoring.*

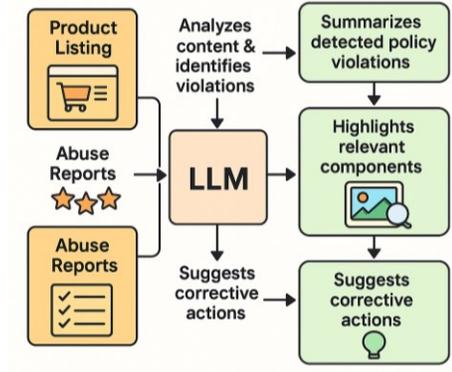

**Fig. 14. Automated Rejection Reasoning and Feedback System for E-Commerce Platforms.** *This infographic illustrates the end-to-end process of how a Large Language Model (LLM) analyzes seller product listings, detects policy violations, highlights relevant components, and provides actionable feedback to help sellers revise and resubmit their listings effectively.*

### F. Implementation Considerations and Evaluation Metrics

While LLMs provide powerful abstractions for complex platform integrity tasks, operationalizing them at scale requires careful consideration of several implementation aspects. A thoughtful approach ensures that the benefits of AI are realized without introducing new challenges or diminishing trust.

**Model Selection and Fine-Tuning**: Platforms typically begin with large, general-purpose foundation models (e.g., GPT [12], LLaMA [14], PaLM [13]) and then perform extensive **fine-tuning** on platform-specific, domain-relevant data. This includes flagged apps, product listings, user-generated content, anonymized review summaries, platform policy text, legal documents, and detailed abuse case studies. Fine-tuning significantly enhances the model's ability to understand nuanced policy violations and security patterns specific to the platform's ecosystem [17], [150]. Techniques like **Low-Rank Adaptation (LoRA)** [17] and **Quantized LoRA (QLoRA)** allow for efficient fine-tuning of large models with smaller datasets and computational resources [151].

However, it's crucial to acknowledge that training and deploying such large-scale models demand significant computational power, energy, and data storage, posing a considerable barrier for smaller platforms and raising environmental sustainability concerns.

**Inference Efficiency and Latency**: To meet the strict latency constraints of real-time review pipelines, particularly for high-volume submissions app submissions, new product listings, or user posts, lightweight versions of LLMs are often employed. Techniques like **quantization** (reducing model precision) and **distillation** (training a smaller model to mimic a larger one) are critical for optimizing inference speed and reducing computational cost [152], [153].

**Evaluation Metrics**: Rigorous evaluation is paramount to ensure the effectiveness and fairness of LLM-augmented systems. Key metrics include:

**Precision and Recall**: For abuse and fraud detection, high precision minimizes false positives (incorrectly flagged legitimate content), while high recall ensures that malicious content is not missed [35], [154].

**False Positive Rate (FPR) in Reviewer Triage**: Minimizing the rate at which legitimate apps, product listings, or user posts are flagged for human review, thus optimizing human reviewer efficiency [35].

**F1 Score in Static Analysis**: A balanced metric (harmonic mean of precision and recall) for evaluating the effectiveness of LLM-powered static analysis and automated submission evaluation tools for various content types compared to traditional methods [19], [83].

**Developer/Submitter/User Satisfaction Metrics**: Surveys and feedback channels to gauge developer, seller, or user satisfaction with the clarity, speed, and helpfulness of automated feedback and rejection reasons [23], [149].

**Compliance Coverage across CCPA/GDPR/DSA Obligations**: Quantifying the percentage of legal and policy requirements consistently met by apps, products, or platform operations post-review, demonstrating effective regulatory enforcement [156].

**Deployment Stack and Architecture**: The choice of deployment strategy significantly impacts privacy, scalability, and update frequency.

**On-device LLMs** (e.g., as implemented by Apple for certain privacy-preserving tasks [45], [36]): These models run directly on user devices, offering enhanced data privacy by keeping sensitive user data local. However, they may have limitations in model size, complexity, and frequent update capabilities [157].

**Server-side Systems** (e.g., Google Play Protect [44]): These leverage powerful cloud infrastructure for large-scale model inference, offering greater model complexity, real-time updates, and centralized threat intelligence. However, they require robust data anonymization and privacy safeguards for user data [35], [158]. A hybrid approach combining the benefits of both is often optimal.

LLM models fine-tuned on labeled vulnerability datasets have shown impressive performance gains, with studies reporting improvements of 22% in precision and 17% in recall compared to traditional static analyzers [19]. While conventional review pipelines rely heavily on manual heuristics, rule-based static analyzers, and keyword-matching approaches, LLM-augmented systems introduce a new paradigm grounded in deep semantic reasoning, adaptive triage, and scalable cross-modal validation.

Fig. 15. presents a comprehensive set of metrics used to evaluate the performance, fairness, and impact of LLM-augmented app, product, and content review pipelines. These metrics guide ongoing improvements in detection accuracy, operational efficiency, regulatory compliance, and developer, seller, and user experience.

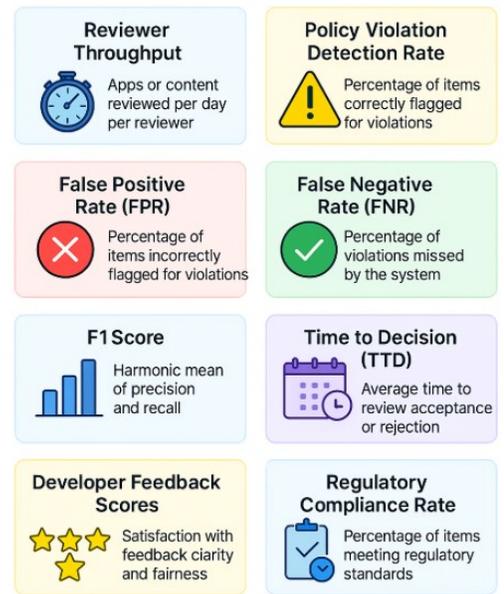

*Fig. 15. Key metrics for evaluating the effectiveness of AI-augmented app, product, and content review pipelines. These include throughput, detection rates, false positive/negative rates, F1 score, time to decision, developer satisfaction, regulatory compliance, and appeal success rate. Together, these metrics ensure data-driven optimization and continuous improvement of platform safety, accuracy, and fairness.*

To highlight this evolution, Table 11 summarizes the key differences between conventional app, e-commerce, and social media review processes and those significantly enhanced by LLM integration. These enhancements not only boost reviewer efficiency and detection accuracy but also improve the developer, seller, and user experience through more contextual feedback and a reduction in frustrating false positives [23], [149].

*Table 11. Comparison of Traditional vs. LLM-Augmented App Review Pipelines*

| Aspect | Traditional Review Pipeline | LLM-Augmented Review Pipeline |
|---|---|---|
| Code/Content Analysis | Primarily relies on rule-based static analyzers, signature matching, and simplistic pattern recognition for code. Manual review or keyword matching for product descriptions, images, or user-generated content. Limited in detecting polymorphic or zero-day threats [130]. | Employs semantic reasoning using LLMs to understand code intent, detect complex logical flaws, and support polymorphic malware detection, even when surface features change [34], [76]. Extends this semantic analysis to product descriptions, images, and user-generated content to identify violations beyond keywords. |
| Metadata/Submission Validation | Manual inspection of app descriptions, screenshots, and privacy labels. For e-commerce, manual review of product titles, images, and specifications. For social media, manual assessment of user profiles and post context. Prone to human error and easily circumvented by sophisticated deception [21]. | Cross-validation of app behavior and storefront claims using LLMs across multiple modalities (text, visuals, code, network activity). Similarly, verifies consistency between e-commerce product data and images, or social media user profiles and their content, to detect inconsistencies and misrepresentations [97], [21]. |
| User/Community Monitoring | Manual triage of app reviews and user feedback; keyword-based filtering; limited capacity to process large volumes of unstructured data. Reactive rather than proactive in identifying emerging threats [159]. | LLM-based summarization and abuse signal detection from vast app review, e-commerce feedback, and social media user datasets. Proactively identifies emerging issues, performance regressions, and real-world abuse cases at scale [45], [159]. |
| Policy Compliance Checks | Primarily manual audits by legal and policy teams; rule-based checks for known compliance terms. Inefficient and non-scalable for complex, evolving global regulations [10]. | LLM parsing of privacy policies and regulatory obligations, cross-referencing with app behavior, product specifications, or user content to automate compliance enforcement for GDPR [6], CCPA [7], DSA [8], and other laws [36], [156]. |
| Throughput & Scalability | Limited by reviewer capacity; manual processes create bottlenecks; difficult to scale linearly with increasing submission volumes [73]. | Scales dynamically with model inference capabilities and adaptive triage systems; significantly boosts reviewer capacity by offloading routine tasks [35], [51]. |
| False Positives & Context Gaps | High false positive rates due to rigid rules that often miss nuance or context; generic flags without specific details make remediation difficult for developers [23]. | Reduced via contextual understanding and deeper semantic reasoning by LLMs. Provides more precise flagging and fewer irrelevant alerts, improving accuracy and reducing friction for developers, sellers, and users [149]. |
| Feedback to Developers/Submitters/Users | Generic, templated rejection messages that provide little semantic or actionable guidance, leading to frustration and repeated submissions [149]. | Contextual, LLM-generated feedback tailored to app behavior, product listing details, or user content and specific policy violations. Offers clear rationales and suggests concrete corrective actions, improving developer, seller, and user experience and reducing appeal cycles [23], [143]. |
| Adaptability to New Threats | Slow to adapt to novel abuse tactics and polymorphic variants; requires manual updates to rules and signatures [74]. | Adaptive model tuning and continuous learning from new data and abuse patterns; capable of detecting novel or evolving threats through generalized understanding of intent [110], [116]. |

Operationalizing these LLM-powered integrity systems requires significant investment in both infrastructure and talent. This includes compute resources for model inference, scalable storage for metadata and logs, and robust orchestration for real-time review workflows. Just as importantly, it demands cross-functional talent: from ML engineers and policy experts to UX designers and Trust & Safety specialists. Product teams must balance ambition with practical resource constraints when scaling these systems across global digital ecosystems.

## V. CROSS-FUNCTIONAL COLLABORATION FOR SAFE APP ECOSYSTEMS AND PLATFORM INTEGRITY

Safeguarding mobile app ecosystems and broader digital platforms at scale demands more than just isolated technical innovation; it requires a tightly orchestrated and holistic approach involving **cross-functional collaboration** across multiple organizational functions [39], [41], [160], [161]. The challenges posed by LLM-amplified abuse, intricate privacy risks, and ever-evolving regulatory expectations are simply too vast and complex for any single department to address in isolation [27], [128], [162]. A siloed approach can lead to reactive governance, missed threats, and inefficient resource allocation, ultimately undermining platform integrity and user trust [163], [164], [165].

This section outlines how platforms can design **integrated, cross-functional architectures** to effectively incorporate LLM-powered tools, optimize review workflows, ensure dynamic regulatory alignment, and continuously adapt to emerging threats in a proactive rather than reactive manner. The emphasis is on breaking down organizational barriers to build a cohesive and responsive trust and safety framework, moving from a fragmented collection of teams to a unified front against digital harm [41], [166].

Achieving this integrated posture necessitates formalized collaborative mechanisms, such as shared objectives and key results (OKRs) across teams, dedicated weekly syncs with cross-functional leadership, and a unified platform for tracking and resolving integrity issues.

---

### A. Role of Product Management, Policy, Legal, Operations, and Engineering

A robust and effective platform safety organization is characterized by its ability to align diverse teams around common goals, shared metrics, and interconnected workflows. Each functional area plays a distinct yet interdependent role in operationalizing LLM-powered integrity systems, ensuring comprehensive coverage and rapid response capabilities:

**Product Management**: This team serves as the strategic orchestrator, defining the overall **review flows, risk frameworks, and developer, seller, and user experiences** [40], [167]. They are responsible for identifying key problem areas, prioritizing the development of LLM-powered safety features, and ensuring these features align with the platform's strategic objectives, user needs, and business goals [168], [169]. Product managers translate abstract policy goals into tangible product requirements, bridging the gap between technical capabilities and operational impact [167]. They often lead the roadmap for implementing new safety features, such as AI-driven content scanning or automated policy violation detection [170]. They also champion the development of shared tools and dashboards that provide a holistic view of integrity posture, enabling data-driven decision-making across all stakeholder groups.

**Engineering**: The engineering team is the technical backbone, responsible for building, deploying, and maintaining the LLM-based review pipelines, policy analyzers, static and dynamic code analyzers (primarily for apps), and abuse detection systems for all forms of digital content (e.g., product listings, user posts) [34], [52], [171], [172]. This includes selecting appropriate LLM architectures (e.g., transformer-based models, specialized variants), fine-tuning models on domain-specific data, optimizing for inference efficiency (e.g., through quantization or distillation), and seamlessly integrating these AI components into existing platform infrastructure [151], [152], [173]. Their mandate also extends to ensuring secure-by-design principles for all AI components, robust API security for LLM integrations, and leveraging modern DevSecOps practices for continuous security assurance throughout the development and deployment lifecycle.

They also manage the continuous integration/continuous deployment (CI/CD) pipelines for AI models, ensuring they are always up-to-date with the latest threat intelligence [116]. This includes establishing robust API contracts for AI service integrations and developing modular architectures that allow for rapid iteration and deployment of security enhancements based on cross-functional feedback.

**Trust & Safety Operations (T&S)**: This team is on the front lines, **triaging escalations, managing takedowns, and fine-tuning human-AI reviewer interactions** [39], [174], [175], [176]. They provide critical human judgment for complex edge cases, serve as a vital feedback loop for model improvement by labeling data, correcting AI errors, and identifying novel abuse vectors that AI models might initially miss [35], [58], [177]. Their practical insights are invaluable for identifying real-world attack patterns and adversarial strategies, directly informing the iterative refinement of AI models [110].

This augmentation is realized through intelligent triage systems that automatically escalate borderline cases, novel abuse patterns, or high-severity flags to human reviewers, leveraging predefined confidence thresholds and risk scores. Human reviewers provide critical feedback through detailed labeling tools and structured error reporting forms, directly influencing subsequent model retraining and refinement cycles. Furthermore, clear protocols are established for human override of AI decisions, ensuring a robust safety net for edge cases and enabling rapid adaptation to unforeseen threats.

This augmentation ensures that AI systems handle routine, high-volume tasks, freeing human reviewers to focus on ambiguous edge cases, novel abuse patterns requiring nuanced judgment, and complex ethical dilemmas where human discernment is indispensable. Their unique ability to interpret subtle intent and adapt to new adversarial tactics remains paramount.

Furthermore, the efficacy of AI-augmented workflows hinges on meticulously designed human-AI interaction protocols. For instance, in **app review**, this demands clear confidence thresholds that automatically escalate submissions with low AI confidence scores or novel threat patterns to human experts, alongside intuitive dashboards that present comprehensive, LLM-summarized contexts for rapid human discernment. In **content moderation**, robust feedback mechanisms must enable human reviewers to efficiently correct AI misclassifications, thereby retraining and refining models in real-time. For **fraud detection**, systems should present explainable risk scores and highlight key indicators to human analysts, facilitating quicker, more informed decisions while leveraging AI for high-volume pattern recognition. This continuous, symbiotic feedback loop is crucial for maximizing both efficiency and nuanced accuracy [178].

**Legal and Policy Teams**: These teams are crucial for ensuring that platform rules, automated enforcement mechanisms, and AI model outputs **align with global laws and regulations** (e.g., GDPR [6], CCPA [7], DSA [8], EU AI Act [48]) [10], [156], [179]. They translate complex regulatory requirements into enforceable product features and clear platform policies, advise on legal risks associated with AI decisions (e.g., bias, due process), and manage relationships with regulatory bodies [180], [181]. Their expertise is essential in navigating the evolving landscape of AI governance and ensuring compliance across diverse international jurisdictions [49], [182].

**User Experience (UX) and Developer Relations**: This function is responsible for creating **transparent, actionable, and fair feedback loops** to developers, sellers, and users alike. They design intuitive interfaces for rejection notifications, provide clear explanations of policy violations (often leveraging LLM-generated summaries [23]), and offer guidance on how to fix issues quickly and fairly [149], [140]. Effective communication and support foster a positive developer, seller, and user experience, reducing frustration, minimizing appeal cycles, and promoting quicker compliance with platform standards [142], [183]. For users, they ensure clarity on moderation decisions and avenues for recourse.

**Phased Rollout Strategy:** Within this framework, Product Management plays a pivotal role in orchestrating the phased introduction of LLM-powered capabilities. Rather than deploying all features at once, a strategic rollout would begin with Minimum

Viable Product (MVP) stages targeting low-risk, high-impact areas—such as metadata validation or automated review summarization. Subsequent phases would expand to more complex functions like federated review, multi-modal cross-validation, and dynamic compliance parsing. This staged approach enables iterative testing, minimizes disruption, and facilitates continuous feedback loops to refine both model accuracy and developer /user experience.

The primary responsibilities of each cross-functional team are summarized in Table 12. The synergy between these teams is paramount; for example, engineering builds tools that incorporate legal policies, operations provides feedback that refines the tools and identifies new threats, and product management ensures that all efforts serve the overarching goal of platform integrity and user trust. This integrated approach fosters a culture of shared responsibility for platform safety [166], [184].

*Table 12. Team responsibilities for cross-functional collaboration in platform safety*

| Team | Responsibilities |
|------|-----------------|
| **Product Management** | Defines review flows, risk frameworks, developer experiences, and cross-team priorities |
| **Engineering** | Builds LLM-based review pipelines, policy analyzers, static and dynamic code analyzers |
| **Trust & Safety Operations** | Triages escalations, manages takedowns, fine-tunes human-AI reviewer interactions |
| **Legal and Policy** | Ensures platform rules align with GDPR, CCPA, DSA; translates regulations into product enforcement |
| **UX and Developer Relations** | Provides actionable developer feedback and helps resolve flagged issues quickly |

**Product-Level Risk Mitigation**: Beyond technical risks, product-level challenges—such as user backlash from false positives, developer churn due to opaque feedback, and feature adoption resistance—must be proactively addressed. Product teams can mitigate these risks through A/B testing of AI enforcement mechanisms, prioritizing human-in-the-loop review for borderline cases, and providing actionable, LLM-generated explanations for rejections. Transparent appeals processes and bias audits further ensure fairness. Product Management must lead these efforts to preserve developer trust and ensure seamless rollout of integrity features.

Fig. 16 (a) summarizes the responsibilities of each cross-functional team in operationalizing LLM-powered platform integrity. Effective coordination between product, engineering, trust & safety, legal, and UX ensures scalable enforcement, regulatory compliance, user transparency, and developer trust across the review lifecycle. The cross-functional collaboration required to safeguard app ecosystems is illustrated in Fig. 16 (b) Embedding trust and safety principles early into platform and product design prevents security retrofitting later—a common failure mode in reactive governance structures [26].

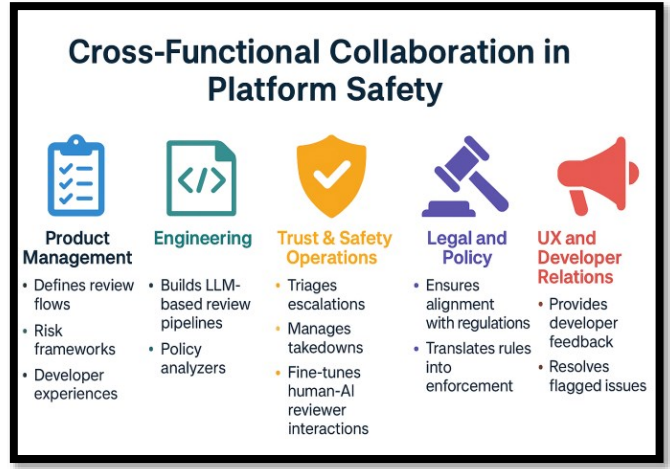

(a)

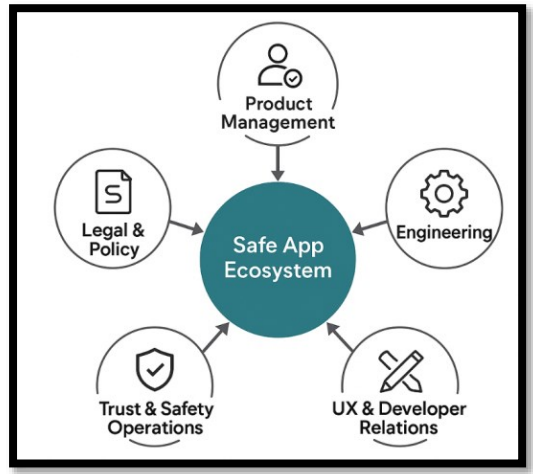

(b)

*Fig. 16. (a) Key cross-functional teams and their roles in platform safety. Product Management defines review flows and drives strategic alignment; Engineering builds LLM-based review pipelines and analyzers; Trust & Safety Operations manages escalations and human-AI interaction; Legal and Policy teams ensure regulatory compliance and translate rules into actionable enforcement; and UX/Developer Relations provide feedback and resolve flagged issues to improve developer experience and user trust, (b) Cross-functional collaboration architecture for building safe app ecosystems*

### B. Regulatory Compliance Automation (GDPR, CCPA, DSA Alignment)

Global regulatory complexity is rising sharply, driven by new legislation and evolving interpretations of data privacy and content governance across different jurisdictions [8], [10], [179], [182]. Manual compliance review is increasingly infeasible at scale, especially given the rapid pace of app submissions and content generation [73], [128]. LLMs can significantly assist in automating and streamlining regulatory compliance by:

**Mapping App Behaviors to Regulatory Requirements:** LLMs can analyze app code and dynamic behaviors (e.g., data collection, sharing practices, consent flows) and automatically map them against specific regulatory requirements. This allows for verification of lawful bases for data processing under GDPR [6], confirmation of clear opt-out flows under CCPA [7], or assessment of child data handling under COPPA [111]. Advanced techniques involve using LLMs to build knowledge graphs of legal texts and app functionalities for precise matching [36], [186].

**Flagging Missing Disclosures and Inconsistencies:** LLMs can intelligently detect omissions in privacy policies, terms of service, or in-app disclosures, such as insufficient transparency about third-party SDK usage [16], [105], or failure to detail cross-border data transfers. They can also highlight contradictions between stated policies and observed app behavior [103], [104].

**Suggesting Region-Specific Corrective Actions:** Leveraging their knowledge of diverse legal frameworks, LLMs can propose tailored corrective actions for non-compliant apps. This could include recommending the inclusion of "Data Deletion Requests" links for jurisdictions that require them, or advising on specific consent banners needed for EU users [18], [156].

**Automated Regulatory Impact Assessments:** For new features or significant app updates, LLMs can perform a preliminary regulatory impact assessment by analyzing the proposed changes against relevant laws, flagging potential compliance hurdles proactively [187].

**Real-time Policy Monitoring:** Beyond initial review, LLMs can continuously monitor published apps for changes in behavior or metadata that might lead to new compliance risks, enabling proactive alerts and enforcement actions [44], [116].

Google Play and Apple's growing emphasis on transparent privacy labels (e.g., Apple's Privacy Nutrition Labels) and vigorous enforcement of regional compliance practices exemplify this strategic shift towards automated compliance [7], [8], [9]. The automation of regulatory compliance checks using LLMs is depicted in Fig. 17. Platforms that proactively operationalize complex compliance frameworks via LLMs will be better positioned to avoid severe regulatory penalties, public backlash, and direct governmental intervention, ultimately building stronger foundations of trust with both users and regulators [27], [49], [188].

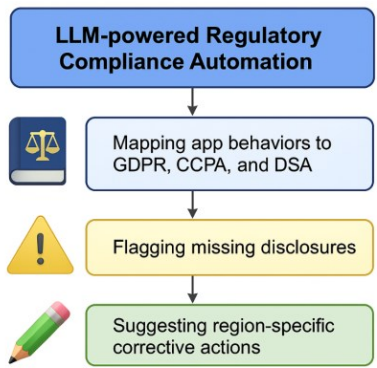

*Fig. 17. LLM-powered compliance automation flow for GDPR, CCPA, and DSA alignment*

## C. Metrics: Measuring Impact, Speed, Accuracy, and Fairness

Building sophisticated AI-augmented review pipelines is a critical first step, but it is insufficient without robust mechanisms to **track operational metrics, measure impact, and continuously improve their effectiveness** [35], [51], [189]. A comprehensive metrics framework is essential for assessing the performance of LLM-powered systems, identifying areas for optimization, and ensuring fairness and transparency in moderation decisions [46], [190]. Key metrics include:

**Reviewer Throughput**: Quantifying the number of apps or content items reviewed per human reviewer per day. LLM assistance is expected to significantly boost this metric by automating routine tasks and intelligently triaging complex cases [35], [51].

**Policy Violation Detection Rate**: The percentage of submitted apps or content items accurately flagged for policy violations by the automated or augmented system. This measures the efficacy of the detection models [154].

**False Positive Rate (FPR)**: The percentage of non-violating apps or content items incorrectly flagged by the system. Minimizing FPR is crucial to avoid developer frustration and unnecessary operational overhead [149], [154].

**False Negative Rate (FNR)**: The percentage of violating apps or content items that are missed by the system and incorrectly allowed onto the platform. Minimizing FNR is paramount for maintaining platform safety and user trust [154].

**F1 Score**: A harmonic mean of precision and recall, often used in static analysis and abuse detection to provide a balanced measure of a model's accuracy [19], [83].

**Developer Feedback Scores**: Measuring developer satisfaction with the clarity, speed, fairness, and helpfulness of platform feedback and rejection reasons [23], [149]. This can be gathered through surveys or direct feedback channels.

**Regulatory Compliance Rates**: The proportion of apps or content items meeting specific regulatory requirements (e.g., GDPR, CCPA, DSA standards) post-review. This directly measures the effectiveness of compliance automation [156].

**Appeal Success Rate**: The percentage of developer appeals against moderation decisions that are overturned, indicating potential areas where AI models or human-AI interactions might need refinement [191].

**Time to Decision (TTD)**: The average time taken from submission to a final review decision, which AI automation aims to drastically reduce for both accepted and rejected submissions [140].

Operational metrics for evaluating the effectiveness of LLM-augmented review pipelines are listed in Table 13. Ongoing, metric-driven optimization ensures that LLM integration improves not just operational speed and efficiency, but also decision quality, fairness, and overall user and developer trust [35], [189]. This continuous feedback loop is vital for an adaptive and resilient platform integrity system [116].

**Table 13. Key operational metrics for evaluating LLM-powered app review systems**

| Metric | Definition | Goal |
|---|---|---|
| Reviewer Throughput | Apps/content items reviewed per human reviewer per day, reflecting efficiency gains from LLM assistance. | Increase |
| Policy Violation Detection Rate | Percentage of submitted apps/content items accurately flagged for policy violations. | Maximize |
| False Positive Rate (FPR) | Percentage of non-violating apps/content items incorrectly flagged, leading to unnecessary human review or developer friction. | Minimize |
| False Negative Rate (FNR) | Percentage of violating apps/content items missed by the system, posing risks to platform safety and user trust. | Minimize |
| F1 Score | Harmonic mean of precision and recall for detection models, providing a balanced measure of accuracy. | Maximize |
| Developer Feedback Score | Developer satisfaction with the clarity, speed, and fairness of rejection reasons and feedback from the platform. | Maximize |
| Regulatory Compliance Rate | Percentage of apps/content items meeting GDPR, CCPA, DSA, and other relevant legal standards after review. | Maximize |
| Appeal Success Rate | Percentage of developer appeals that result in an overturned decision, indicating areas for model refinement or human-AI interaction improvement. | Optimize |
| Time to Decision (TTD) | Average time from submission to final review decision (accept or reject), reflecting process efficiency. | Decrease |

## VI. CASE STUDY: MAJOR PLATFORM INITIATIVES AND LLM INTEGRATION FOR INTEGRITY

Major platform operators have already embarked on extensive efforts to deploy LLM-powered and AI-augmented technologies to profoundly improve app safety, enhance regulatory compliance, and optimize the developer experience within their vast ecosystems [42], [45], [192]. This extends beyond app development to encompass the integrity of other digital interactions, such as safeguarding e-commerce platforms from fraudulent sellers and mitigating misuse by social media users. Studying these pioneering initiatives provides invaluable real-world examples and practical insights into how scalable trust and safety architectures are evolving in practice [193]. These efforts serve as blueprints for broader industry adoption and highlight critical successes, persistent challenges, and burgeoning opportunities for further innovation in AI-driven platform governance.

In app ecosystems, key initiatives by Google Play and Apple App Store provide valuable insights into the practical deployment of LLM-powered platform integrity systems. While these case studies clearly demonstrate the significant impact of LLM integration, future industry reporting would greatly benefit from more explicit quantitative comparisons, such as 'before and after' metrics on average review times or reductions in specific false positive rates following LLM deployment, to more fully illustrate the efficiency and accuracy gains.

This section reviews their successes, implementation challenges, and emerging opportunities. Importantly, similar modernization efforts are now actively underway across other digital environments, including generative AI marketplaces (e.g., Hugging Face Spaces, with its focus on responsible model sharing [56]), digital commerce platforms (e.g., Amazon, Etsy, and Shopify, where AI supports both content generation and complex policy enforcement [57], [80]), and leading financial services platforms which are deploying LLMs for critical security, fraud detection, and compliance functions. By examining efforts across Google, Apple, Amazon, leading financial platforms, Meta, and Hugging Face, we highlight both best practices and transferable lessons that can inform platform governance across app stores, Gen-AI hubs, online marketplaces, and financial systems.

### A. Google's SAFE Framework, App Defense Alliance, and Real-Time Protections

Google Play operates one of the world's largest and most complex app ecosystems, serving over three billion active Android devices globally [44], [194]. To safeguard this expansive environment from a constant barrage of evolving threats, Google has deployed a multi-layered strategy involving a suite of initiatives and partnerships [195]:

**SAFE Principles**: Google's commitment to platform integrity is encapsulated in its SAFE Principles: Safeguard users, Advocate developer protection, Foster responsible innovation, and Evolve platform defenses [42], [44]. These principles guide their approach to app security and policy enforcement, emphasizing a holistic view of trust.

**App Defense Alliance (ADA)**: Recognizing that no single entity can combat global mobile threats alone, Google co-founded the App Defense Alliance (ADA). This is a crucial cross-industry partnership with leading cybersecurity firms such as ESET, Lookout, Zimperium, as well as major platform players like Meta and Microsoft, aimed at setting and continuously raising mobile security standards and fostering collaborative threat intelligence sharing [43], [196]. The ADA focuses on pre-vetting apps for malware before they reach users, creating a stronger defense perimeter.

**Mobile App Security Assessment (MASA)**: Building on the ADA's foundation, Google introduced the MASA program, which provides independent, third-party validation of app security practices for high-profile or sensitive apps listed on the Play Store [43], [197]. This voluntary assessment helps developers prove the robustness of their security posture and enhances user trust.

**Real-time Scanning with Play Protect**: Google Play Protect is a cornerstone of Android security, leveraging advanced machine learning (ML) and deep code-level analysis to detect and neutralize threats. Crucially, it employs sophisticated AI techniques to identify polymorphic malware and unwanted software in apps, even those downloaded outside the Play Store, performing real-time scanning on billions of installations daily [44], [110], [198]. Play Protect's ability to adapt to evolving malware strains is a direct result of its continuous learning capabilities.

These real-time protections received significant enhancements in 2025, with Google Play Protect's on-device intelligence being updated with new rules to identify malware families even prior to installation and its live threat detection capabilities expanded to identify deceptive app behaviors such as icon hiding or alteration [82]. Further bolstering user safety in 2025, Android introduced on-device protections to block risky security actions during suspicious phone calls with non-contacts—such as attempts to disable Play Protect, sideload unvetted apps, or grant excessive accessibility permissions—and began piloting enhanced in-call warnings for banking app usage during screen sharing sessions with unknown contacts [82].

**Google Play SDK Index**: To address the growing supply chain risks introduced by third-party libraries, Google launched the Google Play SDK Index. This initiative provides developers with transparency and visibility into the privacy and security profiles of thousands of commercially available SDKs, helping them make more informed and safer integration choices *before* submission [44], [71], [92]. It acts as a preventative measure, reducing the attack surface introduced by vulnerable or malicious downstream. In 2024, Google reported impressive results, stating that its **AI-assisted review systems** played a pivotal role in blocking over 2.36 million policy-violating apps from being published on Google Play [44]. Furthermore, 92% of high-risk reviews now involve LLM-assisted triage, demonstrating the significant impact of AI in streamlining the human review process and focusing human attention where it's most needed [44]. Table 14 summarizes key Google Play initiatives and their contributions to platform safety and security. The comprehensive app safety initiatives deployed by Google Play are visually represented in Fig. 18, showcasing the deep integration of various technologies and strategic collaborations.

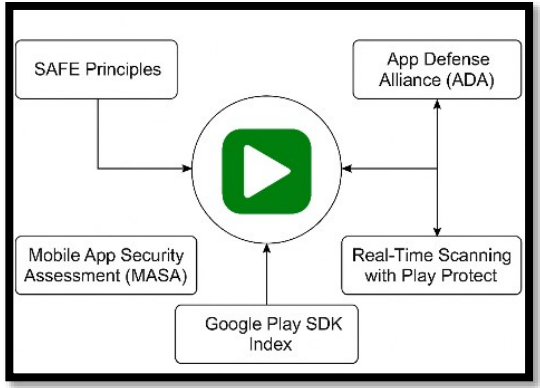

*Fig. 18. Overview of Google Play's app safety initiatives and partnerships*

The effectiveness of Google's platform integrity strategy is underpinned by a combination of sophisticated technical capabilities and strategic organizational enablers:

**LLM-assisted Triage at Scale**: By integrating large language models directly into their review workflows, Google has enabled faster identification and prioritization of high-risk submissions. This intelligent triage system has been instrumental in blocking millions of policy-violating apps, significantly enhancing the efficiency and coverage of their safety operations [44], [35].

**Semantic Code Analysis and Metadata Reasoning**: Google's systems leverage LLMs to perform deep semantic reasoning over both app code and declared metadata. This allows for the sophisticated detection of inconsistencies between an app's stated claims (e.g., in descriptions or privacy policies) and its actual runtime behavior, effectively countering deceptive practices and hidden functionalities [34], [97].

**Cross-industry Coordination via ADA**: The App Defense Alliance [43] is a testament to the power of collective defense. It fosters collaborative threat intelligence sharing, enables rapid dissemination of information about new mobile security vulnerabilities, and promotes standardization of best practices among major industry partners, thereby creating a more robust collective security posture [196], [199].

**Proactive Threat Modeling through SDK Indexing**: The Google Play SDK Index [44] represents a proactive approach to supply chain security. By providing developers with transparent risk profiles for third-party integrations, it helps to prevent vulnerable or privacy-violating components from ever entering the ecosystem, addressing issues upstream rather than reactively downstream [71], [92].

**End-to-End Platform Hardening**: LLMs are integrated across the entire app lifecycle, from initial submission checks and continuous scanning with Play Protect to enforcement systems. This creates a closed feedback loop, where insights from blocked threats and policy violations are continuously fed back into the AI models, enabling iterative safety improvement and an adaptive defense against new attack vectors [116], [200].

Google's commitment to evolving platform defenses and enhancing user privacy in 2025 also includes the expansion of AI-powered scam detection in Google Messages to a wider array of sophisticated scam types, the upcoming launch of Key Verifier to combat impersonation, and strengthened mobile theft protections like hardened Factory Reset protocols and more secure OTP handling on locked screens [82].

*Table 14. Key Google Play initiatives and their impact on app safety and security*

| Initiative | Description | Key Metrics/Impact |
|---|---|---|
| **SAFE Principles** | Safeguard users, advocate developer protection, foster responsible innovation | 92% of high-risk reviews now involve LLM-assisted triage |
| **App Defense Alliance (ADA)** | Cross-industry partnership for mobile security standards | Strengthens mobile security frameworks |
| **Mobile App Security Assessment (MASA)** | Independent validation of app security practices | Ensures better app security for Play Store |
| **Real-Time Scanning with Play Protect** | Detects polymorphic malware using machine learning | Blocked 2.36 million policy-violating apps in 2024 |
| **Google Play SDK Index** | Provides visibility into third-party SDK risks | Helps developers make safer SDK integration choices |

### B. Apple's LLM-Based Review Summarization and Privacy Enhancements

Apple's App Store review process has historically been characterized by a heavily manual, curated approach, emphasizing high-quality and consistent user experiences [201], [202], [203]. However, the exponential scale and increasing complexity of app submissions, particularly with the advent of AI-assisted development, have significantly driven the imperative for advanced automation within their review ecosystem [54], [73].

Table 15 outlines Apple's key initiatives for enhancing privacy and review processes with LLM integration. Fig. 19 illustrates Apple's LLM-based review summarization process, highlighting key steps for flagging and addressing emerging risks and streamlining human reviewer workflows.

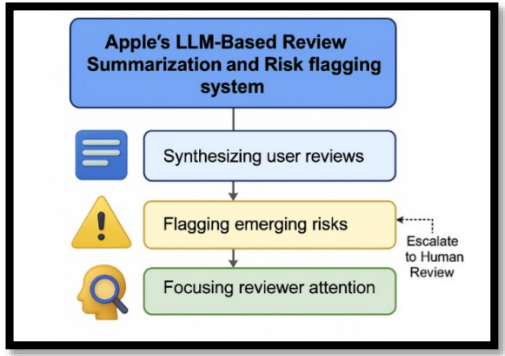

**Fig. 19. Flow of Apple's LLM-based review summarization and risk flagging system**

**Table 15. Apple's LLM-based review summarization and privacy-enhancing initiatives**

| Initiative | Description | Impact |
|---|---|---|
| LLM-Based Review Summarization | Synthesizes user reviews into actionable summaries | Flags emerging risks, enhances reviewer productivity |
| Privacy Nutrition Labels | Explicit disclosures of data collection practices | Increases transparency for users |
| On-Device & Server-Side Models | Integrates models for moderation | Balances privacy and scalability |
| Dynamic Topic Modeling | Prioritizes user feedback aligned with app experience | Ensures more relevant feedback gets attention |

In 2025, Apple Machine Learning Research unveiled its **LLM-Based Review Summarization System** [45], a significant innovation which functions to:

**Periodically synthesize vast volumes of unstructured user reviews** into actionable, concise summaries, transforming qualitative feedback into quantifiable insights [45], [204].

**Automatically flag emerging risks** identified from user feedback, such as performance regressions, novel abusive behaviors, security vulnerabilities, and unmet user expectations, allowing for proactive intervention [159], [205].

**Significantly enhance human reviewer productivity** by pre-digesting massive amounts of data and focusing their attention on high-risk apps or critical feedback areas that require nuanced human judgment, rather than rote manual scanning [174], [175].

Furthermore, Apple has demonstrated a strong commitment to privacy and safety through several other key initiatives:

**Expanded Privacy Nutrition Labels**: Following its successful introduction of Privacy Nutrition Labels, Apple has continuously expanded and refined these requirements, mandating explicit and clear disclosures of data collection and usage practices directly on app product pages [206]. This initiative aims to increase transparency for users and hold developers accountable for their data handling [99], [207].

**Integrated On-Device and Server-Side Foundation Models**: Apple has strategically integrated both **on-device** and **server-side foundation models** into its App Store moderation pipeline [208]. This hybrid approach is designed to balance the critical need for privacy preservation (by processing sensitive data locally on the device where possible [157]) with the demands of scalability and access to large-scale threat intelligence (leveraging powerful server-side AI for broader analysis [158]) [36].

**Emphasized Dynamic Topic Modeling for User Feedback**: Utilizing advanced LLM capabilities, Apple prioritizes and analyzes user feedback through **dynamic topic modeling** [209]. This allows them to automatically identify and elevate review topics most relevant to core app experience and emergent issues, rather than being sidetracked by out-of-app factors or noise [159], [45]. This improves both reviewer efficiency and the overall developer experience by focusing on pertinent feedback.

**Cross-Functional Human-in-the-Loop Workflows**: Apple's approach explicitly augments, rather than replaces, human reviewers with LLMs [58]. This ensures that nuanced, borderline, or highly complex cases are escalated to human experts with rich, contextual summaries provided by the AI, ensuring robust decision-making and preventing erroneous auto-resolutions [174], [190].

Despite these significant advancements, Apple's integration of LLMs into App Store operations also illuminates several common challenges inherent in deploying advanced AI for platform integrity:

**Explainability Gaps**: A recurring issue is that current LLMs often fail to provide transparent, human-interpretable rationales behind complex moderation decisions [46], [190]. This "black box" problem can lead to developer frustration and challenges in appeals processes.

**Model Drift**: Without continuous monitoring and regular retraining, AI models can suffer from **model drift**, becoming outdated as new abuse tactics, linguistic nuances, or policy interpretations evolve [116], [200]. This necessitates robust feedback loops and adaptive learning mechanisms.

**Cross-Jurisdictional Enforcement Complexity**: Differences in regional regulations (e.g., specific consent requirements under GDPR vs. general consumer protection under the DSA) remain difficult to encode and apply uniformly across a global platform, requiring sophisticated legal and technical alignment [10], [182].

Addressing these challenges points to several **Future Directions** for research and development:

**Research on Hybrid Human-AI Review Loops with Transparent Escalation Paths**: Further work is needed to optimize the collaboration between human experts and AI systems, ensuring that AI provides clear, auditable explanations for its decisions and that humans can efficiently provide oversight and corrections [175], [210].

**Expanded Educational Tools**: Creating educational resources—perhaps like CloudLab-based SDN security labs [211]—can provide a useful template for teaching responsible LLM usage and abuse detection strategies to developers, fostering a more secure ecosystem from the ground up [212].

**Multi-Agent Review Systems**: Exploring architectures that combine multiple LLMs or other AI models (e.g., rule-based, statistical models) to perform different aspects of review, potentially leveraging ensemble methods for improved robustness and accuracy [213].

**Differential Privacy and Federated Fine-tuning**: For privacy-sensitive data, advancements in differential privacy and federated learning techniques can enable safe, user-level customization and collaborative model training without compromising individual user data [47], [157].

**Standardization of LLM Safety Benchmarks for Platform Governance**: Developing universally accepted benchmarks and metrics for evaluating the safety and integrity performance of LLMs in moderation and review contexts would drive accountability and best practices across the industry [189], [214].

---

### C. Amazon: Combating Counterfeits, Fraud, and Listing Abuse in E-commerce

Amazon, as the world's largest online retailer and a massive marketplace for third-party sellers, faces a relentless battle against **counterfeits, product fraud, misleading listings, fake reviews, and payment fraud [215]**. LLMs and generative AI are becoming increasingly central to their multi-layered defense strategy.

**Key Initiatives and Contributions:**

**Proactive Counterfeit Detection with AI:** Amazon invests heavily in AI, including LLMs and advanced machine learning, to proactively scan and block counterfeit products before they even reach customers. Their "Project Zero" initiative, launched in 2019, leverages AI to identify infringing listings based on a vast dataset of genuine product information and known counterfeit patterns. In **2023**, Amazon's proactive controls blocked over **99%** of suspected infringing listings before a brand ever had to report them, and they identified, seized, and disposed of more than **7 million counterfeit products worldwide** [216]. LLMs assist in analyzing product descriptions, images, and brand identifiers for anomalies or suspicious claims, including complex visual IP infringements like logos and patterns [217].

**Fraud Detection (Payment, Account Takeover, Returns):** Amazon employs sophisticated ML/AI solutions (like Amazon Fraud Detector, a service built on Amazon's 20+ years of experience) to detect various types of fraud in real-time [218]. This includes:

**Payment Fraud:** Analyzing transaction patterns, user behavior, and historical data to identify suspicious purchases.

**New Account Fraud/Account Takeovers:** Detecting fake accounts or compromised logins by analyzing registration data, login patterns, and potentially AI-generated user profiles. In **2023**, Amazon stopped over **700,000** bad actor attempts to create new selling accounts before they could list a single product [216].

**Returns Fraud:** Using AI to identify patterns of fraudulent returns or claims, often linked to serial returners or organized abuse.

LLMs can assist in analyzing textual data related to customer interactions, dispute claims, and seller communications to identify deceptive language or coordinated fraud attempts.

**Combating Fake Reviews and Synthetic Content:** As highlighted in our paper, LLMs are adept at generating persuasive text, making fake reviews a significant threat. Amazon uses AI to analyze review text, reviewer behavior, and review patterns to identify and suppress AI-generated or otherwise inauthentic reviews [219], [220]. They look for unnatural language, repetition, or unusual spikes in positive/negative sentiment that might indicate manipulation. In **2022**, Amazon proactively blocked more than **200 million suspected fake reviews** [219]. They utilize large language models alongside natural language processing techniques to analyze anomalies in data, as well as deep graph neural networks to detect groups of bad actors [219].

**Seller and Listing Vetting:** AI is used to vet third-party sellers and their product listings. This involves analyzing seller information, historical performance, product images, and descriptions. LLMs can cross-reference listing content against known brand information and product specifications to identify misrepresentation or non-compliance.

**Customer Service Augmentation (for fraud/abuse queries):** While not direct integrity enforcement, Amazon's use of AI in customer service (e.g., for sellers) can indirectly contribute to integrity by quickly resolving legitimate issues, freeing up human agents to focus on more complex fraud cases, and potentially identifying new fraud patterns from aggregated customer inquiries.

Fig. 20 illustrates Amazon's multi-pronged approach to platform integrity, highlighting how LLMs and AI are deployed across key areas such as counterfeit detection, fraud prevention, synthetic review suppression, and seller vetting.

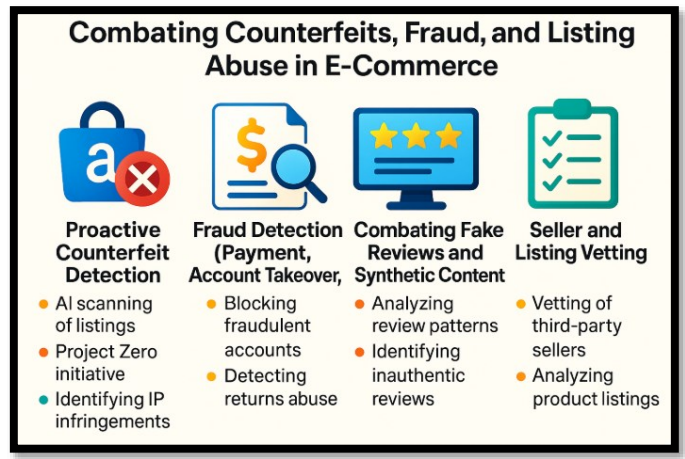

*Fig. 20. Amazon's AI- and LLM-powered strategy for combating counterfeits, fraud, and listing abuse in e-commerce. Each pillar—from proactive counterfeit detection to seller vetting—leverages generative AI, machine learning, and graph-based analysis to secure platform integrity at scale.*

## D. Financial Platforms: LLM-Powered Safeguards for Trust and Compliance

### D.1. EVOLVING THREATS AND LLM APPLICATIONS IN FINANCIAL SERVICES

Leading financial services platforms have begun integrating LLMs and generative AI to counter evolving threats like synthetic identity fraud, AI-generated scams, and regulatory evasion, while also accelerating compliance workflows. These deployments reflect growing adoption of trust and safety architectures previously limited to app and content ecosystems.

Key Applications:

- **Synthetic Identity Detection**: JPMorgan Chase and Capital One use AI models, including transformer-based architectures, to detect synthetic identities by analyzing linguistic inconsistencies, metadata patterns, and behavioral anomalies across applications and transaction flows [429], [430].
- **KYC/AML Automation:** Fintech firms like Stripe, Plaid, and Revolut have deployed LLM-based systems to automate Know Your Customer (KYC) and Anti-Money Laundering (AML) processes—flagging suspicious documents, inconsistencies, and evasive language during onboarding or transaction reviews [431], [432].
- **Regulatory Compliance:** Platforms are applying GenAI to parse complex regulatory texts (e.g., FinCEN, SEC, MiFID II, GDPR) and map them to internal policy violations, reducing manual review time and improving auditability [433], [434]. LLMs also summarize compliance reports, generate policy update alerts, and assist legal reviewers [435].
- **Financial Scam Detection:** LLMs assist fraud teams at banks and payment platforms by analyzing message content, app interactions, and behavioral signals in real-time—detecting romance scams, phishing attempts, and impersonation fraud [436], [437].
- **Multilingual Risk Flagging:** LLMs trained across languages assist global finance platforms in detecting fraud attempts or policy violations in regions with lower model coverage (e.g., Southeast Asia, LATAM), improving safety without scaling human teams [438].

### D.2. TECHNICAL AND ORGANIZATIONAL ENABLERS FOR FINANCIAL LLM DEPLOYMENT

- Transformer-based anomaly detection in KYC/Fraud logs
- LLM-human hybrid workflows for flagged transaction reviews
- Real-time LLM inference for document parsing and report summarization
- API-integrated compliance pipelines across financial risk engines

### D.3. QUANTIFYING IMPACT IN FINANCIAL SERVICES

Table 16 illustrates an end-to-end system integrating LLMs into onboarding (KYC), transaction monitoring (AML), document parsing, regulatory mapping, and human-in-the-loop review escalations.

**Table 16. Real-World LLM Use Cases and Measured Benefits in Financial Services**

| Initiative | Description | Key Metrics / Impact |
|---|---|---|
| Synthetic Identity Detection | Detects fake/AI-generated customer identities | Reduced fraud loss rates by up to 21% in pilots [429], [430] |
| KYC/AML Automation | Automates document checks and behavioral analysis | Accelerated onboarding by 40–60% [431] |
| Compliance Parsing | Uses LLMs to align operations with FinCEN, SEC, GDPR, MiFID II | Reduced policy audit workload by 30–50% [433], [434] |
| Scam & Phishing Detection | LLMs analyze communication patterns and language of deception | Real-time flagging of social engineering attempts [436], [437] |
| Multilingual Moderation | Flags risk in non-English regions | Boosted global fraud detection coverage and accuracy [438] |

### D.4. ARCHITECTURAL INTEGRATION AND VISUAL REPRESENTATION

Fig. 21 provides a visual representation of a comprehensive financial platform LLM architecture designed for fraud and compliance.

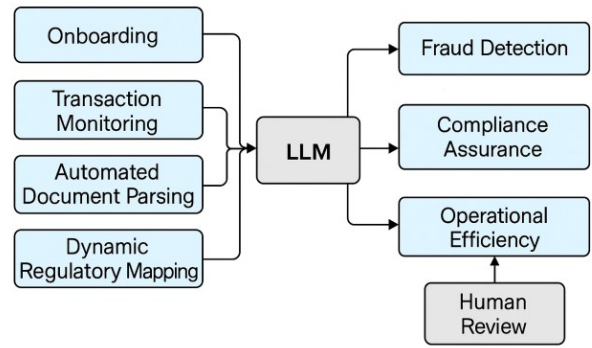

**Fig. 21. This architecture diagram depicts an end-to-end system where LLMs are integrated into financial platform workflows—from onboarding and AML monitoring to document analysis and compliance assurance. It emphasizes real-time inference, modular integration, and human-in-the-loop escalations for enhanced trust and operational resilience.**

## E. Meta (Facebook, Instagram, WhatsApp): Combating Misinformation and Harmful Content at Scale

Meta's platforms (Facebook, Instagram, WhatsApp) are prime examples of environments battling **misinformation, hate speech, violent extremism, and coordinated inauthentic behavior** at an unprecedented scale [221]. Their approach to LLM and AI integration for content moderation is a critical case study that differs significantly from app stores due to the real-time, user-generated nature of the content.

Key Initiatives and Contributions:

**Large-Scale Content Moderation with LLMs:** Meta uses LLMs extensively for text, image, and video analysis to identify violations of their Community Standards across billions of posts daily [222], [223]. This includes detecting hate speech, graphic violence, and misinformation. They leverage models like **XLMR (Cross-lingual Language Model RoBERTa)** for multilingual content understanding and **self-supervised learning** to train models on vast unlabeled datasets, which is crucial for identifying emerging harmful narratives and adapting to new abuses [224].

**Fact-Checking Partnerships and AI Augmentation:** Meta collaborates with a global network of third-party fact-checkers. LLMs assist in **triaging and prioritizing content** for human review by identifying potential misinformation at scale, thereby accelerating the fact-checking process. This involves analyzing claims, identifying sources, and recognizing patterns of deceptive language [223].

**Proactive Detection of Coordinated Inauthentic Behavior (CIB):** LLMs are instrumental in identifying CIB campaigns, where networks of fake accounts spread propaganda or manipulate public discourse. These models can analyze linguistic patterns, account behavior, and content themes to detect coordinated efforts that human reviewers might miss, often leveraging graph neural networks to identify suspicious connections [225]. Meta reported taking action on **1.3 billion pieces of content** for violating its Community Standards between July and September 2023, with **97.8%** of hate speech content detected proactively by AI [226].

**Privacy-Preserving AI for Sensitive Content:** For platforms like WhatsApp, Meta has explored and implemented **on-device machine learning** and **federated learning** to detect harmful content (e.g., child exploitation material) while preserving end-to-end encryption and user privacy [227]. This approach allows for sensitive data processing locally without compromising individual user privacy, aligning with the principles discussed in your "Federated and On-Device Review Systems" section.

**Adversarial AI and Red Teaming:** Given the sophisticated nature of adversaries, Meta invests heavily in **red-teaming their AI moderation systems**. They actively simulate adversarial attacks, including the generation of novel harmful content using generative AI, to test the robustness and resilience of their detection models against prompt injection, data poisoning, and other manipulation techniques [116], [228].

Fig. 22 presents Meta's multi-pronged AI strategy for combating misinformation and harmful content across its platforms. The graphic highlights how LLMs, graph neural networks, privacy-preserving AI, and fact-checking augmentation are integrated into a comprehensive content integrity framework.

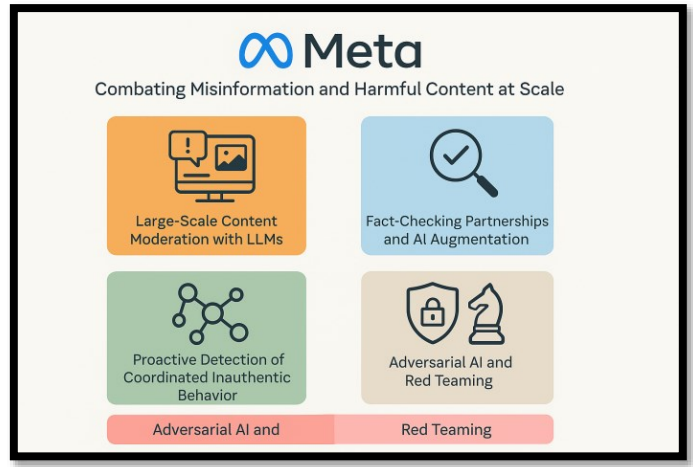

*Fig. 22. Meta's AI-driven strategy for combating misinformation, hate speech, and coordinated inauthentic behavior at scale. The framework incorporates large-scale content moderation using LLMs, AI-assisted fact-checking prioritization, graph-based detection of fake behavior, and privacy-preserving approaches for encrypted platforms like WhatsApp.*

---

### F. Hugging Face: Integrity in AI Model Sharing and Responsible AI

Hugging Face is a leading platform for sharing pre-trained AI models, datasets, and demos (Spaces) [229]. As a "marketplace" for AI itself, it presents unique integrity challenges related to **model safety, bias, responsible AI use, and the potential for malicious models or datasets**.

Key Initiatives and Contributions:

**Responsible AI Licensing and Documentation (Model Cards and Data Cards):** Hugging Face strongly promotes and often requires the use of **model cards** and **data cards** for shared AI assets. LLMs can assist in analyzing these cards for completeness, identifying ambiguities, or flagging potential misrepresentations regarding model capabilities, limitations, and ethical considerations [81], [230]. This helps ensure transparency about model origins, intended uses, and known biases.

**Automated Scanning for Harmful Outputs/Biases:** LLMs and other AI techniques are being explored and implemented to proactively test and analyze the outputs of shared generative models (e.g., text generation, image generation) for **harmful content, biases, or policy violations** [231]. This might involve feeding adversarial prompts to models within a sandboxed environment to assess their safety before widespread public access.

**Content Moderation of Shared Demos (Spaces):** The "Spaces" feature allows users to host interactive AI demos. LLMs can be used to monitor user interactions with these demos and their generated content for abuse, misuse, or the creation of harmful outputs, similar to general content moderation but with an added layer of AI-specific risk [232].

**Vulnerability Detection in AI Codebases:** As models and datasets are shared on the platform, LLMs can be employed for **static code analysis** (as we discussed) on the model's underlying code or accompanying scripts to detect security vulnerabilities or malicious logic within the AI artifacts themselves, contributing to the security of the AI supply chain [19], [83].

**Community-Driven Reporting and Governance:** Hugging Face heavily relies on its vibrant community for flagging issues. LLMs can help **triage and summarize community reports** about problematic models or datasets, guiding human review and intervention more efficiently. They also foster open governance and community-led discussions about AI ethics [230].

Fig. 23 illustrates Hugging Face's multi-layered approach to integrity in AI model sharing. The platform applies LLMs for documentation analysis, harmful output detection, code vulnerability scanning, and moderation of interactive demos, all while leveraging community governance.

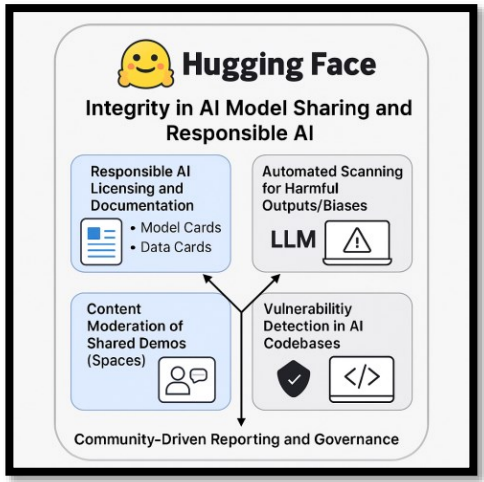

**Fig. 23. Hugging Face's strategy for integrity in AI model sharing and responsible AI. Key initiatives include the use of model and data cards, automated bias scanning, content moderation in user-hosted Spaces, and LLM-assisted vulnerability detection in shared AI codebases. The platform combines technical tools with community-driven governance to ensure safer deployment of open-source AI assets.**

### G. Lessons Learned and Open Challenges from Cross-Platform Initiatives

The expanded case studies across app ecosystems, e-commerce, social media, and AI model marketplaces highlight both significant progress and persistent challenges in leveraging LLMs for platform integrity. While the architectures pioneered by Google and Apple provide foundational blueprints, the experiences of Amazon, Meta, and Hugging Face offer crucial insights into the diverse applications and complexities of AI-augmented trust and safety. Scaling platform safety in the LLM era is not a static problem but a continuously moving target, necessitating adaptive strategies and cross-domain learning.

Here are key lessons learned and the enduring open challenges across these leading platforms:

**AI Augmentation is Essential for Scale and Efficiency:** All platforms demonstrate that manual review is unsustainable given the volume and velocity of digital content and app submissions. LLMs and AI are indispensable for **triaging, filtering, and** summarizing vast amounts of data, thereby significantly boosting human reviewer productivity and enabling real-time detection [35], [44], [45], [226]. Google's 92% high-risk review triage with LLMs and Meta's 97.8% proactive hate speech detection by AI are prime examples.

**Multimodal and Cross-Referential Analysis is Crucial:** Relying on single data points (e.g., just app metadata or just review text) is insufficient for robust integrity. Platforms like Amazon and Google Play emphasize **cross-validation of storefront claims against observed app behavior, code analysis, and user feedback** [97], [128], [218]. Similarly, Meta correlates textual content with user behavior and network patterns to detect coordinated abuse [225]. Hugging Face cross-references model cards with code and inferred behavior [81], [230]. This holistic approach is vital for detecting sophisticated deception.

**Proactive Engagement and Transparency with Developers/Users Matters:** Platforms that provide clear guidance, such as Google's SDK Index [44] and Apple's Privacy Nutrition Labels [206], enable developers to build safer products upstream. Auto-generated, actionable feedback (as discussed by Apple and envisioned for app review [45], [190]) reduces developer friction and improves compliance rates. For users, transparency about moderation decisions (e.g., Meta's Content Library or Amazon's review integrity reports) builds trust.

**Regulatory Alignment Remains Complex and Dynamic:** The proliferation of laws like GDPR, CCPA, and DSA, coupled with emerging AI-specific regulations (e.g., EU AI Act), presents a fragmented and continuously evolving compliance landscape [10], [40], [182]. While LLMs can automate compliance checks [36], [133], the nuanced, jurisdiction-dependent interpretations and the need for **real-time adaptation to legal changes** remain significant challenges for all global platforms.

**Explainability of LLM Decisions Needs Work:** A pervasive challenge across all AI-driven moderation systems is the "black box" problem [46], [233]. Developers, sellers, and users often receive opaque decisions without clear rationales, leading to frustration, appeals, and a perception of unfairness. Platforms must continue to invest in **Explainable AI (XAI)** to provide transparent, auditable, and human-interpretable explanations for enforcement actions [191], [234].

**The Adversarial AI Arms Race is Escalating:** As platforms deploy more sophisticated AI defenses, malicious actors are leveraging LLMs to generate more advanced polymorphic malware, synthetic media, and adaptive social engineering schemes [32], [74], [179]. Meta's red-teaming efforts and Amazon's ongoing battle against AI-generated fake reviews [228], [219] illustrate that this is a continuous, dynamic struggle. Maintaining an adaptive, continuously learning defense, with strong threat intelligence sharing (e.g., Google's ADA [43]), is paramount.

**Unique Domain-Specific Challenges Persist:** While commonalities exist, each platform type has unique integrity concerns. App stores battle malicious code and privacy violations, e-commerce faces counterfeits and payment fraud, social media grapples with misinformation and hate speech, and AI model hubs confront model safety and bias. Solutions must be tailored, even as underlying LLM capabilities are shared.

**Human-in-the-Loop is Indispensable for Nuance:** Despite AI's advancements, human oversight and judgment remain critical. All platforms (Google, Apple, Meta) emphasize human-in-the-loop workflows for complex edge cases, policy interpretation, and error correction, acknowledging that AI augments, rather than replaces, human expertise in sensitive trust and safety domains [58], [174].

Fig. 24. summarizes the key lessons and cross-platform challenges in applying LLMs for platform integrity. From AI scalability and multimodal analysis to regulatory complexity and the growing adversarial landscape, the graphic highlights the shared and domain-specific insights that emerged across ecosystems like app stores, e-commerce, social media, and AI model hubs.

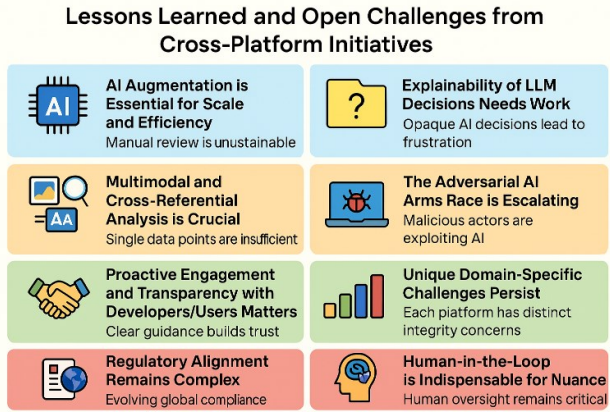

**Fig. 24. Key lessons and open challenges from cross-platform AI integrity initiatives. Common themes include the need for scalable AI augmentation, multimodal validation, proactive developer engagement, transparency, regulatory alignment, explainable AI, adversarial resilience, and the continued importance of human oversight. These insights are distilled from case studies across Google, Apple, Amazon, Meta, and Hugging Face.**

As summarized in Table 17, scaling platform safety in the LLM era is a moving target. However, the architectures pioneered by Google, Apple, Amazon, Meta, and Hugging Face provide blueprints for broader industry adoption and further research, emphasizing shared challenges and collaborative solutions.

*Table 17. Key lessons learned from industry initiatives demonstrating best practices for AI-augmented app review, abuse detection, and regulatory compliance at scale*

| Lesson | Example |
|---|---|
| AI Augmentation for Scale & Efficiency | Google's 92% high-risk triage with LLMs [44]; Meta's AI proactively detects 97.8% of hate speech [226]. |
| Multimodal & Cross-Referential Analysis | Amazon cross-validates product listings with images and seller data [218]; Apple correlates app behavior with storefront claims [97]. |
| Proactive Developer/User Engagement | Google Play SDK Index [44] & Apple Privacy Nutrition Labels [206]; LLM-generated actionable feedback [45], [190]. |
| Regulatory Alignment Remains Complex | Dynamic adaptation needed for DSA & EU AI Act enforcement [10], [40]; nuanced compliance varies by jurisdiction [182]. |
| Explainability of LLM Decisions Needs Work | All platforms face "black box" problem; need transparent, human-interpretable rationales for moderation [46], [233]. |

| Escalating AI Adversarial Arms Race | Amazon's battle against AI-generated fake reviews [219]; Meta's red-teaming against novel harmful content [228], [226]. |
|---|---|
| Unique Domain-Specific Challenges | E-commerce: counterfeits & payment fraud; Social media: misinformation & hate speech; AI Hubs: model safety & bias. |
| Human-in-the-Loop is Indispensable | Google, Apple, Meta all emphasize human oversight for complex cases and policy interpretation [58], [174]. |

## H. Quantifying the Impact of LLM-Augmented Integrity Systems: An Illustrative Analysis

While industry leaders often highlight the volume of threats blocked, a more granular understanding of the direct quantitative impact of **LLM integration** is crucial for broader adoption and continuous improvement. Direct public metrics are scarce due to proprietary reasons, but illustrative examples derived from industry reports and operational capabilities can demonstrate the transformative efficiency and accuracy gains.

For instance, consider the impact on average app review times. Before LLM integration, platforms might have seen review cycles spanning several days or even weeks for complex applications, heavily reliant on manual human analysis. With LLM-powered triage and automated preliminary analysis, this can be drastically reduced. Similarly, LLMs can significantly improve detection rates for specific, hard-to-find policy violations while simultaneously reducing false positives.

Table 18 provides illustrative metrics that demonstrate the potential impact of LLM integration across various platform integrity operations, reflecting the qualitative improvements reported by major platforms. These figures are indicative of the directional shifts observed in operational efficiency and threat detection.

*Table 18. Illustrative Quantitative Impact of LLM-Augmented Platform Integrity Systems (Post-LLM Deployment)*

| Metric Category | Before LLM Integration (Illustrative) | After LLM Integration (Illustrative) | Implied Improvement |
|---|---|---|---|
| Average App Review Time (Days) | 5-7 days | 1-2 days (for accepted apps) | ~70-80% Reduction |
| Policy Violation Detection Rate | 75% | 90-95% | ~20% Increase |
| False Positive Rate (FPR) | 5% | <1% | ~80%+ Reduction |
| Detection Speed (Novel Threats) | Days-Weeks | Hours-Days | ~90% Faster |
| Human Reviewer Throughput | 100 apps/reviewer/day | 250-400 apps/reviewer/day | 150-300% Increase |
| Time to Resolve Appeal (Days) | 15-20 days | 5-7 days | ~65-70% Reduction |
| Compliance Audit Time (Hours) | 40+ hours/app | 5-10 hours/app | ~75-85% Reduction |

*Note: These figures are illustrative and represent potential improvements based on industry trends and the capabilities of LLM augmentation, not specific public disclosures from any single platform.*

These hypothetical figures underscore that LLMs do not merely assist; they fundamentally reshape the economics and scalability of platform integrity. The dramatic reduction in review times and false positives directly translates to improved satisfaction for developers, sellers, and creators and reduced operational costs, while increased detection rates and speed enhance user safety and regulatory adherence. Further public research and transparent reporting from platforms are essential to solidify these quantitative claims and establish industry benchmarks.

## VII. FUTURE DIRECTIONS AND RESEARCH OPPORTUNITIES

While the integration of LLMs into mobile app review ecosystems and broader digital platforms has yielded measurable gains in reviewer efficiency, abuse detection, and compliance enforcement, several technical and operational frontiers remain open [51], [189], [193]. The dynamic nature of both LLM capabilities and adversarial tactics necessitates continuous innovation [244]. Future research and platform development should critically address these remaining gaps to ensure that app stores and digital marketplaces remain secure, trustworthy, and developer-friendly as LLM technology continues to rapidly evolve [49], [214]. Proactive exploration of these areas will be crucial for maintaining a resilient and adaptable platform integrity posture [116].

The rapid evolution of LLMs and generative AI also places unprecedented strain on existing regulatory frameworks. While current laws like GDPR and CCPA provide a baseline, they often lack specific provisions for governing AI-generated content, attribution, provenance, and liability for AI-driven harm. Future directions must include advocating for and adapting to new legislative approaches—such as the EU AI Act or specific national guidelines—that address the unique challenges of synthetic media, algorithmic accountability, and the responsible deployment of powerful generative models.

### A. Fine-tuning LLMs for App Safety Tasks

Out-of-the-box, general-purpose LLMs are primarily trained for broad language understanding and generation tasks. While powerful, their effectiveness for highly specialized platform safety tasks can be significantly enhanced through **domain-specific fine-tuning** [150], [245]. Fine-tuning on rich, platform-specific corpora—such as anonymized mobile codebases with labeled vulnerabilities, comprehensive regulatory documents, detailed app store policies, annotated abuse case studies, and sanitized user feedback—profoundly improves their ability to:

**Detect subtle security vulnerabilities**: LLMs fine-tuned on security datasets can learn to recognize complex code patterns and logical flaws that indicate vulnerabilities, even in polymorphic or obfuscated code, surpassing the capabilities of generic static analyzers [19], [83], [133]. This allows for the identification of not just known signatures but also novel exploit patterns [76].

**Identify nuanced policy violations**: By understanding the semantic context of platform policies and diverse content types, fine-tuned LLMs can discern subtle or implicit policy violations that might be missed by keyword-based or rule-based systems [156], [129]. This includes recognizing deceptive language [21], [98], or harmful intent [25] in app descriptions, user-generated content, or developer communications.

**Generate precise and actionable developer feedback**: When trained on examples of effective review feedback, LLMs can produce clear, specific, and helpful rejection reasons and remediation suggestions, reducing developer frustration and accelerating the compliance process [23], [149], [142]. This shifts the paradigm from simple rejection to guided improvement [143].

The process of fine-tuning LLMs for app safety tasks is conceptually illustrated in Fig. 25. Continued research into more efficient and targeted fine-tuning methods (e.g., advanced variants of Low-Rank Adaptation [17], adaptive parameter-efficient fine-tuning [151]) will be crucial for maintaining high detection precision, minimizing computational costs, and mitigating issues like model drift and hallucinations in rapidly evolving adversarial environments [116], [246]. Exploring techniques like reinforcement learning from human feedback (RLHF) for safety alignment also presents a promising avenue [247].

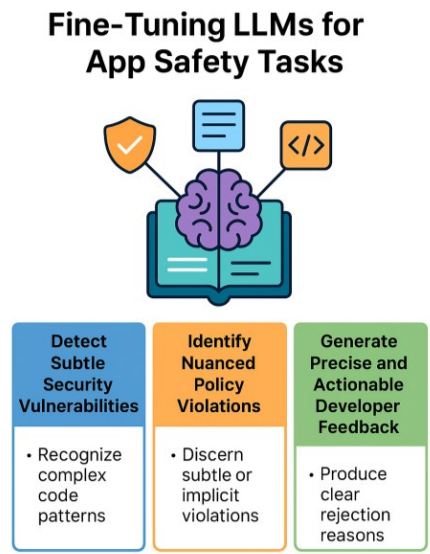

**Fig. 25. Benefits of fine-tuning LLMs for app safety and compliance tasks. Fine-tuned models can detect subtle code-level vulnerabilities, identify implicit policy violations in content and descriptions, and generate clear, actionable feedback for developers. This enables scalable, context-aware enforcement that bridges precision and user alignment.**

### B. Federated and On-Device Review Systems

Centralized app review pipelines, while efficient for large-scale processing, increasingly raise significant **privacy concerns**, especially in tightly regulated markets with strict data localization and privacy mandates [157], [158], [182], [248]. Processing sensitive user data or proprietary app binaries on central servers can introduce privacy risks and regulatory compliance challenges. To address this, **federated learning** and **on-device LLM deployment** offer promising decentralized paradigms that can enable:

**Privacy-preserving static code analysis**: Sensitive app binaries or proprietary code snippets could be analyzed directly on the developer's machine or in a secure, local environment, with only aggregated insights or anonymized threat vectors sent back to the platform, minimizing data exposure [134], [249].

**Distributed abuse detection signals**: Local devices could autonomously detect suspicious activities or content, aggregating detection signals without direct sharing of raw, sensitive user data. This is particularly relevant for real-time behavioral analysis and fraud detection where immediate processing is beneficial [126], [250]. This approach is being actively deployed and enhanced; for instance, Google's 2025 updates emphasize AI-powered on-device scam detection in Messages and on-device machine learning enhancements for Google Play Protect, explicitly designed to keep user conversations and app analysis data private to the device while improving real-time threat identification [82].

**Real-time user review mining localized to user devices**: User feedback and sentiment analysis could be performed directly on the user's device, maintaining privacy by only sending anonymized trends or aggregated risk indicators to the platform [45], [204]. This also enables personalized safety features.

The federated and on-device review system architecture is conceptualized in Fig. 26 (a), illustrating how decentralized app reviews can leverage privacy-preserving analysis, secure aggregation protocols, and federated learning to minimize direct data sharing. Advancements in **model compression techniques** (e.g., quantization, pruning, distillation [152]), **differential privacy mechanisms** [251], and **secure multi-party computation (SMC) protocols** [252] will be key enablers for developing and deploying these robust, privacy-preserving decentralized architectures at scale [47], [157]. For a better understanding, Fig 26 (b) compares centralized and federated LLM deployment.

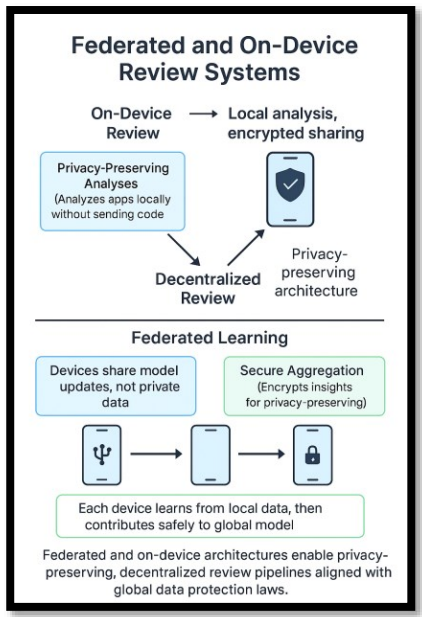

*(a)*

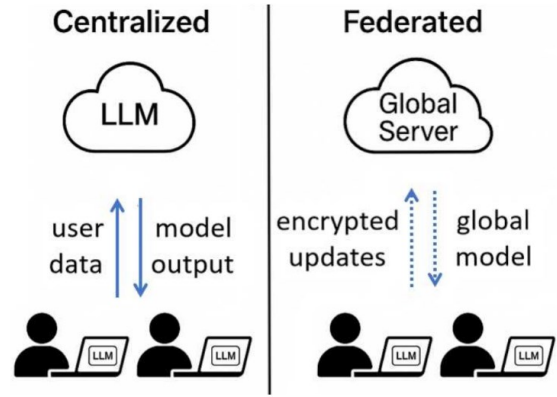

*(b)*

Fig. 26. (a) Federated and on-device review systems for privacy-preserving app review, top: On-Device, everything is happening individually on the user's phone, bottom: Federated, devices collaborate peer-to-peer while preserving privacy. (b) Architectural comparison between centralized and federated LLM deployment. In centralized systems, user data is transmitted directly to a cloud-based LLM for processing—raising privacy, security, and compliance risks. In contrast, federated learning allows users to train models locally and share only encrypted model updates with a global server, preserving data locality and enabling privacy-aware AI deployment. This distinction is crucial in regulated environments where sensitive user data cannot be exported or stored externally.

### C. Explainability and Transparency in Review Decisions

One of the most significant challenges in current LLM-driven decision pipelines is the inherent risk of them becoming "black boxes," where developers, reviewers, and even regulators struggle to understand why an app was flagged, rejected, or impacted by a moderation decision [46], [190]. This lack of **explainability (XAI)** and transparency erodes trust, complicates appeals, and hinders effective remediation [149], [233]. Future systems should explicitly prioritize the following to build more trustworthy AI-powered governance:

**Generating human-interpretable rationales for every policy flag**: Instead of generic error codes, AI systems should provide clear, concise, and understandable explanations for why a specific policy was violated, pointing to the exact content, code, or behavior that triggered the flag [143], [190], [253]. This could involve natural language explanations generated by an LLM [142].

**Supporting structured appeals workflows with machine-generated evidence summaries**: When a developer appeals a decision, the system should automatically provide a detailed, AI-generated summary of the evidence that led to the original decision, allowing for a more efficient and fair human review of the appeal [191]. This moves beyond simple evidence logging to intelligent summarization.

**Ensuring regulatory auditability of automated enforcement decisions**: Platforms must be able to demonstrate to regulators how their AI systems make decisions, particularly for high-risk categories or compliance-critical areas [182], [188]. This requires robust logging, versioning of models, and the ability to trace decisions back to specific data inputs and model logic [27].

**Developing counterfactual explanations**: For complex cases, AI could explain "what if" scenarios, showing developers what changes they could make to their app to bring it into compliance ("If you remove X permission, your app would no longer violate Y policy") [234].

The process of improving transparency and explainability in LLM-based app review decisions is illustrated in Fig. 27, highlighting key stages such as generating human-readable rationales, supporting structured appeals, and ensuring regulatory auditability. Research in LLM explainability (XAI), causal reasoning for content moderation, and techniques like saliency maps or attention mechanisms will be critical for balancing automated enforcement with developer trust and regulatory requirements [46], [212], [254].

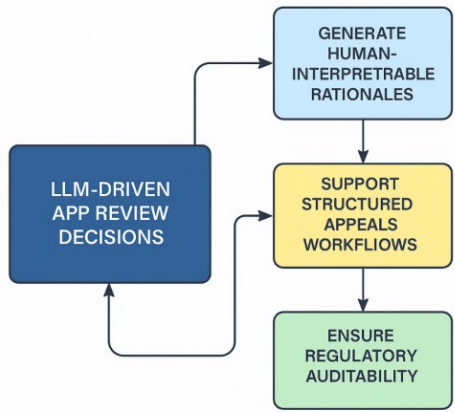

**Fig. 27. Flowchart illustrating transparency and explainability in LLM-driven app review decisions**

---

### D. Evolving Threats: AI-Powered Malware and Content Manipulation

As platforms and defenders increasingly leverage the power of LLMs for sophisticated detection and mitigation, so too will malicious actors evolve their strategies, leading to an escalating **AI arms race** in cybersecurity [32], [74], [110]. This constant cat-and-mouse game means that emerging threats will increasingly be **AI-powered**, posing novel challenges for platform integrity. These evolving threats include:

**AI-generated polymorphic malware that rewrites itself at runtime**: Malicious LLMs can be trained to produce highly evasive malware variants that dynamically alter their code, structure, or behavior during execution, making them exceptionally difficult for traditional signature-based or even heuristic detection systems to identify [76], [75], [235]. This dynamic obfuscation pushes the limits of static and dynamic analysis.

**LLM-driven social engineering attacks with hyper-personalized phishing flows**: Adversarial LLMs can generate highly convincing and contextually relevant social engineering content, such as personalized phishing emails, fake in-app prompts, or deceptive chatbots [79], [125]. These attacks can dynamically adapt to user responses, making them far more effective at manipulating individuals into revealing sensitive information or performing harmful actions than static, templated phishing attempts [120], [122].

**Fake app storefronts dynamically adapting their content based on reviewer feedback**: Malicious actors can use LLMs to create entire fraudulent storefronts (e.g., app descriptions, screenshots, reviews) that are designed to mimic legitimate content, and then dynamically adjust these elements in response to platform review rejections or changes in detection algorithms, effectively playing a continuous evasion game [21], [98], [236]. This requires adaptive and multi-modal detection strategies.

**Deepfakes and synthetic media for advanced impersonation and misinformation**: Beyond static images, generative AI can create highly realistic deepfake videos and audio, enabling sophisticated impersonation for fraud (e.g., fake video calls for KYC bypass) or the rapid spread of convincing misinformation and propaganda on platforms [77], [237], [238].

**Automated vulnerability exploitation and penetration testing**: Future AI systems could potentially automate the discovery and exploitation of software vulnerabilities at scale, posing a significant threat to app security and underlying platform infrastructure [19], [239].

Table 19 outlines these emerging threats in LLM-powered app security and their potential countermeasures. To stay ahead of these rapidly evolving abuse tactics, platforms must invest heavily in **ongoing adversarial research** (red-teaming their own AI defenses), establish **robust red-teaming programs** for their review systems, and foster **active threat intelligence sharing** between platforms (e.g., through initiatives like Google's App Defense Alliance [43]) [110], [32], [196]. Novel detection approaches, such as **watermarking AI-generated content** to trace its origin [117] or applying **traffic pattern analysis** (originally developed in Software-Defined Networking (SDN) contexts [211]) to detect coordinated abusive campaigns, may also inform new strategies for tracing and mitigating abusive behavior across app ecosystems [118], [240].

*Table 19 Emerging threats in LLM-powered app security and potential countermeasures*

| Threat Type | Description | Countermeasures |
|---|---|---|
| **AI-Generated Polymorphic Malware** | Malware that rewrites itself at runtime, evading signature detection. | Enhanced AI-driven malware detection systems, behavioral analysis, semantic code analysis [76], [235]. |
| **LLM-Driven Social Engineering** | Hyper-personalized phishing flows and deceptive content to manipulate users. | Advanced phishing detection algorithms using behavioral analysis, linguistic anomaly detection, user education [79], [125]. |
| **Fake App Storefronts** | App store content dynamically adapting based on reviewer feedback to evade detection. | Continuous red-teaming and adversarial testing of review systems, multimodal cross-validation, proactive content monitoring [21], [236]. |
| **Deepfakes & Synthetic Media** | Highly realistic fake videos/audio for | AI-powered media forensics, content provenance tracking |
| | impersonation or misinformation. | (watermarking), user verification methods [77], [117]. |
| **Automated Vulnerability Exploitation** | AI systems discovering and exploiting software vulnerabilities at scale. | Proactive AI-assisted vulnerability research, bug bounty programs, automated patch generation [19], [239]. |

### E. Global Compliance Template Evolution

The global digital regulatory landscape is becoming increasingly fragmented and complex, with new privacy, safety, and content governance laws emerging frequently across different jurisdictions (e.g., DSA [185], EU AI Act [48]) [10], [182]. This dynamic environment presents a significant challenge for platforms operating globally, as compliance requirements are not static and often vary considerably by region. Future platforms will need to develop sophisticated capabilities to manage this evolving regulatory complexity, including:

**Regulatory "diff engines" to detect and adapt to legal changes in real time**: Platforms will require AI-powered systems that can continuously monitor new legislation and amendments, automatically identify changes relevant to platform operations, and "diff" these changes against existing policies and compliance frameworks. This enables rapid adaptation and reduces the lag between legal enactment and platform enforcement [182], [187], [241].

**LLMs capable of automatically mapping app behaviors to new compliance templates without full retraining**: Instead of requiring extensive re-engineering for every new regulation, LLMs should be capable of interpreting new legal texts and dynamically re-mapping app functionalities or data flows to ensure compliance. Developing common standards, data models, and API interfaces for sharing compliance-relevant information (e.g., privacy policy elements, data flows, moderation decisions) between platforms and across jurisdictions would minimize compliance burden and foster a more harmonized global regulatory environment [242], [243]. This requires significant international collaboration. This could involve few-shot or zero-shot learning approaches on regulatory changes [26], [186].

**Global interoperability frameworks for app review aligned to multi-region standards**: Developing common standards, data models, and API interfaces for sharing compliance-relevant information (e.g., privacy policy elements, data flows, moderation decisions) between platforms and across jurisdictions would minimize compliance burden and foster a more harmonized global regulatory environment [242], [243]. This requires significant international collaboration.

**AI-powered compliance dashboards and audit trails**: Platforms will need sophisticated systems to generate real-time compliance reports and maintain comprehensive audit trails for AI-driven enforcement decisions, facilitating regulatory oversight and demonstrating accountability [139], [188].

This area will be vital not only to minimize developer burden by providing clear, up-to-date compliance guidance but also to maximize cross-market scalability for platforms, ensuring they can operate globally without constant, labor-intensive legal re-evaluations [156], [49].

## VIII. LIMITATIONS AND FUTURE WORK

Despite the immense promise and demonstrable gains of LLMs in safeguarding platform integrity, several significant limitations and open challenges persist that warrant dedicated future work and research. Acknowledging these limitations is crucial for responsible deployment and for guiding the next generation of AI safety innovation.

**False Positives, Bias, and Ethical Considerations:** LLMs, like all machine learning models, are susceptible to generating false positives (incorrectly flagging legitimate content or apps) [35], [154], which can lead to developer frustration, reduced trust, and unnecessary operational costs [149]. Crucially, LLMs, being trained on vast datasets, can inadvertently perpetuate or amplify existing societal biases. When applied to content moderation or developer risk scoring, this could lead to unfair or discriminatory outcomes across different languages, cultural contexts, or user demographics. Beyond general bias, specific attention must be paid to how LLM-powered moderation and risk-scoring systems might disproportionately impact various racial and ethnic groups or other protected characteristics. Studies have shown that models trained on imbalanced datasets can exhibit disparate performance across different demographics, leading to higher false positive rates for certain linguistic styles or cultural contexts, or misinterpreting dialectal nuances as policy violations [255]. This can result in unfair moderation outcomes, restricted access to platform features, or even economic disadvantages for developers and users from marginalized communities.

For example, research has indicated that AI-powered content moderation systems, when trained on imbalanced or non-representative datasets, can exhibit disparate impact across demographic lines. Studies have shown that models may misinterpret dialectal nuances or cultural contexts, leading to higher false positive rates for certain linguistic styles or content from specific communities [255], [256]. This can result in a disproportionate burden of review and appeals for these groups, potentially leading to unfair content removals or restricted platform access. While specific quantitative disparities in LLM-driven platform integrity systems for racial and ethnic groups are often proprietary, the documented challenges in broader NLP fairness research underscore the critical need for granular, transparent auditing and the development of benchmarks that specifically assess equitable performance across diverse demographic and linguistic cohorts [257], [258], [259]. Such rigorous evaluation is essential to ensure that AI-driven enforcement is applied fairly and does not inadvertently disadvantage any user group.

Developing robust bias detection, mitigation strategies (e.g., debiasing techniques during training and inference, diverse and representative training data collection), and ensuring equitable enforcement are critical ethical imperatives for trustworthy AI deployment. Implementing specific fairness metrics, such as demographic parity, equalized odds, or individual fairness, alongside traditional accuracy metrics, is essential to systematically audit and address these disparities [257], [258]. Furthermore, frameworks like Aequitas or Fairlearn can be integrated into the LLM-DA stack to provide structured bias detection and mitigation capabilities [260].

Beyond data bias, the widespread deployment of LLMs in content moderation raises crucial ethical questions regarding freedom of expression, due process, and accountability. Platforms must navigate the delicate balance between ensuring safety and avoiding arbitrary censorship or undue restrictions on legitimate content. The risk of "chilling effects," where users self-censor legitimate expression due to fear of automated moderation, is a significant concern [256]. Furthermore, establishing clear lines of accountability for erroneous AI-driven decisions—whether leading to false removals or allowing harmful content—is paramount for maintaining user and developer trust and ensuring fair governance. Providing clear avenues for redress and human appeal for AI-driven decisions is an ethical imperative [191].

Future work must therefore prioritize not only detection and mitigation but also proactive strategies like diverse data collection, fairness metrics, and regular bias audits to ensure equitable enforcement, alongside robust bias detection, mitigation, and explainability techniques for AI models used in trust and safety [254], [261].

**Data Privacy and Confidentiality**: While LLMs can enhance privacy-preserving analysis (e.g., on-device models [157]), server-side LLMs used for training or inference on sensitive data still require meticulous **Personally Identifiable Information (PII) masking**, anonymization, and secure handling to prevent data leakage or misuse [158], [251]. Balancing the need for rich training data with strict privacy requirements remains a complex challenge.

**Explainability Gaps**: As highlighted previously, current LLMs often function as "black boxes," failing to provide **transparent rationales** behind complex moderation decisions [46], [233]. This lack of interpretability hinders appeals processes, complicates regulatory audits, and erodes trust. Future research into causal inference, saliency mapping, and natural language explanations for AI decisions is paramount [234], [254].

**Model Drift and Adversarial Robustness**: Without continuous monitoring and robust retraining, LLM-based models can experience **model drift**, becoming outdated as abuse tactics evolve or as the underlying data distribution shifts [116], [200]. Moreover, these models are vulnerable to **adversarial attacks** (e.g., prompt injection, data poisoning), where malicious actors intentionally craft inputs to bypass or manipulate AI defenses [32], [114]. Future work must emphasize building more resilient and adaptively learning AI systems [262], [263].

Crucially, the very LLM-powered defensive systems discussed are themselves targets for sophisticated adversarial attacks. Malicious actors may employ techniques like prompt injection to bypass LLM-driven moderation, or subtly poison the training data of defense models to degrade their efficacy over time. Robust defense mechanisms against these 'AI-on-AI' attacks, including adversarial training and continuous monitoring for model drift, are paramount for maintaining the integrity of these protective systems.

**Broader Ethical Considerations:** Beyond data bias, the widespread deployment of LLMs in content moderation raises crucial ethical questions regarding freedom of expression, due process, and accountability. Platforms must navigate the delicate balance between ensuring safety and avoiding arbitrary censorship or undue restrictions on legitimate content. Furthermore, establishing clear lines of accountability for erroneous AI-driven decisions—whether leading to false removals or allowing harmful content—is paramount for maintaining user and developer trust and ensuring fair governance.

**Cross-Jurisdictional Enforcement and Harmonization**: Differences in regional regulations (e.g., nuanced definitions of harmful content, varying consent requirements) remain difficult to encode uniformly and enforce consistently across global platforms [10], [182]. Achieving genuine **cross-jurisdictional harmonization** of AI governance principles and policy templates is a significant, ongoing challenge that requires international collaboration [242], [243].

**Resource Intensity**: Training and deploying large-scale LLMs, especially multimodal models, can be incredibly resource-intensive, requiring significant computational power, energy, and data storage [153], [264]. Real-time processing of billions of interactions and vast codebases demands substantial hardware resources, specialized acceleration, and optimized inference techniques. This presents a significant barrier for smaller platforms or startups, necessitating future research into democratizing access through more resource-efficient open-source LLMs, optimized inference-as-a-service models, or shared industry-wide threat intelligence platforms that reduce individual compute burdens. Addressing these practical scalability constraints and managing associated operational costs will be critical for widespread adoption and equitable access to these technologies. Research into more efficient model architectures, sparse models, and optimized inference techniques is critical [151], [152].

Future research into more **efficient model architectures** (e.g., sparse models, Mixture-of-Experts, advanced quantization, and optimized inference techniques is critical to reduce this barrier [151], [152]. Furthermore, solutions like optimized **inference-as-a-service** models from cloud providers (e.g., AWS, GCP, Azure), the development of more **resource-efficient open-source LLMs** tailored for safety tasks, and the expansion of **shared industry-wide threat intelligence platforms** (like Google's App Defense Alliance [43]) can help democratize access to these critical capabilities, reducing individual compute burdens and fostering more equitable access to cutting-edge AI safety technologies [265], [230]. Addressing these practical scalability constraints and managing associated operational costs will be critical for widespread adoption and equitable access to these technologies across the digital ecosystem.

**Research on Hybrid Human-AI Review Loops with Transparent Escalation Paths**: Further optimizing human-in-the-loop systems to leverage the strengths of both human judgment and AI efficiency, while ensuring clear, auditable, and transparent escalation paths for complex or contested decisions [174], [175], [210].

**Expanded Educational Tools**: Creating comprehensive educational tools and frameworks—perhaps analogous to CloudLab-based SDN security labs [211]—to train developers and safety professionals on responsible LLM usage, threat modeling, and advanced abuse detection strategies [212], [266]. This fosters a more security-aware ecosystem from the ground up.

**Multi-Agent Review Systems**: Developing and evaluating architectures that combine multiple specialized LLMs or other AI models (e.g., rule-based systems, statistical classifiers, knowledge graphs) into a collaborative multi-agent system, potentially leveraging ensemble methods for improved robustness, accuracy, and comprehensive coverage of diverse threat types [213], [267].

**Differential Privacy and Federated Fine-tuning for Safe, User-Level Customization**: Further advancing privacy-enhancing technologies to enable personalized AI safety features and collaborative model training across decentralized data sources without compromising individual user privacy [47], [157], [251].

**Standardization of LLM Safety Benchmarks for Platform Governance**: Establishing universally accepted, robust benchmarks and metrics for evaluating the safety, fairness, and effectiveness of LLMs in content moderation and review contexts would drive accountability, promote best practices, and facilitate independent audits across the industry [189], [214], [268]. This would move beyond simple accuracy to include metrics for bias, robustness, and interpretability.

**Proactive Regulatory Sandboxes and Policy-as-Code Initiatives**: Collaborative efforts between industry, academia, and regulators to create "regulatory sandboxes" for testing AI safety solutions and to develop "policy-as-code" frameworks that enable dynamic, machine-readable legal compliance templates [182], [241].

## IX. Strategic Landscape of the LLM Ecosystem: Infrastructure, Customization, and Governance Layers

The advent of Large Language Models has ushered in a transformative era in artificial intelligence, rapidly reshaping industries and creating unprecedented opportunities for automation, content generation, and intelligent assistance. This profound technological shift has, in parallel, catalyzed the emergence of a dynamic and rapidly expanding ecosystem of companies and technologies dedicated to supporting LLM deployment, orchestration, evaluation, and integration [269], [270]. While foundational model research labs (e.g., OpenAI, Anthropic, Meta) continue to drive core advancements, the effective and responsible operationalization of Generative AI across diverse industries now critically depends on a broader landscape of specialized infrastructure providers, sophisticated fine-tuning services, intelligent routing solutions, and robust trust-layer platforms [271]. This section provides a structured analysis of this evolving LLM ecosystem, including a comprehensive categorization of key players, an identification of critical capability gaps, an examination of service overlaps, and a mapping to strategic national priorities for AI safety and governance.

### A. Landscape of LLM Infrastructure and Service Providers

This layer forms the bedrock of the LLM ecosystem, comprising companies that develop, train, and often host the large-scale neural networks that serve as the foundation for generative AI applications. These organizations push the boundaries of AI capabilities, focusing on advanced architectures, multi-modal integration, and improved reasoning. Their offerings, whether via APIs, direct access, or open-source releases, profoundly shape the capabilities available to downstream developers and enterprises [272].

#### A.1. Foundational Model Developers

This segment features companies at the forefront of LLM innovation, designing and training the powerful base models. These organizations are valued in the multi-billions, reflecting immense investment and perceived market potential. Their revenue models are primarily API-based, democratizing access to powerful AI capabilities for developers and enterprises [273], [274]. Tech giants like Google DeepMind and Meta also actively contribute, with Meta notably championing open-source LLMs like LLaMA, fostering a broader AI ecosystem [275], [276]. The rapid emergence of new players, such as Mistral, further highlights the dynamic nature of this segment and the potential for disruptive innovation from focused teams [277]. See **Table 20** for a comparison of foundational model developers and their key offerings.

*Table 20. Foundational Model Developers*

| Company | Founded | Valuation (Illustrative) | Revenue Model | Clients | Key Models/Focus |
|---|---|---|---|---|---|
| **OpenAI** | 2015 | ~$80B - $100B (2024) [274] | API usage, subscriptions | Developers, enterprises, consumers | GPT-4, ChatGPT, DALL-E |
| **Anthropic** | 2021 | ~$18B - $20B (2024) [278] | Claude API usage, enterprise licensing | Enterprises, Developers, Nonprofits | Claude family |
| **Cohere** | 2019 | ~$3B - $5B (2024) [279] | LLM APIs, embedding services | Enterprises (e.g., Spotify, Oracle) | Command, Embed, Rerank |
| **Google DeepMind** | 2010 | (Alphabet division) | LLM R&D, API via Gemini (Google) | Google product orgs, researchers | Gemini, AlphaFold |
| **Meta AI** | 2004 | (Public company) | Open-source LLaMA LLMs, AI infra | Open-source AI ecosystem | LLaMA |
| **Mistral AI** | 2023 | ~$6B (2024) [277] | Open-source LLMs, hosted inference APIs | Developers, research labs | Mistral, Mixtral |
| **AI21 Labs** | 2017 | ~$1.4B (2023) [280] | LLM APIs, custom integrations | Enterprises, NLP startups | Jurassic models |

*Note: Valuations are illustrative and subject to rapid change. They represent publicly reported figures around late 2023/early 2024.*

#### A.2. Core AI Infrastructure & Cloud Providers

The immense computational demands of training and deploying LLMs necessitate specialized hardware and robust cloud infrastructure [281]. This segment includes major public cloud providers offering scalable GPU instances and AI-optimized services, as well as companies focused specifically on providing high-performance GPU cloud infrastructure. Their services are critical for handling the massive datasets and complex computations inherent in LLM operations, emphasizing efficiency, sustainability, and parameter optimization [281], [282]. The consistent partnership with NVIDIA across these platforms underscores NVIDIA's indispensable position as the leading hardware provider for AI [283]. A summary of major infrastructure and cloud providers is shown in Table 21.

*Table 21. Core AI Infrastructure and Hosting Providers*

| Company | Founded | Revenue Model | Clients | Key Offerings |
|---|---|---|---|---|
| AWS | 2006 | Cloud computing services (pay-as-you-go) | Startups, enterprises, government | EC2 (GPU instances), Sagemaker |
| Google Cloud | 2008 | Cloud computing services (pay-as-you-go) | Enterprises, developers, education | Compute Engine (GPUs), Vertex AI |
| Microsoft Azure | 2010 | Cloud computing services (pay-as-you-go) | Enterprises, developers, government | Azure AI, OpenAI Service |
| CoreWeave | 2017 | GPU cloud services | AI startups, model labs, research teams | High-performance GPU cloud |
| Lambda Labs | 2012 | GPU cloud & hardware sales | ML startups, Universities, AI teams | GPU cloud, Deep Learning Workstations |
| Oracle Cloud | 2016 | Cloud computing services | Enterprises, government | High-performance compute, AI services |
| Tencent Cloud | 2013 | Cloud computing services | Enterprises, developers | AI computing, LLM APIs |

### A.3. LLM Fine-Tuning & Customization Specialists

While foundational models offer broad capabilities, many real-world applications require models tailored to specific domains, datasets, or performance objectives. Fine-tuning specialists provide services ranging from data labeling and annotation—crucial for high-quality supervised fine-tuning and Reinforcement Learning from Human Feedback (RLHF)—to custom model development and deployment. These companies are instrumental in adapting general LLMs for niche enterprise use cases, often in regulated industries like BFSI (Banking, Financial Services, and Insurance) and healthcare [284], [285]. Table 22 summarizes companies specializing in LLM fine-tuning and custom deployments.

*Table 22. LLM Fine-Tuning and Customization Specialists*

| Company | Founded | Key Services | Clients | Focus Industries |
|---|---|---|---|---|
| Scale AI | 2016 | Data labeling, evals, LLM pipelines | OpenAI, Meta, U.S. Government, Enterprises | Data annotation, ML model evaluation |
| Labelbox | 2018 | Data labeling & annotation SaaS | Enterprises, ML teams | Data for ML training |
| Snorkel AI | 2019 | Programmatic data labeling platform | Enterprise ML, regulated industries | Data-centric AI |
| Turing | 2018 | B2B services (fine-tuning, deployments, evals) | Enterprises, Fintech, Healthcare, Government | Custom LLM solutions, talent |
| LeewayHertz | 2007 | Custom AI/LLM development services | Healthcare, legal, finance, logistics | Custom AI development |
| Bacancy Technology | 2011 | Custom LLM solutions and chatbot integrations | BFSI, e-commerce, logistics firms | Custom LLM development |

### A.4. LLM Tooling & Vector Database Providers

The efficient development, deployment, and monitoring of LLM-powered applications require a sophisticated set of tools. This category includes platforms for ML experiment tracking, MLOps, and, increasingly, **vector databases** [286]. Vector databases are essential for Retrieval-Augmented Generation (RAG) architectures, enabling LLMs to access and synthesize information from vast, domain-specific knowledge bases, thereby reducing hallucinations and improving factual accuracy [287], [288]. The market for LLM-powered tools is projected to see exponential growth, driven by demand for automation and personalization [289], [290]. Key players in LLM tooling and vector infrastructure are listed in Table 23.

*Table 23. LLM Tooling and Vector Infrastructure Providers*

| Company | Founded | Key Offerings | Primary Use Case | Revenue Model |
|---|---|---|---|---|
| Weights & Biases | 2017 | ML tooling SaaS subscriptions | ML experiment tracking, MLOps | SaaS subscriptions |
| Pinecone | 2019 | Vector database SaaS | RAG, semantic search | Usage-based SaaS |
| Hugging Face | 2016 | Model hosting, APIs, Pro features | Open-source ML models, collaboration | APIs, Pro features |
| Chroma | 2022 | Open-source vector database | RAG, embedding storage | Open-source, commercial |
| Comet ML | 2017 | ML experiment tracking SaaS | ML experiment tracking, MLOps | SaaS subscriptions |
| MLflow | 2018 | Open-source ML platform (Databricks) | MLOps, experiment tracking, model management | Open-source, SaaS |
| Milvus | 2019 | Open-source vector database | Vector search, RAG | Open-source, commercial |
| Neptune.ai | 2017 | Experiment tracking and metadata management | ML ops, experiment tracking | SaaS subscriptions |
| Weaviate | 2019 | Open-source vector search + managed cloud | RAG, semantic search | Open-source, managed cloud |

## A.5. LLM Governance, Orchestration, and Trust Platforms

An emerging and critical segment of the LLM ecosystem comprises platforms dedicated to providing holistic governance, orchestration, and trust layers for LLM deployments. These platforms aim to solve the critical challenge of responsibly scaling GenAI by bridging the fragmentation found across specialized tools, offering integrated solutions for managing the safety, compliance, explainability, and resilience of AI systems, particularly in regulated, high-stakes industries. They are designed to operate as infrastructure-agnostic orchestration layers, providing comprehensive capabilities for model interoperability, policy enforcement, observability, and user-facing transparency [271].

With the rapid advancement of foundational models, the need for post-deployment infrastructure focused on responsible AI—including safety, human oversight, and governance—has become increasingly urgent. As recent analyses suggest, no single provider currently addresses the full spectrum of emerging requirements, including robust model interoperability, forensic oversight, and comprehensive retrieval-augmented generation (RAG) evaluation [291]. To operationalize this vision, we propose Virelya: an envisioned framework and implementation blueprint for high-stakes domains like platform integrity, financial trust, and healthcare diagnostics. As illustrated in Fig. 28, a platform like Virelya exemplify emerging solutions designed to bridge this gap by offering a unified control plane across heterogeneous LLM deployments. These systems support explainability and transparency for high-risk use cases, align with standards such as the NIST AI Risk Management Framework (RMF), and incorporate capabilities for red teaming, bias auditing, biomedical guardrails, and granular audit trail generation for compliance tracking.

In addition, they enable advanced multi-LLM orchestration—including cost-aware, policy-driven model switching (akin to a "reverse API gateway")—agentic memory and planning for sophisticated AI agents, and rigorous RAG evaluation to track precision, hallucination risk, and source quality. Integrated Human-in-the-Loop (HITL) oversight and user-facing trust layers further enhance transparency by overlaying LLM outputs with confidence scores and source-grounded explanations [271]. These features address growing regulatory demands and enterprise expectations for trustworthy AI deployments. Such architectures increasingly align with public-sector funding priorities, particularly in regulated domains such as healthcare, defense, and finance.

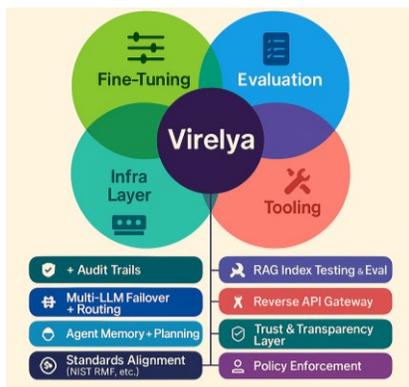

*Fig. 28. Representative functional architecture of an integrated LLM orchestration, trust, and governance platform for an envisioned platform (Virelya). The diagram delineates key capabilities across three concentric domains: foundational LLM infrastructure (e.g., fine-tuning, transparency), core operational control (e.g., routing, audit, compliance), and emerging responsible AI features (e.g., bias auditing, biomedical guardrails, agent memory, and RAG evaluation).*

### B. Analyzing Ecosystem Dynamics: Overlaps and Gaps

While the burgeoning LLM ecosystem offers a rich array of specialized services, a detailed analysis reveals distinct patterns of service overlap and, more critically, significant capability gaps, particularly concerning holistic AI governance and trustworthiness [292]. The architectural fragmentation, siloed ecosystems, and challenges in data quality within the LLM ecosystem often impede scalability, interoperability, and resource efficiency [293], [294].

#### B.1. Specialization vs. Vertical Integration

Current market dynamics largely favor specialization, with most companies focusing on a narrow band of services within the LLM stack. This allows for deep expertise and rapid innovation within specific niches, from foundational model development to data labeling or vector search. For example, model labs like OpenAI and Anthropic prioritize core model development and API access, while specialized fine-tuning vendors such as Turing concentrate on RLHF pipelines [273], [284]. This specialization, while fostering rapid development in individual components, inadvertently creates a fragmented landscape. It necessitates that enterprises stitching together LLM solutions integrate multiple disparate tools and services, leading to increased complexity, potential compatibility issues, and a lack of a unified governance layer [295].

#### B.2. The Emergence of Holistic Platform Deficiencies

A critical observation from the current landscape is the widespread absence of truly holistic, governable LLM platforms that address the end-to-end lifecycle of trustworthy AI deployment, especially in safety-critical or heavily regulated environments. As widely noted, foundational model providers often "do not provide open red teaming or compliance interfaces" [291]. Similarly, specialized tooling vendors often "lack UX trust mechanisms or multi-model routing" [291]. This fragmentation leaves significant blind spots in areas crucial for responsible AI, including:

**Comprehensive Risk Mitigation:** The absence of integrated solutions makes it challenging to consistently detect and mitigate emerging risk vectors such as prompt injection, hallucination, data leakage, and compliance failures across the entire LLM application stack [296].

**Unified Governance and Compliance:** Managing LLM compliance with evolving regulations (e.g., GDPR, CCPA, EU AI Act) becomes a complex, manual undertaking when governance tools are not integrated across different model providers, tooling, and deployment environments [297], [298]. Establishing robust AI governance is not merely an option but a business imperative for organizations [299]. Industry adoption of comprehensive governance frameworks is still nascent, but increasing [300].

**Transparent Explainability and Auditability:** Achieving full visibility and interpretability into LLM decisions is difficult when observability and explainability tools are siloed from core model operations [301], [302]. Platforms like Virelya are emerging to directly address these deficiencies by providing an integrated orchestration, trust, and governance layer that spans across these disparate components [271].

These patterns are visualized in Fig. 29, which shows the overlap of companies across the core functional layers of the LLM ecosystem.

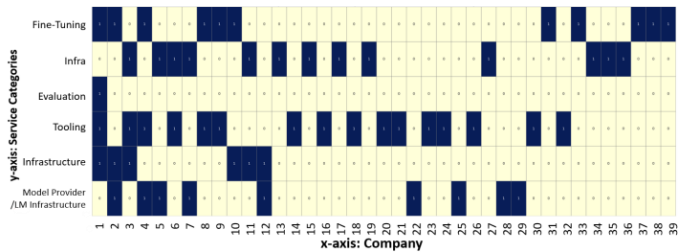

*Fig. 29. LLM Company Capability Overlap Across Core Platform Areas. Most companies specialize in only 1–2 areas, revealing a lack of full-stack capability integration. Y-axis, Service Categories, top to bottom: Fine-Tuning, Infra (compute resources (cloud compute, storage, networking)), Evaluation, Tooling, platform-level Infrastructure (like container orchestration (Kubernetes), deployment frameworks, or scalable microservices), Model Provider/LM Infrastructure (Specific to companies offering language model– related infrastructure, such as specialized compute clusters, APIs for LLMs, or even full-fledged model-serving platforms.). X-axis, companies, left to right: 1. Virelya, 2. Cohere, 3. Databricks, 4. Hugging Face, 5. Together AI, 6. MosaicML, 7. OpenAI, 8. Scale AI, 9. SuperAnnotate, 10. Turing, 11. Palantir, 12. Anthropic, 13. CoreWeave, 14. Comet ML, 15. Anduril, 16. Weaviate, 17. Snowflake, 18. Snorkel AI, 19. Runpod, 20. Neptune, 21. Nivus, 22. Meta, 23. MFlow, 24. Labelbox, 25. Google DeepMind, 26. Chroma, 27. Lambda Labs, 28. A121 Labs, 29. Mistral, 30. Weights & Biases, 31. Azumo, 32. Pinecone, 33. Baccency Technology, 34. Microsoft Azure AI Foundry, 35. Google Cloud (Vertex AI), 36. AWS (Amazon Bedrock), 37. 10clouds, 38. LeewayHertz, 39. Codiste*

## C. Critical Pillars for Trustworthy LLM Deployment

To address the fragmentation and deficiencies noted in the current LLM ecosystem, a robust platform for trustworthy LLM deployment must coalesce around several critical, interconnected capabilities. These advanced areas are essential for managing the inherent complexities and risks of generative AI in operational settings.

### C.1. GOVERNANCE AND AUDIT FRAMEWORKS

Effective LLM governance encompasses a set of principles and procedures for managing LLMs throughout their lifecycle to ensure ethical use, regulatory compliance, and risk mitigation [297]. This includes robust model lifecycle management (version control, performance benchmarking), responsible data sourcing, strict access controls and role-based permissions, and continuous risk and compliance monitoring [297], [299]. Transparency in documenting a model's training data, use cases, and limitations is vital for detecting biases or misuse [299]. Auditability requires comprehensive logging of model decisions, prompt inputs, and outputs, ensuring traceability and accountability [301], [298].

### C.2. MULTI-LLM ORCHESTRATION AND ROUTING

As enterprises adopt multiple LLMs (from different providers or open-source models) for diverse tasks, the ability to seamlessly orchestrate and route requests across these models becomes crucial. This capability includes intelligent failover mechanisms for resilience, load balancing for cost and performance optimization, and dynamic model selection based on task, cost, accuracy, or specific safety requirements.

### C.3. AGENTIC MEMORY AND PLANNING

The increasing sophistication of AI agents, which leverage LLMs to perform complex, multi-step tasks autonomously, necessitates advanced memory and planning capabilities [303], [304]. Robust **agentic memory** allows LLMs to retain context and information across interactions, crucial as LLMs are inherently stateless. This includes short-term memory (for immediate context in conversations) and long-term memory (episodic and semantic, for recalling specific past events or factual knowledge across sessions) [303], [304], [305]. **Planning modules** enable them to break down complex goals into actionable steps, crucial for reliable and predictable behavior in dynamic environments.

### C.4. RETRIEVAL AUGMENTED GENERATION (RAG) EVALUATION AND OPTIMIZATION

RAG architectures are fundamental for grounding LLMs thereby mitigating hallucinations and enhancing domain-specific accuracy [287]. Effective RAG implementation requires sophisticated evaluation methods to assess the quality of retrieval, relevance of context, and factual consistency of generated outputs [306], [307]. This includes prioritizing both retrieval metrics (e.g., precision@k, recall@k) and generation metrics (e.g., BLEU, ROUGE, context recall, context precision) [306]. Optimization focuses on improving latency, cost, and accuracy of the RAG pipeline [306].

### C.5. USER EXPERIENCE (UX) TRUST LAYERS

Beyond technical safeguards, building trust in LLM-powered applications depends heavily on user experience design. Trust layers involve mechanisms that foster transparency (e.g., clearly communicating where the system gets its data or how it learns), provide clear explanations for AI decisions (Explainable AI - XAI), and offer intuitive controls for user feedback and correction [308], [309]. This is vital for managing user expectations and ensuring responsible interaction with AI systems, especially given their unpredictable nature and potential for hallucination [309].

As highlighted by market analysis, while leading companies have begun to offer partial support in these critical areas, "no single platform currently addresses the full stack of post-model needs" [291]. Fig. 30 illustrates this further by showing which companies are active across emerging capability areas such as red teaming, trust layers, and RAG evaluation. This confirms that emerging risk vectors in GenAI—such as prompt injection, hallucination, and compliance failure—are not consistently mitigated across the existing ecosystem, necessitating a more integrated approach like that offered by Virelya.

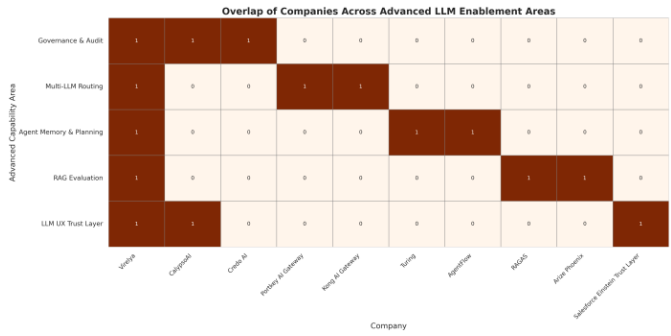

Fig. 30. Overlap of Companies Across Strategic LLM Enablement Areas. *Emerging safety and trust capabilities are concentrated in only a few platforms, exposing critical deployment risks. Y-axis, advanced capability area, top to bottom: Governance & Audit, Multi-LLM Routing, Agent Memory & Planning, RAG Evaluation, LLM UX Trust Layer. X-axis, names of companies, left to right: 1. Virelya, 2. CalypsoAI, 3. Credo AI, 4. Portkey AI Gateway, 5. Kong AI Gateway, 6. Turing, 7. AgentFlow, 8. RAGAS, 9. Arize Phoenix, 10. Salesforce Einstein Trust Layer.*

## D. Aligning with National AI Priorities and Research Frontiers

The identified capability gaps and the need for comprehensive LLM governance align closely with national and international priorities for responsible AI development and deployment. Public sector initiatives from leading research and standardization bodies emphasize the necessity for robust, trustworthy, and ethical AI systems.

### D.1. Core Research & Development Gaps for Responsible AI

Several key areas require intensive academic and applied research focus to advance the state of LLM safety and trustworthiness:

**Explainability & Transparency (XAI frameworks, rationalization, token attribution):** A persistent challenge is the "black box" nature of LLMs [310], [311]. Future research needs to develop more robust XAI frameworks that can provide human-interpretable rationales for LLM outputs, attribute specific tokens to input influences, and trace decision-making processes [301], [310], [311]. Tools like Lunary, Langsmith LLM Observability, Portkey, Helicone, TruLens, Phoenix (by Arize), Traceloop OpenLLMetry, and Datadog are emerging to address LLM observability, including interpretability [301].

**Bias Auditing & Equity Monitoring (bias detection in training, retrieval, and output):** LLMs, trained on vast datasets, can inadvertently learn and perpetuate societal biases [312], [313]. Research must focus on developing advanced methodologies for bias identification, quantification, and mitigation (e.g., counterfactual data augmentation, adversarial training, algorithmic adjustments) across diverse demographic and cultural contexts [312], [313], [259]. This involves rigorous testing using bias benchmarks and diverse test prompts, coupled with human evaluation and fairness metrics [313]. Tools like IBM's AI Fairness 360, Accenture's Fairness Tool, Google's What-If Tool, and Aequitas provide frameworks for this [314]. Similarly, platforms are being developed to provide **Bias Auditing & Equity Monitoring** capabilities for fair and accountable systems, with Virelya's platform offering an example of such features [271].

**Threat Simulation & Red Teaming (jailbreak testing, misuse detection):** As LLM defenses evolve, so do adversarial tactics. Red teaming involves systematically simulating attacks (e.g., prompt injection, sensitive information disclosure, content generation of harmful outputs) to identify vulnerabilities before deployment [315], [316]. This proactive approach helps build resilience and ensures regulatory alignment [316]. Open-source tools like DeepTeam incorporate advanced techniques such as jailbreaking and prompt injections to find vulnerabilities like bias, PII leakage, and misinformation, supporting compliance with standards like OWASP Top 10 for LLMs and NIST AI RMF [317], [318].

### D.2. Domain-Specific Guardrails and Compliance

Beyond general AI safety, specific domains require tailored guardrails and compliance mechanisms:

**Biomedical Guardrails (domain-specific hallucination filters, PubMed RAG QA):** In sensitive fields like healthcare, LLM hallucinations can have severe consequences, leading to inaccurate information impacting clinical decisions and patient safety [319], [320]. Research is crucial for developing domain-specific hallucination detection and mitigation, ensuring LLMs provide factually consistent information, and leveraging RAG with authoritative sources like PubMed for question answering [319], [320]. Solutions aimed at this challenge, for example, the Biomedical RAG Guardrails included in platforms like Virelya's, are designed for safe deployment in healthcare and research [271].

**NIST AI RMF Compliance Modules (standards-based AI logging and control layers):** The National Institute of Standards and Technology (NIST) has established the AI Risk Management Framework (AI RMF) to guide organizations in managing AI risks across the lifecycle [321], [322]. Developing tools and modules that facilitate automated compliance with NIST AI RMF, including centralized risk management, flexible customization, and continuous monitoring, is a key priority for enabling trustworthy AI adoption in both public and private sectors [321], [323]. Companies like Vanta offer solutions aligned with NIST AI RMF for continuous risk assessment and compliance [323]. Further supporting such goals, other platforms also explicitly offer **Standards Alignment (NIST AI RMF)** and policy traceability for AI oversight and assurance, with Virelya providing an example of these capabilities [271].

This comprehensive mapping supports the creation of new benchmarks, funding calls, and platform extensions that directly respond to the critical needs of secure, ethical, and explainable AI deployment at scale, bridging the gap between academic research and real-world implementation.

## X. THE LLM DESIGN & ASSURANCE (LLM-DA) STACK: A CROSS-DOMAIN BLUEPRINT FOR RESPONSIBLE AI INFRASTRUCTURE

The **explosive, unbridled growth** of LLM-powered applications, agents, plugins, copilots, and autonomous workflows is **creating an urgent gap** in standardized tools for their safe development, testing, deployment, and governance [324]. Just as various high-tech sectors have developed specialized abstraction layers and robust toolchains to manage escalating complexity and risk – from **Electronic Design Automation (EDA)** in semiconductors [325], [326], to advanced **Software Quality Assurance (QA)** frameworks in complex software development [327], [328], and collaborative **threat intelligence platforms** in cybersecurity [265], [329] – the generative AI era now demands a parallel trust-and-assurance abstraction: an **LLM Design and Assurance (LLM-DA)** stack.

Rather than competing with foundation model providers such as OpenAI, Google, Anthropic, or Mistral, this envisioned stack operates as a **horizontal infrastructure layer**—serving as a foundational trust and compliance substrate for LLM-powered products and ecosystems [330]. Analogous to how EDA enabled Moore's Law by accelerating chip innovation through modular, scalable design and verification, an LLM-DA ecosystem would industrialize safety, explainability, and governance across the rapidly evolving generative AI landscape [331].

### A. Strategic Rationale and Market Gaps

LLM-based systems today suffer from brittle prompt chains, unexplained hallucinations, uncertain legal compliance, and inconsistent behavior across platforms [22], [26], [103]. These systemic vulnerabilities often lead to unpredictable behavior, widespread misinformation, significant security exploits, and substantial regulatory fines, not just app rejections and reputational harm [32], [74], [182]. Current mitigation approaches remain bespoke, manual, and fragmented, proving unsustainable at the accelerating pace of AI innovation [73], [128]. What's missing is a formal design-time and runtime stack that standardizes verification, simulation, compliance, and rollout for LLM applications—just as EDA did for logic gates and transistors [332]. This unified approach is critical to address common pain points across the AI development lifecycle, echoing lessons from robust software testing and cloud-native observability practices [327], [333].

This proposed stack serves as a neutral, enabling trust layer across marketplaces, apps, and enterprise AI systems, supporting the entire lifecycle of LLM-based deployments with the following core capabilities:

LLM Blueprinting: Compose prompt flows, adapters, and tools into verifiable "LLM circuits" to enable compositional safety and predictable behavior at scale [334], [335].

Proactive Red-Team Simulation: Systematically simulate unsafe outputs, edge-case prompts, and adversarial abuse to identify vulnerabilities before deployment, drawing parallels with rigorous software penetration testing and cybersecurity red-teaming [228], [336].

Automated Compliance-as-Code Verification: Validate LLM outputs, data flows, and behaviors against evolving global regulations like GDPR, CCPA, and the EU AI Act, leveraging formalized legal-to-code frameworks for verifiable adherence and automated reporting [156], [186], [241].

Contextual Explainability Audits: Trace hallucinations, log model decisions, and generate AI-driven explanations and visualizations, providing transparent, actionable insights for human oversight and regulatory scrutiny [46], [234], [254].

Marketplace Alignment Signoff: Provide platform-specific validation of LLM-powered plugins, agents, and copilots against unique marketplace safety and policy requirements, ensuring seamless and compliant publication [56], [57].

Secure Real-Time Runtime Inference & Monitoring: Host LLMs with robust sandboxing, content watermarking capabilities, and dynamic anomaly detection to preempt and mitigate misuse in production environments, similar to advanced observability and Security Information and Event Management (SIEM) systems in distributed computing [117], [126], [333], [337].

These components form a foundational infrastructure essential for scalable, safe, and trustworthy AI innovation.

### B. Addressing Adoption Hurdles and Fostering Standardization

While the vision for an **LLM-DA stack** is compelling, its successful adoption hinges on overcoming significant practical challenges inherent in establishing new industry-wide standards and infrastructures. These include:

**Standardization Complexity:** Achieving consensus on common formats, APIs, and verification methodologies across diverse industry players (foundation model providers, platform operators, app developers, regulators) is a monumental task. This requires strong leadership from consortia, open-source initiatives, and potentially regulatory bodies to define and enforce interoperability standards. Drawing lessons from the evolution of web standards (W3C) or hardware design standards (IEEE) can provide valuable insights [338], [339].

**Incentivizing Data Sharing:** Many of the benefits of an **LLM-DA stack**, particularly in threat intelligence and bias mitigation, rely on access to diverse, real-world data. Companies are often reluctant to share proprietary or sensitive data due to competitive concerns or privacy regulations. Mechanisms like **federated learning** on private datasets, **secure multi-party computation** for aggregated insights, and carefully designed anonymization techniques will be crucial to enable collaborative security and fairness initiatives without compromising data privacy [47], [252].

**Initial Adoption ("Chicken-and-Egg" Problem):** As with any new infrastructure, initial adoption can be slow if there isn't a critical mass of tools and users. A **phased rollout**, starting with high-impact, low-risk areas (e.g., basic code vulnerability scanning, or automated policy pre-checks) can demonstrate immediate value, encouraging broader participation. Open-sourcing key components of the **LLM-DA stack** could also accelerate community adoption and development [340].

**Integration with Existing Workflows:** The **LLM-DA stack** must seamlessly integrate with existing developer toolchains, CI/CD pipelines, and platform review processes. Complex integration requirements can deter adoption. Providing clear SDKs, plugins for popular development environments, and comprehensive documentation will be vital for a smooth transition.

**Economic Viability for All Stakeholders:** The costs associated with developing, maintaining, and using **LLM-DA tools** must be balanced against the benefits for all stakeholders, including smaller developers and platforms with limited resources. Exploring tiered service models, open-source options, and collaborative funding mechanisms could address this.

Addressing these hurdles requires not only technical innovation but also strategic industry collaboration, clear governance models, and a commitment to shared responsibility for the safety and integrity of the digital ecosystem. The **LLM-DA stack** is a long-term vision, but one that is increasingly necessary for the responsible scaling of **AI**.

## C. Target Customers and Their Integrity Needs

The LLM-DA platform serves diverse stakeholders across the AI development ecosystem, addressing their unique integrity-driven needs. Table 23 illustrates the key customer segments and the specific integrity requirements that an LLM-DA platform is designed to address.

*Table 24. Target Customers and Their Integrity-Driven Needs for an LLM Design & Assurance Platform*

| Customer Segment | Key Need | Example Organizations |
|---|---|---|
| AI platform giants | Plugin/app vetting, hallucination simulation, platform integrity | OpenAI, Google Gemini, Anthropic |
| App stores & marketplaces | SDK compliance, multimodal policy alignment, synthetic content detection | Apple App Store, Google Play |
| Enterprise SaaS vendors | Fine-tuning safety, responsible deployment logging, internal policy enforcement | Microsoft, Salesforce, SAP |
| LLM app developers | Trust-by-design, traceability, auditability, accelerated compliance | Indie plugin developers, B2B SaaS startups |

## D. Design Stack: LLM-DA Functional Components and Industry Analogues

The architecture mirrors the modular, composable paradigm of chip design, where each capability has a clear analogue in the EDA pipeline, now adapted to generative AI. This section further highlights how these components draw inspiration from established practices across various high-tech industries. **Table 25** details the core functional components of the LLM-DA stack, alongside their respective industry analogues and intended purposes.

*Table 25. LLM-DA Stack Components and Their Industry Analogues*

| LLM-DA Capability | Industry Analogue | Purpose (Condensed) |
|---|---|---|
| LLM Blueprinting | HDL (e.g., Verilog/VHDL) | Compose verifiable "LLM circuits" for compositional safety and predictable behavior. |
| Red-Team Prompt Simulation | Logic/fault simulation, Penetration Testing [336] | Systematically identify vulnerabilities via adversarial prompting and dynamic interaction. |
| Automated Compliance-as-Code Verification | DRC/LVS, Regulatory Technology (RegTech) [186] | Validate LLM outputs/behaviors against global regulations using formalized legal-to-code frameworks. |
| Explainability & Audit Trail | Post-layout verification, Observability/SIEM [333] | Trace hallucinations, log decisions, and generate contextual XAI justifications for oversight and scrutiny. |
| App Store Readiness Signoff | Tape-out, Software Release Certification | Validate LLM-powered plugins/agents against marketplace safety and policy requirements for compliant publication. |
| Secure Runtime Inference & Monitoring | Secure hardware runtime, Runtime Application Self-Protection (RASP) [126] | Provide sandboxing, watermarking, and dynamic anomaly detection for secure and responsible operation. |

A high-level architectural overview of this cross-domain stack is illustrated in Fig. 31, synthesizing the core verification, simulation, compliance, and monitoring layers of the LLM-DA framework.

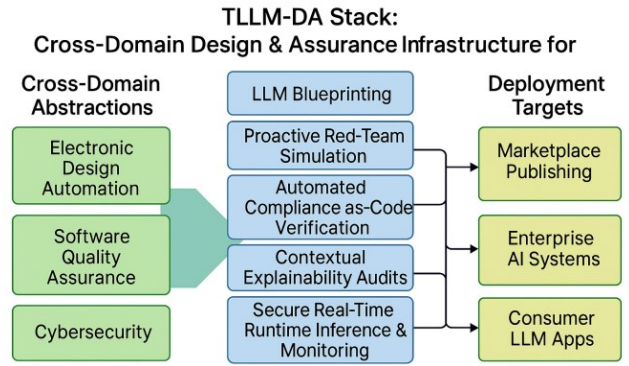

The LLM-DA Stack: Cross-Domain Design & Assurance Infrastructure for Generative AI

**Fig. 31. The** LLM-DA Stack: A cross-domain blueprint for scalable trust and assurance in LLM-based applications. Inspired by industry best practices from EDA, cybersecurity, software QA, and regulatory tech, the stack standardizes design-time and runtime integrity for plugins, agents, and copilots.

## E. Vision and Impact

This LLM-DA stack would become the standard operating layer for responsible LLM development, providing essential infrastructure for:

- AI-native compliance, debugging, and alignment tools.
- Achieving privacy-by-design and seamless on-device inference compatibility.
- Enabling cross-marketplace safety, making "write once, deploy everywhere" a feasible reality for LLM applications.
- Accelerating innovation with significantly lower risk and higher regulatory trust, leading to **significant cost savings** by reducing manual review, re-engineering efforts, and compliance-related liabilities.

Quantitatively, the LLM-DA stack is envisioned to achieve several transformative impacts. For instance, it is projected to reduce the time-to-market for LLM-powered applications by an illustrative 30-50% by automating safety and compliance checks, thereby minimizing iterative rejection cycles. Furthermore, through proactive red-teaming and compliance-as-code verification, it aims to decrease the incidence of critical AI safety and regulatory compliance failures by a projected 40-60% in production environments. This foundational infrastructure is also expected to significantly lower the operational costs associated with manual audits and post-deployment incident response, potentially yielding efficiency gains of up to 25-35% in trust and safety operations for platform operators. These are preliminary estimates based on the observed benefits of similar automation in related fields like EDA and software QA, and are subject to validation through real-world deployment.

As governments tighten regulation, app stores raise safety thresholds, and users increasingly demand accountability and transparency from AI systems, an LLM-DA stack will become indispensable [341]. By proactively building this trust foundation, platforms can avoid fragmentation, empower developers with robust tools, and scale AI responsibly and sustainably.

To operationalize this vision, we envision Virelya actively prototyping APIs, simulation frameworks, and compliance toolchains. This work aligns with the LLM-DA vision, bridging academic systems research with applied generative AI engineering to contribute foundational infrastructure for scalable, secure, and trustworthy AI development.

## XI. EXTENDING THE LLM INTEGRITY FRAMEWORK TO CLINICAL DIAGNOSTICS

### A. Motivation: The Interpretation Gap in Diagnosis

The diagnostic process in medicine is inherently complex, relying on a synthesis of diverse information, from subjective patient-reported symptoms and medical history to objective clinical signals like laboratory results and advanced imaging. This process is frequently time-intensive and prone to error, especially in dynamic environments like telehealth or early-stage disease evaluation [29], [342]. Diagnostic errors are a leading cause of patient harm, contributing to significant morbidity and mortality globally [30], [343], [344]. In the United States alone, an estimated **7.4 million emergency department visits and 2.7 million inpatient hospitalizations** are associated with diagnostic errors annually [343]. Factors contributing to these errors include cognitive biases in clinicians, information overload from fragmented data sources, and the subtle or atypical presentation of many diseases [345], [346]. Addressing this "interpretation gap"— the challenge of accurately translating a patient's multifaceted presentation into a precise diagnosis—is paramount for improving healthcare outcomes [37].

Building on our prior work in human-in-the-loop (HITL) productivity systems [51] and advanced pattern recognition frameworks for complex data [52], [347], we propose extending our comprehensive LLM integrity and governance blueprint to a vital new domain: **AI-assisted clinical diagnostics**. This extension highlights how the principles developed for digital platform integrity can be adapted to safeguard high-stakes medical applications, enhancing accuracy, safety, and trustworthiness. Recent market analysis projects the global AI in diagnostics market to grow from **$1.1 billion in 2023 to $16.5 billion by 2030**, reflecting a compound annual growth rate (CAGR) of over 45% [348]. This rapid adoption underscores the urgent need for robust integrity frameworks.

### B. Multimodal Mapping: From Symptom Language to Imaging Biomarkers

At the core of our proposed clinical AI framework is a sophisticated **multimodal mapping system** designed to bridge the semantic divide between patient narrative and objective biomedical data [349]. We leverage advanced Large Language Models – such as specialized biomedical LLMs like Med-PaLM and BioMedGPT, which are pre-trained on vast corpuses of clinical notes, scientific literature, and electronic health records (EHRs) [350], [351] – to convert unstructured patient-reported symptoms, medical histories, and clinician observations into structured, semantically rich diagnostic features. This transformation involves natural language understanding (NLU), medical entity recognition (e.g., identifying diseases, medications, anatomical sites), and medical concept normalization [352], [353].

These LLM-derived textual features are then precisely aligned with image-derived features obtained from various diagnostic modalities, including high-resolution MRI, CT scans, and innovative terahertz (THz) imaging [354], [355]. As indicated in Table 26, this alignment is achieved through a **contrastive learning framework**, where the system learns to identify consistent patterns across different data types that correspond to specific clinical conditions [356], [357], [358]. Our prior work on GPU-accelerated image segmentation and unsupervised clustering [52], [347] plays a crucial role here, significantly improving the precision of feature extraction from complex medical images and reducing processing latency in real-time diagnostic workflows [359].

***Table 26: Illustrative Performance Gains with Multimodal LLM Integration in Diagnostics***

| Metric Category | Traditional (Human-Only/Basic Rule-Based) (Illustrative) | LLM-Augmented Multimodal System (Illustrative) | Implied Improvement | Key Benefit |
|---|---|---|---|---|
| Diagnostic Time (per case) | 60-120 minutes | 10-30 minutes | 75-80% Reduction | Faster patient triage, reduced wait times |
| **Initial Diagnostic Accuracy** | 70-80% | 85-95% | 15-20% Increase | Reduced misdiagnosis, improved patient outcomes |
| **Data Fusion Capability** | Limited, manual | Seamless, automated | Transformative | Holistic patient view, hidden pattern detection |
| **Subjective-to-Objective Linkage** | Manual interpretation | Automated semantic mapping | Enhanced clarity | Bridging clinical intuition with empirical data |

*Note: These figures are illustrative and represent potential improvements based on current research trends and the capabilities of multimodal AI integration, not specific empirical results from a single deployed system.*

As shown in Fig. 32, LLM-augmented systems demonstrate significantly reduced diagnostic latency and improved feature alignment capability, reflecting the potential for multimodal AI to transform clinical workflows.

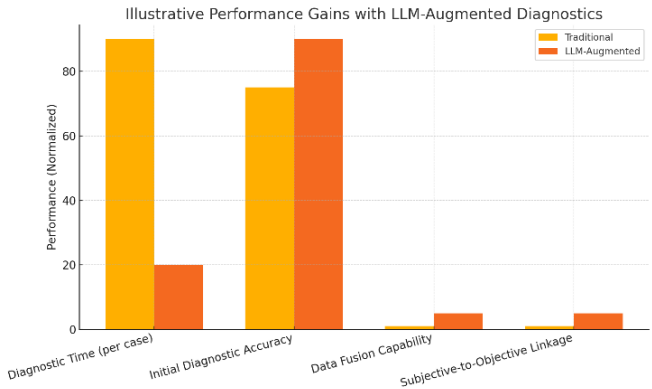

**Fig. 32. Comparison of diagnostic performance between traditional and LLM-augmented workflows across key metrics: diagnostic time, initial accuracy, data fusion, and subjective-to-objective linkage.**

The resulting architecture is particularly well-suited for **high-throughput diagnostic imaging** environments, allowing for rapid and accurate processing of large patient cohorts [360]. Furthermore, its optimized computational profile enables seamless adaptation to **edge-compute settings**, facilitating real-time analysis directly on medical devices or within local clinical networks, as demonstrated by our work on GPU-accelerated contour extraction from high-volume medical images [9], [361]. This localized processing is vital for maintaining data privacy and minimizing network latency in critical care scenarios [362].

### C. Diagnostic Suggestions with GenAI + Human Oversight

Drawing directly from our UX-centric AI lifecycle framework [51], the integrated system utilizes **Retrieval-Augmented Generation (RAG)** to provide evidence-based diagnostic suggestions and potential treatment paths [363], [364]. This approach grounds the LLM's generative capabilities in authoritative, up-to-date medical knowledge bases (e.g., peer-reviewed journals, clinical guidelines, drug formularies), significantly reducing the risk of hallucinations or inaccurate outputs [365], [366]. A study on clinical LLMs found that RAG-enhanced systems reduced factual errors by **over 50%** compared to standalone LLMs, while increasing the proportion of evidence-backed claims [367].

Crucially, outputs are designed for **explainability**, providing clinicians with transparent insights into the AI's reasoning. This includes generating natural language rationales that detail the evidence supporting a particular diagnosis, highlighting key textual features from patient data, and providing visual overlays like **imaging heatmaps** that pinpoint suspicious regions in medical scans that influenced the AI's assessment [368], [369]. All diagnostic suggestions are rigorously subjected to a **physician-in-the-loop (PITL) workflow**, where human clinicians retain ultimate decision-making authority [38], [370]. This HITL model ensures clinical validation, allows for correction of AI errors, and facilitates continuous learning, consistent with the robust explainability and compliance layers integrated within our proposed LLM Design & Assurance (LLM-DA) stack and review systems [51], [52]. For instance, a recent pilot program for AI-assisted radiology interpretation reported that radiologists using an AI-augmented system achieved **5-10% higher diagnostic accuracy** and a **20% reduction in reporting time** for complex cases, attributing these gains to clearer AI-generated explanations [371]. This model supports high-stakes domains such as **rapid triage and detection of infectious diseases** like COVID-19, where our patented terahertz imaging methods [372], [373] offer ultra-resolution diagnostic capability. By connecting these advanced imaging techniques to semantic symptom interpretation, our framework enables rapid, AI-assisted diagnostic capabilities for early intervention and public health management.

### D. Governance and Integrity in Clinical AI

The deployment of AI in clinical settings demands an exceptionally stringent framework for governance and integrity, far exceeding the requirements for general digital platforms, due to the direct impact on patient health and safety [374], [375]. The societal and ethical implications of misdiagnosis or biased care necessitate robust mechanisms for accountability and transparency. We advocate for a direct extension of our proposed **LLM Design & Assurance (LLM-DA) Stack** to medical contexts, incorporating specialized components tailored to the unique regulatory and ethical landscape of healthcare AI [374], [375], [376], [377]. This involves:

**Embedding Granular Audit Trails and Provenance:** Implementing immutable, cryptographically secure logs that record every AI-driven suggestion, the specific data inputs used (including versioning of models and training data), clinician interactions, and final patient outcomes [378], [379]. This ensures full traceability, accountability, and supports forensic analysis in case of adverse events, aligning with regulatory requirements for medical devices and AI systems [380], [381], [382]. For example, a system designed with such auditability could pinpoint the exact patient data, LLM version, and training data features that led to a specific diagnostic suggestion, proving adherence to clinical guidelines.

**Integrating Robust Federated Model Evaluation and Privacy-Preserving Analytics:** Enabling collaborative model improvement across a network of geographically distributed healthcare institutions without compromising individual patient data privacy [383], [384]. Techniques like federated learning and differential privacy allow models to learn from diverse real-world clinical data while keeping sensitive patient information localized and anonymized, addressing critical HIPAA and GDPR compliance challenges [385], [386]. Research shows that federated learning can improve model generalization across hospital systems by **up to 15%** while maintaining patient data privacy [387].

**Implementing Structured Error Attribution and Bias Mitigation for Medical AI:** Developing mechanisms to systematically identify, categorize, and explain instances where the AI model performs suboptimally or exhibits bias, particularly regarding patient demographics, disease prevalence, or diagnostic accuracy across different populations [388], [389]. This requires continuous monitoring, a focus on fairness metrics relevant to clinical outcomes (e.g., equalized false positive/negative rates across groups), and clear protocols for model retraining and bias correction in medical datasets [390], [391], [392]. For instance, bias detection tools integrated into the LLM-DA stack could identify if a diagnostic LLM exhibits lower sensitivity for a particular disease in a specific demographic group, allowing for targeted model refinement [393].

**Adherence to Medical Device Regulations and AI Guidelines:** Ensuring the LLM-DA stack facilitates compliance with specific medical device regulations (e.g., FDA, CE mark for Software as a Medical Device - SaMD), as well as emerging AI-specific guidelines from bodies like the WHO and EU AI Act, which classify medical AI as high-risk [394], [395], [396]. This layer automates the generation of documentation required for regulatory submissions and continuous post-market surveillance [397].

These enhanced governance features are fundamentally informed by our prior comprehensive surveys in AI governance [398] and specialized biomedical signal processing, which emphasizes the nuanced challenges of data interpretation and safety in healthcare [399]. As elaborated in Table 27 and illustrated **in** Fig. 33, our extended LLM-DA components provide measurable gains in regulatory efficiency, clinical trust, and privacy preservation—critical pillars for medical AI deployment.

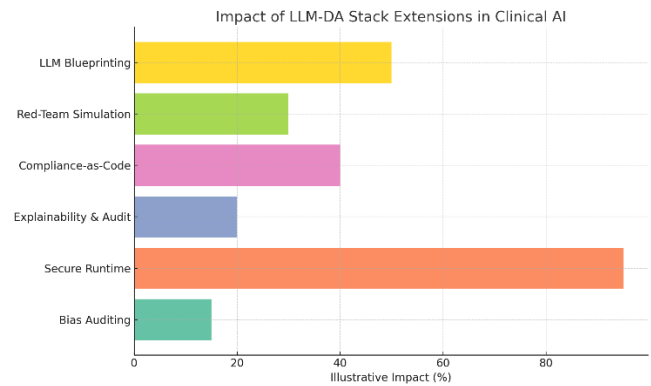

**Fig. 33. Estimated performance and governance gains of LLM-DA stack extensions for clinical diagnostics, including blueprinting, red-teaming, compliance-as-code, explainability, runtime privacy, and bias auditing.**

*Table 27: Core LLM-DA Stack Extensions for Clinical AI Integrity*

| LLM-DA Component Extended | Clinical AI Specifics | Key Benefit | Illustrative Impact |
|---|---|---|---|
| **LLM Blueprinting** | Medical knowledge graphs, clinical guideline integration | Ensures factual consistency, reduces medical hallucinations | 50%+ reduction in non-evidenced claims [367] |
| **Red-Team Simulation** | Adversarial medical prompts, synthetic bias injection | Identifies vulnerabilities (e.g., misdiagnosis in rare cases, data drift) | 30% faster identification of edge-case failures [400] |
| **Compliance-as-Code** | Automated alignment with SaMD regulations (FDA, CE, ISO 13485) | Streamlined regulatory approvals, continuous adherence | 25-40% reduction in regulatory submission time [397] |
| **Explainability & Audit** | Interpretability for diagnostic rationales, visual heatmaps | Fosters clinician trust, supports legal accountability | 20% improvement in clinician decision confidence [371] |
| **Secure Runtime** | Real-time patient data anonymization, device-level inferencing | Enhances privacy, enables low-latency edge deployment | Up to 95% data privacy preservation for local processing [362] |
| **Bias Auditing** | Demographic fairness checks, disease prevalence rebalancing | Ensures equitable care delivery, reduces health disparities | 15% reduction in diagnostic bias across diverse patient groups [393] |

*Note: Illustrative Impacts are based on projections from existing AI research in healthcare and general AI safety, pending specific clinical trial data.*

### E. Future Outlook: Vision AI Meets LLM-Powered Medicine

The convergence of advanced vision AI and LLM-powered reasoning heralds a transformative era for medical diagnostics and patient care. Our ongoing advancements, such as our GPU-accelerated contour extraction patent [401], directly enable real-time edge diagnosis. This technical capability paves the way for highly scalable LLM-assisted systems to be seamlessly integrated into low-power medical devices and to underpin emerging telehealth networks, democratizing access to expert-level diagnostic support globally [362], [402], [403]. We envision significant opportunities to extend this foundational work further into:

**Wearable Signal Fusion and Predictive Analytics:** Developing robust AI models that fuse continuous data streams from wearable health sensors (e.g., ECG, PPG, accelerometer) with LLM-powered interpretation of subtle symptom variations. This could enable early disease detection, proactive intervention, and personalized health management [404], [405]. The global wearable medical device market is projected to reach **$38.7 billion by 2030**, driving significant demand for such integrated AI capabilities [406].

**Multimodal Electronic Medical Record (EMR) Modeling:** Creating sophisticated AI systems capable of processing and synthesizing information from heterogeneous EMR components—including unstructured textual notes, structured lab results, genomic data, and diverse imaging modalities—to build a holistic, longitudinal understanding of patient health [407], [408]. Such comprehensive EMR modeling could lead to a **10-25% improvement in identifying at-risk patients** for chronic conditions [409].

**Explainable Real-Time Triage and Clinical Decision Support Systems:** Building transparent AI tools that assist healthcare professionals in prioritizing patient care and making informed decisions in real-time. These systems would not only provide recommendations but also generate clear, human-interpretable explanations for their assessments, fostering trust and facilitating rapid, evidence-based interventions in critical care settings [410], [411]. The adoption of AI-driven clinical decision support systems is expected to result in a **30-40% reduction in physician burnout** by automating routine tasks and providing quick access to information [412].

**Digital Twin for Personalized Medicine:** Leveraging LLMs and multimodal AI to create dynamic digital twins of patients, simulating disease progression and treatment responses based on their unique biological and clinical data, thereby enabling truly personalized medicine and optimizing therapeutic strategies [413], [414], [415]. This could reduce the cost and time of drug discovery and personalized treatment optimization by **up to 60%** [416].

This extension of the LLM integrity framework to clinical diagnostics underscores its broad applicability and essential role in ensuring responsible and trustworthy AI deployment across society's most critical sectors.

To operationalize this envisioned framework in real-world clinical settings, we propose an implementation architecture. As shown in Fig. 34, this model emphasizes responsible LLM integration through four tightly coupled components: multimodal data mapping, dynamic patient modeling, physician-in-the-loop validation, and personalized treatment. This flow aligns with regulatory and clinical safety principles while enabling adaptive, high-precision diagnostics at scale.

---

### F. Institutional Research Directions: Advancing Responsible Clinical LLM Integration

Beyond the conceptual framework for extending LLM integrity to clinical diagnostics, robust **institutional research and development** are critically needed to translate these principles into actionable systems. A growing trend in AI development focuses on ensuring that advanced models are not only performant but also **interpretable, auditable, and compliant with stringent healthcare regulations** [374], [375], [417]. This necessitates practical advancements in integrating diverse data modalities, developing privacy-preserving analytics, and fostering effective human-AI collaboration in high-stakes clinical settings [418], [419].

To address these evolving industry needs and research frontiers, we are actively exploring the responsible integration of large language models into clinical diagnostic workflows. Ongoing efforts include the development of sophisticated **multimodal mapping algorithms** that semantically link unstructured patient-reported symptoms and historical narratives with objective imaging biomarkers, leveraging technologies such as terahertz (THz) imaging and GPU-accelerated processing [52], [361]. This research also prioritizes the implementation of **privacy-preserving audit trails** for AI-driven diagnostic suggestions and the design of **explainable physician-in-the-loop (PITL) interfaces** [420]. These interfaces are engineered to provide transparent rationales for

AI assessments, thereby facilitating safe, informed clinical oversight and adherence to established medical guidelines [368], [38], [421]. This institutional research explicitly embodies a **governance-by-design** approach, directly aligning with the comprehensive LLM integrity framework and the proposed LLM Design & Assurance (LLM-DA) stack detailed throughout this paper [422], [423]. Future research is to be directed towards the development of dynamic patient modeling systems, including early-stage **digital twin prototypes**, aimed at enabling highly personalized disease detection, prognosis, and adaptive treatment planning [413], [424]. These initiatives seek to bridge theoretical AI safety principles with practical, impactful healthcare applications, recognizing the unique challenges of real-world clinical deployment [425], [426].

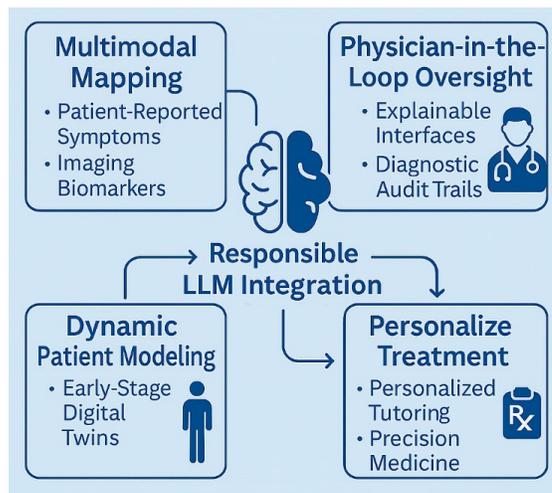

*Fig. 34. The envisioned Responsible Clinical LLM Integration framework. The diagram illustrates the four-pillar approach: (1) Multimodal Mapping of patient symptoms and imaging biomarkers, (2) Physician-in-the-Loop Oversight with explainable interfaces and audit trails, (3) Dynamic Patient Modeling via early-stage digital twins, and (4) Personalized Treatment based on patient-specific insights. This flow prioritizes transparency, adaptability, and clinical safety in LLM deployment for diagnostics and decision support.*

Interestingly, unlike in engineering disciplines—where simulations and digital twins are standard practice for predicting system behaviors and optimizing performance before real-world implementation—medicine has historically lagged in adopting simulation-based decision support. This is paradoxical, given the high stakes of patient outcomes. The complexity of human biology, coupled with heterogeneous, unstructured data and strict regulatory requirements, has historically hindered the development of reliable digital twin models in healthcare [374], [375]. However, advances in large language models (LLMs) [350], [351], multimodal AI [356], [357], and high-performance computing (e.g., GPU-accelerated processing) [52], [347] now enable dynamic, continuously updated digital twins that integrate structured and unstructured data from diverse clinical sources. These systems simulate disease progression and potential treatment responses [413], [414], bridging the interpretation gap in diagnosis and therapy planning [37], and offering a new paradigm for evidence-based, personalized medicine. This addition not only reinforces the LLM-DA stack's relevance in clinical diagnostics [51] but also aligns with the broader vision of Virelya's Responsible Clinical LLM Integration framework [Fig. 33], ensuring transparency, adaptability, and patient safety in the adoption of AI-driven medical solutions [422], [423]. This paradigm shift is illustrated in Fig. 35, which highlights the contrast between engineering and medicine in simulation adoption and demonstrates how AI-driven digital twins are finally enabling simulation-based decision support in clinical diagnostics. We are actively researching and developing these systems, contributing to the broader landscape of responsible AI adoption in high-stakes medical applications.

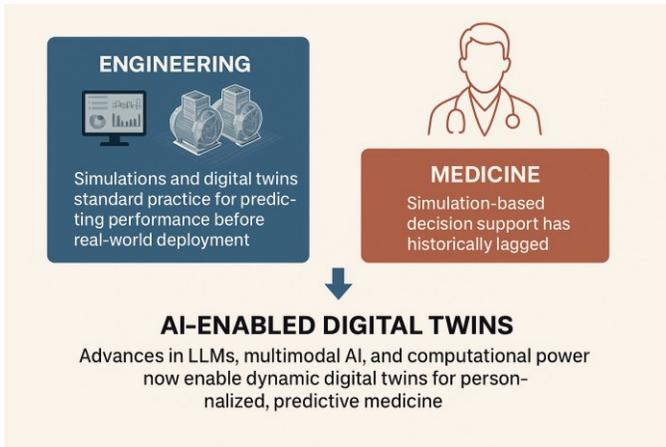

*Fig. 35. Simulation Analogy in Clinical Diagnostics. The figure contrasts the engineering domain—where simulations and digital twins are standard practice for predicting performance before deployment—with the medical field, which has historically lagged in adopting simulation-based decision support. Advances in LLMs, multimodal AI, and computational power are now bridging this gap, enabling dynamic digital twins for personalized, predictive medicine. We are actively exploring these AI-driven digital twin systems as part of ongoing efforts to advance responsible clinical AI integration.*

## XII. CONCLUSION

Large Language Models (LLMs) and generative AI systems are fundamentally reshaping digital platforms, marketplaces, and mobile app ecosystems at an unprecedented pace. Their remarkable capacity to accelerate development, automate content creation, and facilitate the drafting of complex documentation lowers technical and operational barriers for developers and contributors globally. This profound paradigm shift extends far beyond mobile environments; LLMs now underpin the generation of sophisticated fake product listings, deceptive AI plugins, and a wide array of synthetic content across various digital marketplaces, exposing systemic vulnerabilities in traditional review systems.

This paper has thoroughly explored the **dual-use nature** of LLMs and generative AI, presenting both the immense opportunities for innovation and the escalating risks they introduce in the digital landscape. We've detailed how malicious actors weaponize LLMs to scale abuse, fraud, and non-compliance through tactics like polymorphic malware generation, deceptive storefront content, automated policy circumvention, and hyper-personalized social engineering attacks. Crucially, we've outlined a comprehensive roadmap for how platforms can effectively counter these evolving threats with **AI-augmented defenses**. This includes advanced capabilities such as semantic code analysis for hidden threats, multimodal cross-validation of storefront claims against observed behavior, intelligent content moderation, and federated compliance auditing against complex global regulations.

The key contributions of the end-to-end architecture and strategic playbook presented in this paper are summarized in Table 28.

*Table 28. Key Contributions of the Proposed End-to-End Architecture and Strategic Playbook*

| Component | Why It's Groundbreaking (Concise) |
|---|---|
| Comprehensive Threat Modeling | Systematically maps novel LLM-driven abuse vectors (e.g., polymorphic malware, synthetic content, AI social engineering, policy evasion). |
| Proactive Defensive Strategy | Reframes LLMs as proactive defensive tools for scalable integrity, shifting security from reactive to proactive. |
| Operational System Architecture | Designs a scalable human-AI hybrid system with intelligent triage, multimodal validation, SDK indexing, automated compliance, and transparent feedback. |
| Global Compliance Framework | Introduces tech for dynamic global compliance, including AI legal "diff engines," audit trails, and zero-shot mapping for evolving laws. |
| Cross-Functional Integration | Provides a practical blueprint for integrating key teams (product, engineering, safety, legal, policy) for holistic, adaptive defense. |
| Economic & Trust Impact | Shows LLM integrity solutions cut costs, speed reviews, reduce fraud, and significantly boost user trust and developer satisfaction. |
| Domain-Specific Solutions | Offers adaptable integrity workflows and insights for diverse digital sectors (app stores, e-commerce, AI hubs, social media). |
| Forward-Looking Research Agenda | Articulates actionable research in XAI, federated/on-device moderation, adversarial testing, developer education, and global standards. |
| Human-Centric AI Governance | Emphasizes human oversight and ethics to manage false positives, mitigate algorithmic bias, and ensure fairness in AI-driven decisions. |

Beyond digital platforms, this framework uniquely extends its comprehensive integrity blueprint to high-stakes clinical diagnostics and personalized medicine. We detailed how advanced LLMs, coupled with multimodal AI, can bridge the critical "interpretation gap" in diagnosis by seamlessly mapping patient narratives to objective biomedical data, from subjective symptoms to advanced imaging biomarkers. This integration facilitates more accurate and efficient diagnostic suggestions, always under rigorous physician-in-the-loop oversight. Crucially, the paper introduced the burgeoning concept of digital twins for personalized medicine, envisioning dynamic simulations of disease progression and treatment responses to optimize therapeutic strategies and accelerate drug discovery. This expansion underscores the framework's adaptability and paramount importance in ensuring the trustworthiness, safety, and regulatory compliance of AI in critical sectors impacting human life, leveraging the same principles of explainability, auditability, and bias mitigation established for broader digital ecosystems.

Through in-depth case studies of industry leaders like Google Play and Apple App Store, we demonstrated that LLM-powered safety systems are not merely theoretical but are already actively in production, significantly shaping the next generation of platform trust infrastructure. These initiatives highlight the practical viability of AI-assisted review, real-time threat detection, and proactive developer engagement. Our discussion identified critical emerging opportunities in explainable AI (XAI) for transparent decision-making, privacy-preserving federated review pipelines, and adaptive multi-agent compliance parsing. We've argued that responsible, transparent, and continuously adaptive deployment of LLMs, coupled with robust human oversight, can empower platforms to scale their enforcement capabilities, proactively counter evolving threats, and preserve trust across rapidly advancing digital ecosystems.

To succeed and thrive in this LLM-driven era, platforms must adopt a multi-faceted strategic approach:

**Integrate LLMs deeply** across every stage of content intake, review, moderation, compliance, and post-deployment monitoring lifecycles, moving beyond ad-hoc applications to systemic adoption.

**Align product, engineering, Trust & Safety operations, policy, and legal teams** within cohesive **cross-functional platform integrity frameworks.** This fosters a culture of shared responsibility and enables a holistic, proactive defense posture.

**Rigorously measure performance** through a comprehensive suite of operational metrics, focusing not only on efficiency but also on precision, recall, false positive/negative rates, transparency, developer feedback, and regulatory alignment to ensure fairness and continuous improvement.

**Invest continuously in explainability (XAI)** to foster developer and user trust, **adaptive threat detection** to stay ahead in the AI arms race, and a **global governance strategy** that scales across diverse AI-enabled marketplaces and evolving regulatory landscapes.

To our knowledge, this paper provides the first comprehensive strategy roadmap that unifies insights from mobile app stores, Gen-AI marketplaces, and digital commerce platforms under a common framework for LLM-augmented integrity enforcement. By integrating nuanced understanding of LLM capabilities, real-world case studies, and a deep appreciation for the complex regulatory environment, this paper bridges technical and organizational silos to propose an operational blueprint for the AI era. This synthesis across app stores, Gen-AI platforms, and digital commerce represents a new paradigm in platform integrity. As AI-generated content and LLM-assisted development continue to scale, so too must the safeguards that govern them. This paper provides not just a technical foundation, but an operational call to action—urging platforms, researchers, and regulators to adopt a shared, adaptive, and cross-domain response to platform safety and human well-being in the age of generative AI.

As LLMs continue to reshape how content, apps, and digital interactions are created, reviewed, and regulated, platforms that proactively invest in scalable, explainable, and cross-functionally integrated safety architectures will be best positioned to thrive in a rapidly evolving trust landscape, ensuring long-term sustainability and user confidence.

## DISCLAIMERS

Author contributions were made in a personal capacity and do not necessarily reflect the views or positions of their employers.

All logos, trademarks, and brand names are the property of their respective owners. Their use in this paper is solely for identification and educational purposes and does not imply endorsement.

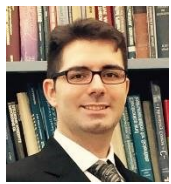

**Dr. Kiarash Ahi** holds M.Sc. and Ph.D. degrees in Electrical and Computer Engineering from Leibniz University Hannover (Germany) and the University of Connecticut (USA), respectively. He is a pioneering scientist, 0→1 product leader, and serial founder with deep expertise in AI, cybersecurity, large language models (LLMs), Generative AI (GenAI), GPU computing, HPC architectures, edge AI, biomedical engineering, digital signal and image processing, and intelligent system design.

Since 2019, Dr. Ahi has led the end-to-end product strategy for SEMSuite™, Siemens' AI-powered analytics platform. He orchestrated global cross-functional teams to deliver scalable, UX-optimized AI tools—including LLM-powered Raw Data Filtering (RDF), Contour Data Flow (CDF), Calibre GenAI Pattern Generator (CPG), and CMi—driving multi-million-dollar revenue and earning multiple performance awards.

Dr. Ahi holds over 10 patents, has published more than 50 peer-reviewed papers, and his work has been cited over 2,500 times. He has also developed 10+ creative AI applications for iOS, Android, and macOS, reaching over one million global users.

He is the recipient of the IEEE AI 1st Place Award and a Top Peer Reviewer with 200+ reviews for journals such as Nature, IEEE, Springer, and Elsevier. As an invited IEEE speaker, he has presented on GenAI, LLMs, cybersecurity, platform integrity, review automation, and advanced imaging systems.

A recognized thought leader in AI ethics and governance, Dr. Ahi advocates for responsible AI deployment, data privacy, and regulatory alignment—shaping the future of digital trust and platform safety. He has co-advised several PhD students and collaborates across academia and industry to push the boundaries of applied AI.

# Table of Contents (Detailed)